\documentclass[11pt]{article}

\usepackage[a4paper,margin=2.5cm]{geometry}
\usepackage[T1]{fontenc}
\usepackage[utf8]{inputenc}
\usepackage{lmodern}
\usepackage{microtype}
\usepackage{amsmath,amssymb}
\usepackage{bm}
\usepackage{graphicx}
\usepackage{overpic}
\usepackage{siunitx}
\usepackage{booktabs}
\usepackage{setspace}
\usepackage[labelfont=bf,font=small]{caption}
\usepackage[section]{placeins}
\usepackage[hidelinks]{hyperref}
\usepackage{pdfpages}

\graphicspath{{figures/}}
\title{\bfseries Layered matter that maintains spacing\\ but loses stacking order}

\author{%
Osvaldo Trigueiro Neto$^{1}$, Paulo Henrique Michels-Brito$^{1}$,\\
Bruno Ceccato Telli$^{1,2}$, Natalie Jasmin Eichstaedt$^{3}$,\\
Leide P.\ Cavalcanti$^{4}$, Tom\'as S.\ Plivelic$^{5}$,\\
Josef Breu$^{3}$, Jon Otto Fossum$^{1,3,6,7,\ast}$\\[0.6em]
\footnotesize $^{1}$Department of Physics, Norwegian University of Science and Technology (NTNU), Trondheim, Norway\\
\footnotesize $^{2}$School of Chemical Engineering, University of Campinas (UNICAMP), Campinas, Brazil\\
\footnotesize $^{3}$Department of Chemistry, University of Bayreuth, Bayreuth, Germany\\
\footnotesize $^{4}$ISIS Neutron and Muon Source, STFC, Rutherford Appleton Laboratory, Didcot, United Kingdom\\
\footnotesize $^{5}$MAX IV Laboratory, Lund University, Lund, Sweden\\
\footnotesize $^{6}$SCML International, Oslo, Norway\\
\footnotesize $^{7}$Laboratoire de Physique et M\'ecanique des Milieux H\'et\'erog\`enes (PMMH), ESPCI Paris, Paris, France\\[0.3em]
\small $^{\ast}$Corresponding authors: Jon Otto Fossum (jo@scml.no);\\
\small Osvaldo Trigueiro Neto (osvaldo.t.neto@ntnu.no)%
}
\date{}

\begin{document}
\maketitle

\begin{abstract}
\noindent
In layered materials, spacing and stacking-order extent are usually locked. Here we show that in swollen suspensions of stiff, charged nanosheets they decouple, and that this defines a distinct regime, apart from the crystalline- and Wigner-swelling regimes such systems usually occupy. The mean spacing stays sharp and salinity-tunable while scattering-weighted stacking spans only two to three layers. We demonstrate this in a near-perfect model material, so the behaviour is intrinsic, not defect-driven. X-ray and neutron scattering, sedimentation and a Donnan analysis show the spacing is held by a parameter-free osmotic restoring slope below one pascal per nanometre. Because the slope is so weak, the spacing sits at equilibrium while faults relax slowly, a quenched metastable registry whose ageing-like relaxation of low-dimensional periodic order has not, to our knowledge, been realised before. The same decoupling is expected across stiff, swollen nanosheets, from clays to oxide nanosheets and graphene oxide.
\end{abstract}

\section{Introduction}\label{sec:intro}

The behaviour of a layered material depends on two independent aspects of how its neighbouring layers stack: the spacing between them, and how far their stacking order (the registry of successive layers) extends. Different functions draw on different aspects of this arrangement. A well-defined local spacing sets optical response, structural colour as in a soap film or opal when it approaches the wavelength of light, together with dielectric behaviour and spacing-selective transport and barrier properties. The extent of stacking order instead governs mechanical reinforcement and long-range electronic transport. Which functions a given material can deliver therefore depends on which of these two it can independently control. Lamellar soft matter runs from surfactant and lipid membranes to colloidal clay and delaminated nanosheets. It is built from quasi-two-dimensional units separated by fluid interlayers, and its behaviour is governed by the interactions between adjacent layers\cite{Kronberg2003,Katsaras2013,Coleman2011,Davidson2004,Michot2006}. Many of these systems swell to large separations, with charged inorganic nanosheets reaching hundreds of nanometres\cite{Geng2013,TrigueiroNeto2025}.

In classical lamellar phases such as lipid stacks and surfactant membranes, thermal fluctuations and the fluctuation-induced Helfrich repulsion dominate\cite{Helfrich1978,Safinya1986PRL,Safinya1990PRL}. The bending moduli are of order \(k_{\mathrm B}T\), and the interlayer spacings lie in the nanometre to tens-of-nanometre range. This produces strong undulations and the quasi-long-range order described by Caill\'e correlations\cite{Caille1972,Nallet1993}. By the Landau--Peierls argument, such soft layered systems cannot sustain true long-range order along the stacking direction. Their positional order decays algebraically with distance\cite{AlsNielsen1980,deGennesProst1993}.

Stiff, strongly swollen nanosheet suspensions occupy a different part of this landscape. These span swelling clays, from fluorohectorite to vermiculite and beidellite\cite{Walker1960,Dudko2022,Pacakova2025,Hotton2025}, delaminated oxide and niobate nanosheets\cite{Geng2013,Osada2012,Kikuchi2025}, and swollen graphene oxide\cite{IakunkovTalyzin2020}, natural and synthetic alike. In clays, swelling proceeds through successive regimes. First comes a crystalline-swelling regime of discrete cation-hydration states. Next is a Wigner-crystal regime, in which electrostatic repulsion between the anionic sheets sets a salinity-independent spacing, evidenced by rational series of up to seven hydration layers\cite{Stevenson2026}. At the largest separations comes the osmotic-swelling regime studied here, where delaminating fluorohectorite near \SI{150}{\nano\meter} already shows lamellar peak broadening comparable to that reported below\cite{Rosenfeldt2016}. The nanosheets are stiff, with in-plane moduli of order \SI{e11}{\pascal}, about \SI{146}{\giga\pascal} for a single lamella and \SI{171}{\giga\pascal} for a double layer, the latter approaching bulk mica\cite{Kunz2013,Stoter2016}. Their bending rigidity therefore exceeds \(k_{\mathrm B}T\) by three to four orders of magnitude, so thermal undulations are suppressed. The interlayer spacing, meanwhile, is one to two orders of magnitude larger than in thermotropic systems, \(d \approx 130\)--\SI{300}{\nano\meter}. At these spacings the van der Waals and electrostatic double-layer forces are small\cite{Derjaguin1941,VerweyOverbeek1948,Israelachvili}, and the swollen state is held mainly by osmotic pressure balanced against weak mechanical confinement (Supplementary Sections~S5--S7). This weak-coupling situation is not specific to fluorohectorite, or even to clays. It arises whenever three conditions hold together: the sheets are stiff (bending rigidity far above \(k_{\mathrm B}T\)), they swell to large interlayer spacings (tens to hundreds of nanometres), and they are of high aspect ratio (micrometre-scale lateral size). Lipid and surfactant lamellae fall outside it because they are floppy, with bending rigidity of order \(k_{\mathrm B}T\); small platelets such as Laponite and montmorillonite fall outside it because their aspect ratio is low\cite{RuzickaZaccarelli2011,Segad2012}. Conductive 2D materials fall outside it for a different reason: pristine graphene is uncharged and does not swell osmotically, while in MXenes the interlayer spacing is set by cation intercalation and redox in a conductive host rather than by a clean Donnan balance\cite{Celerier2019}. Vermiculite, beidellite and large delaminated oxide nanosheets meet all three, so the same decoupling is expected there\cite{Dudko2022,Pacakova2025,Osada2012}; graphene oxide, which swells osmotically with an ionic-strength- and humidity-dependent spacing, is a further member of the class, though it swells to smaller spacings than the clays studied here\cite{IakunkovTalyzin2020}. Resolving it cleanly requires a layered material with low structural and electrostatic heterogeneity, so that the intrinsic behaviour is not masked by defects. Synthetic fluorohectorite provides such a platform. Unlike natural clays, it has exceptionally uniform charge and high crystalline purity\cite{Breu2001,Stoter2013,Kalo2010}.

In this swollen state the spacing is set by a Donnan balance. Classically this is the equilibrium across a semipermeable membrane that traps a charged species on one side\cite{Donnan1924}. Here there is no physical membrane. Its role is played by the immobile structural charge of the sheets, with the gallery--reservoir boundary acting as a virtual semipermeable membrane\cite{ElRifaii2022}. The mobile ions partition between the charged interlayer galleries and the external salt reservoir until their electrochemical potentials are equal. The resulting excess ion concentration inside the galleries produces an osmotic pressure that pushes the layers apart. Because this balance is governed by exchange with the reservoir, the swelling is tuned by the reservoir salinity as well as by the volume fraction.

The present work follows a sequence of studies on this material. An early study showed that fluorohectorite undergoes a temperature-induced swelling transition to large interlayer spacings\cite{Hansen2012}. Michels-Brito et al.\ reported bright structural colour in double-layer suspensions. They noted, qualitatively, that the colour shifts with salinity in the freshly prepared homogeneous state. That observation was not made quantitative, however, and was not connected to a specific underlying mechanism\cite{MichelsBrito2022}. El Rifaii et al.\ characterised the same family of clays at smaller interlayer spacings. There a combined DLVO and Donnan description is necessary, in which van der Waals and electrostatic double-layer contributions both enter alongside the osmotic term\cite{ElRifaii2022}. In that regime the stacking stays coupled and does not break up into the short domains found in the present, larger-spacing osmotically swollen state, and the restoring response opposing spacing mismatch was not quantified. Trigueiro Neto et al.\ subsequently used small-angle X-ray scattering across a volume-fraction series. They linked the structural colour to lamellar \(d\)-spacings and established the empirical relation \(d^{\ast} \propto \phi^{-x}\) between spacing and clay volume fraction (Supplementary Section~S2). Their peak widths already correspond to domains of only a few registered layers in the swollen state, essentially the same short domains reported here. However, they did not vary salinity or osmotic pressure, and did not quantify the mechanical restoring forces\cite{TrigueiroNeto2025}. The freshly prepared, homogeneous stock suspensions used here (Methods) already display a structural colour that varies with the prepared volume fraction. Because this colour is set by the interlayer spacing, it shows that a well-defined spacing is already established at preparation, before any sedimentation. A recent study of swollen beidellite notes few-layer stacking domains (about \num{2.5} registered layers) in that sister clay, without quantifying the restoring mechanism\cite{Hotton2025}. A further study clarified that optical and X-ray probes interrogate different platelet populations, and that wall anchoring extends over a finite penetration length near container walls\cite{Pacakova2026GeometricDomain}. These studies, taken together, establish that osmotically swollen smectite suspensions break up into domains of only a few registered layers. But none isolated the pure-Donnan regime by tuning reservoir salinity, none quantified the restoring slope opposing spacing mismatch, and none addressed the timescale over which stacking order recovers after a mechanical perturbation. The present work pins all three and traces them to a single cause.

We use salinity as the control variable here, at fixed volume fraction, to set the swollen state of fluorohectorite nanosheets and to quantify the forces that fix its stacking order. We combine small-angle X-ray scattering (SAXS), small-angle neutron scattering (SANS) and sedimentation with a pressure-based reconstruction of the swelling balance. We find that the mean interlayer spacing remains well defined and salinity-tunable. The stacking domains, however, stay below two to three layers. The osmotic restoring slope \(k(d)\) opposing spacing mismatch then follows from independent input parameters: the empirical swelling exponent (\(x = 1/\alpha\)), the measured Bragg spacing \(d\), and the sedimentation-constrained confinement pressure \(\Pi\). This gives the expression \(|k(d)| = \alpha\Pi/d\), which contains no adjustable parameters (\(k(d)\) is defined below as the curvature of the free-energy well; Methods). The slope turns out to be small. Once layers are knocked out of register they therefore heal only very slowly. In this regime, spacing selection and stacking-order recovery are effectively decoupled, with the slow recovery, we propose, proceeding as an athermal ageing of this stacking order.

\section{Results}\label{sec:results}

\subsection{Two nanosheet architectures from one parent clay}
\begin{figure}[p]
\centering
\captionsetup{font=footnotesize}
\includegraphics[width=0.72\linewidth]{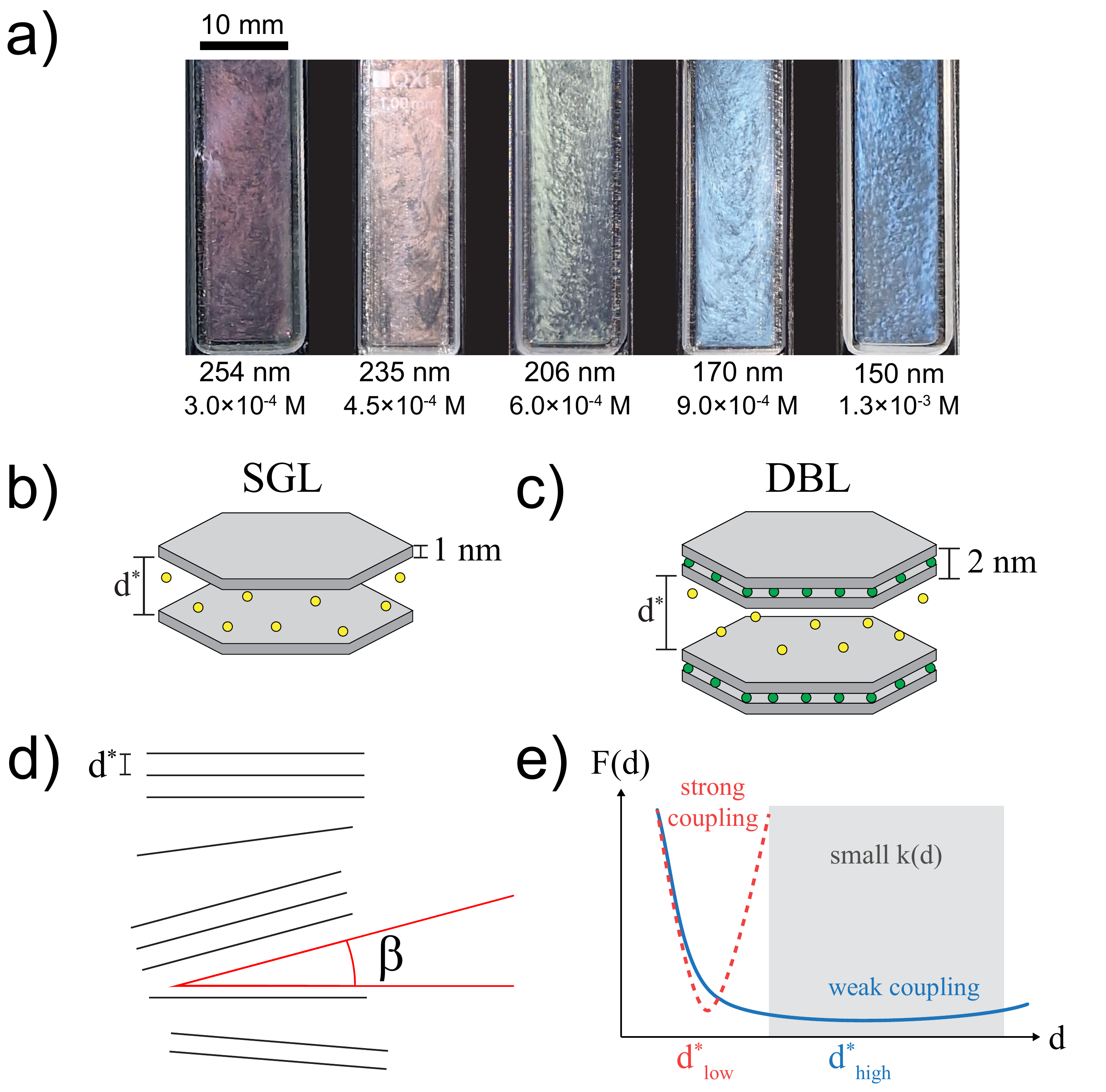}\\[0.5em]
\begin{overpic}[width=0.56\linewidth]{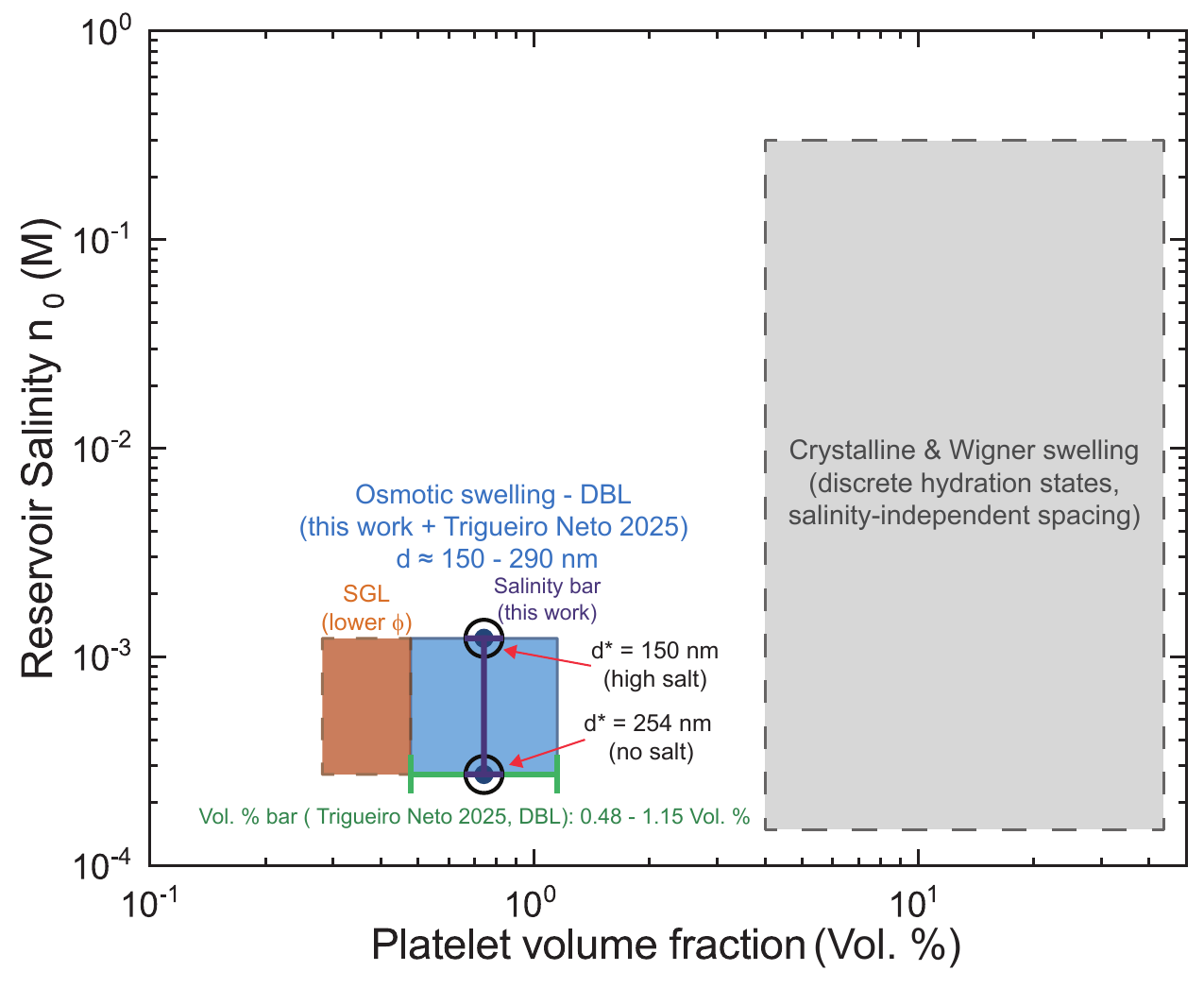}\put(-3,80){\textbf{\LARGE f)}}\end{overpic}
\caption{\textbf{The swollen, structurally coloured state, its building blocks, and where it sits in the nanosheet phase diagram.}
\textbf{a)} Freshly prepared DBL fluorohectorite suspensions (\SI{1.00}{\milli\meter} path) at five reservoir salinities \(n_0\), before sedimentation; structural colour shifts purple to blue as \(n_0\) rises and the mean spacing \(d^{\ast}\) falls (values beneath each cuvette). Same samples as SANS (Supplementary Fig.~S5); on standing they sediment into a vertical colour gradient (Fig.~\ref{fig:fig4}, Supplementary Fig.~S7). Scale bar, \SI{10}{\milli\meter}. \textbf{b,c)} SGL and DBL building blocks from the same parent clay. \textbf{d)} Real-space schematic: short, finite domains around a pressure-selected spacing \(d^{\ast}\) (singlets, dimers, trimers and short stacks; Supplementary Fig.~S17); each interruption is a domain wall, and neighbouring domains are misaligned by an angle \(\beta\) (mosaicity). \textbf{e)} Restoring landscapes, from strong coupling under stronger confinement to a swollen weak-coupling regime where \(d^{\ast}\) sits in a broad, shallow, asymmetric minimum (steep osmotic wall under compression; pressure decaying as \(d^{-\alpha}\) under expansion), giving a small restoring slope. \textbf{f)} Control plane spanned by platelet volume fraction \(\phi_{\mathrm{prep}}\) and reservoir salinity \(n_0\). The swollen sector studied here (blue, \(d\approx150\)--\SI{290}{\nano\meter}) is traversed along salinity in this work (purple bar) and along loading by Trigueiro Neto \emph{et al.}\cite{TrigueiroNeto2025} (green bar); SGL nanosheets swell at lower loading (dashed box). It lies apart from the crystalline- and Wigner-swelling regimes (grey), defining a distinct weak-coupling regime in which a sharp, salinity-set spacing coexists with a short, slowly-relaxing stacking registry. Ringed points: no-salt (\(d^{\ast}=\SI{254}{\nano\meter}\)) and high-salt (\(d^{\ast}=\SI{150}{\nano\meter}\)) samples whose wells are reconstructed in Supplementary Fig.~S16.}
\label{fig:fig1}
\end{figure}

We study two lamellar building blocks derived from the same parent fluorohectorite, shown schematically in Fig.~\ref{fig:fig1}b,c. Single-layer (SGL) nanosheets are obtained by osmotic delamination\cite{GardolinskiLagaly2005} into isolated silicate sheets separated by hydrated interlayers. Double-layer (DBL) nanosheets instead retain a non-swelling caesium interlayer that pairs two sheets into a composite unit\cite{TrigueiroNeto2025,Stoter2016,MichelsBrito2022}. The two architectures share the same composition. Only the effective stacking unit differs (Methods). The nanosheets have an average lateral size of about \SI{10}{\micro\meter}, giving a high aspect ratio. The structural layer charge of the parent clay is fixed by composition at \num{0.5} per formula unit. Pairing in the DBL system lowers the charge presented to the surrounding electrolyte, through a mechanism analogous to that established for rectorite\cite{Moller2010} and recently demonstrated for ammonium-paired double layers\cite{Uhlig2026}. Electrophoresis bears this out: the external surfaces of the paired DBL units carry a lower-magnitude effective zeta potential than single sheets (Methods).

\subsection{Finite stacking domains in the swollen state}
\begin{figure}[t]
\centering
\includegraphics[width=0.92\linewidth]{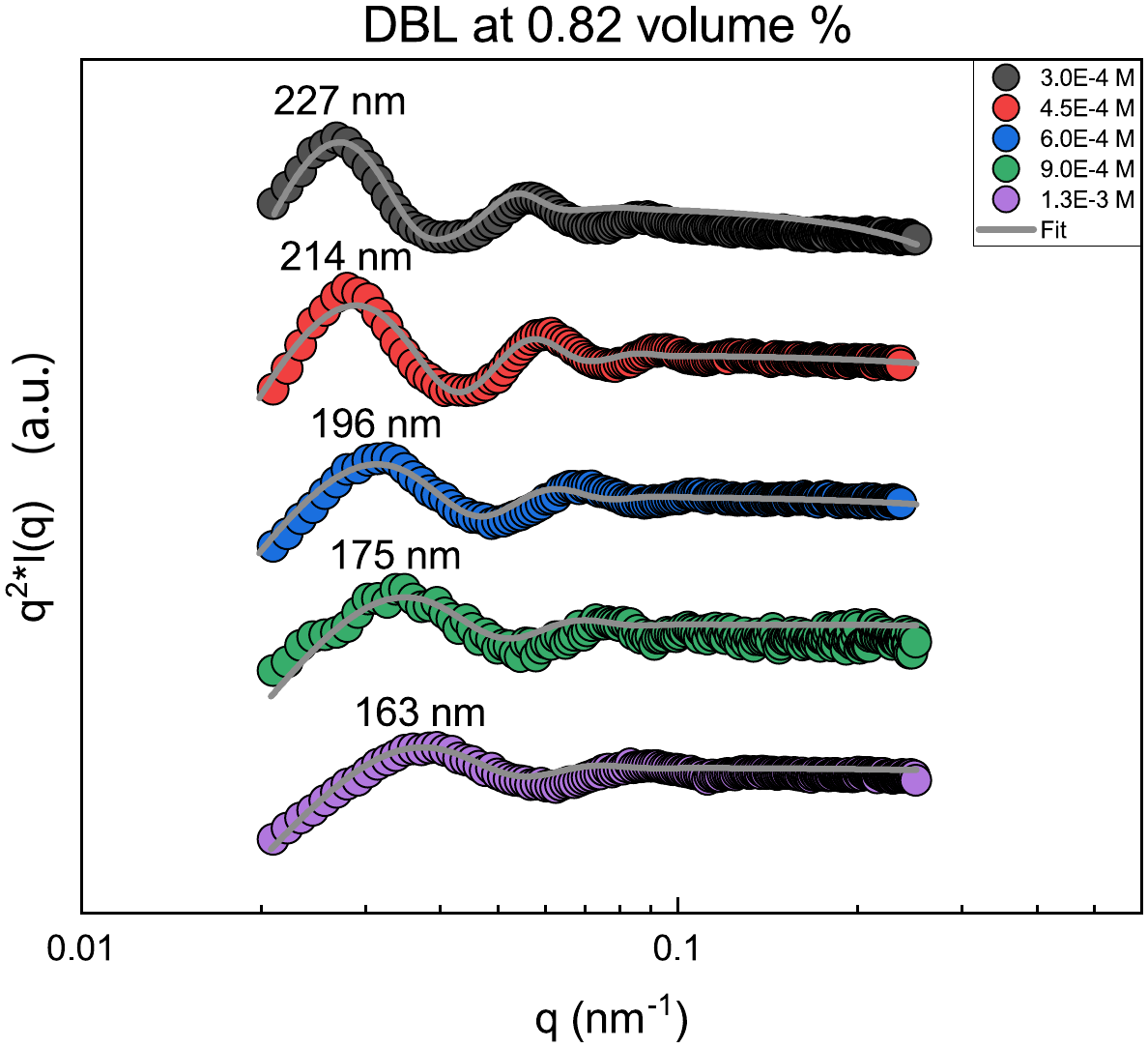}
\caption{\textbf{Scattering signature of finite-domain lamellae.}
Representative small-angle X-ray scattering profiles, plotted as \(q^{2}I(q)\) versus \(q\) for visibility of the lamellar features, for swollen fluorohectorite DBL suspensions at \(\phi = \SI{0.82}{\percent}\) by volume across five reservoir salinities (\SI{3.0e-4}{}, \SI{4.5e-4}{}, \SI{6.0e-4}{}, \SI{9.0e-4}{} and \SI{1.3e-3}{\mole\per\liter}). Each profile shows a pronounced first-order lamellar peak together with weak, broad higher-order harmonics, indicating short stacking domains at a well-defined spacing that shifts with salinity. The mean interlayer spacing \(d^{\ast} = 2\pi/q_{\mathrm{peak}}\) is shown beside each curve, decreasing from \SI{227}{\nano\meter} to \SI{163}{\nano\meter} with increasing salinity. The relative spacing disorder \(\Delta d/d\) and the finite stacking-domain size follow from the peak width, fitted with a lamellar model (grey lines; Methods). The instrumental \(q\)-resolution at CoSAXS is well below \SI{1}{\percent}, far narrower than the intrinsic peak width, so the widths are sample-limited and provide the primary evidence for finite stacking domains. Full SAXS and SANS datasets and fits are in Supplementary Figs.~S4 and~S5.}
\label{fig:fig2}
\end{figure}

In the strongly swollen state the lamellae do not form extended stacks. As illustrated in Fig.~\ref{fig:fig1}d,e, they organise into finite domains a few layers thick, with substantial interlayer displacement disorder. Neighbouring domains are mutually misaligned by an angle \(\beta\) of typically \SIrange{5}{30}{\degree}. Direct evidence comes from SAXS on both architectures (Fig.~\ref{fig:fig2}), cross-validated by SANS on DBL samples through a different contrast mechanism. Full datasets are in Supplementary Figs.~S4 and~S5. The profiles are dominated by a first-order lamellar peak, while higher-order harmonics are weak and broad.

We separate two quantities from the peak shape. The peak \emph{position} gives the mean interlayer spacing, \(d^{\ast} = 2\pi/q_{\mathrm{peak}}\), which is sharply defined. The peak \emph{width} decomposes into the two contributions familiar from powder diffraction: a microstrain part, here the relative spacing disorder \(\Delta d/d\), and a finite-size part, here the stacking-domain size, the number of consecutive layers that remain aligned. These are two distinct kinds of imperfection: a \emph{distortion} of the spacing (\(\Delta d/d\)) and a \emph{breaking} of the stacking (finite domains bounded by stacking faults). Keeping them separate is central to what follows. Fitting with a lamellar model\cite{Forster2010} gives \(\Delta d/d \approx 0.2\)--\num{0.3} and domains averaging only two to three registered layers (the fitted quantity is the stacking-domain thickness in units of the spacing, \(L_{\mathrm{dom}}/d^{\ast} = N-1 \approx 1.3\)--\num{2.3} interlayers; throughout, the quoted layer count is \(N\) = interlayers + 1; Methods). Although the two contributions broaden the peak jointly, hard interference bounds pin them to this corner of the \(N\)--\(\Delta d/d\) space rather than to a fit-statistics ridge (Methods; Supplementary Table~S2; Supplementary Section~S3). Because coherent scattering is weighted towards the larger domains in the population, this value is a scattering-weighted average rather than a statement that every region is two to three layers thick (Supplementary Section~S10). The number distribution behind this average is bottom-heavy: registry breaking with per-junction probability \(p \approx 0.3\)--\num{0.5} gives a geometric stack-size distribution in which the singlet, a platelet carrying no gallery at all, is the single most probable species, and the number-averaged gallery count per object falls below one (Supplementary Sections~S3 and~S10). The colour-generating element is the gallery, and the minimal coloured object is therefore the dimer, whose single gallery sits at the pressure-selected spacing \(d^{\ast}\); the suspension is a sea of optically silent singlets punctuated by these short multilayer stacks, and sedimentation, by preferentially removing the gallery-free singlets, enriches the gallery-bearing fraction and brightens the colour (Supplementary Sections~S4 and~S11). The same domain size is obtained from SAXS and from SANS, which rely on different contrast mechanisms and instrumental resolutions, so it reflects a property of the sample rather than of one technique. The same finite domain size recurs across different synthesis batches and across samples prepared by different group members, indicating a robust and reproducible feature of the swollen state (Supplementary Section~S3).

\subsection{Salinity selects spacing but not stacking order}
\begin{figure}[t]
\centering
\includegraphics[width=0.85\linewidth]{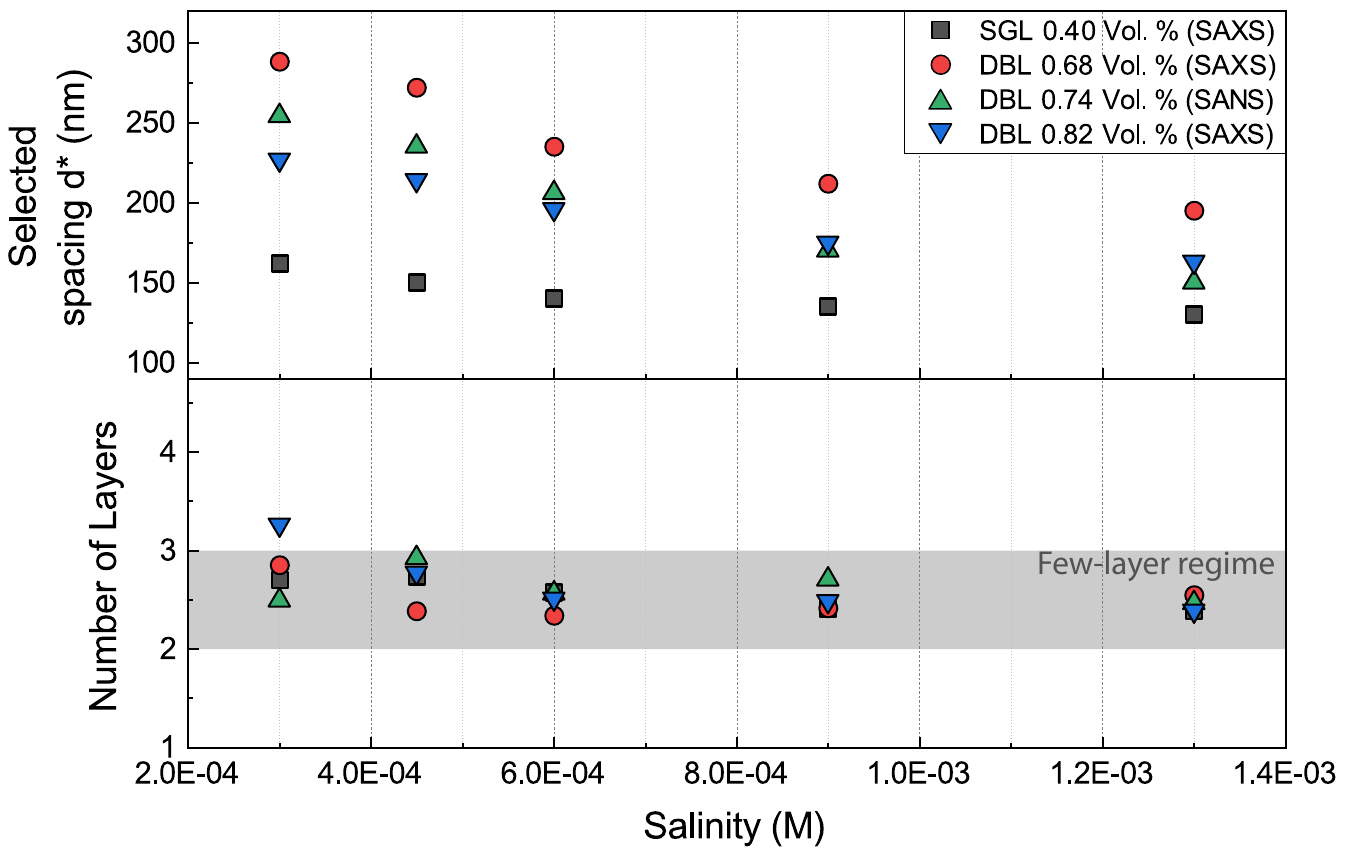}
\caption{\textbf{Salinity selects spacing but not stacking order.}
\textbf{Top}, the mean interlayer spacing \(d^{\ast}\) decreases monotonically with increasing reservoir salinity.
\textbf{Bottom}, the stacking-domain size, reported as the scattering-weighted number of registered layers (interlayers + 1; Methods), stays within the few-layer band (\(N\approx2\)--3, shaded) across the salinity range, while relative displacement distortions remain of order \(\Delta d/d \approx 0.2\)--\num{0.3} (the full \(\Delta d/d\) data are in Supplementary Fig.~S6). Symbols denote the four loadings spanning the SGL and DBL architectures (\num{0.40}, \num{0.68}, \num{0.74} and \SI{0.82}{\percent} by volume; see legend for the scattering method of each series). Error bars for the SAXS series are inside the plotted symbols. They include the spread across three independently prepared replicates (a triplicate; Methods): below \SI{2}{\percent} on \(d\) and on the number of registered layers, with the single-fit precision smaller (\(\approx\SI{1}{\percent}\)). SANS and reflectivity points are single measurements and are not replicated. Full parameter sets are in Supplementary Fig.~S6 and Tables~S2 and~S3.}
\label{fig:fig3}
\end{figure}

The mean spacing decreases monotonically with increasing reservoir salinity (Fig.~\ref{fig:fig3}, top). The lamellar repeat is therefore set by the external ionic conditions. The domain size shows no comparable trend: it stays limited to a few layers (scattering-weighted) across the salinity range, and the relative displacement disorder stays large throughout (Fig.~\ref{fig:fig3}, bottom; full parameters in Supplementary Fig.~S6 and Tables~S2 and~S3). Reservoir salinity therefore controls the spacing but does not extend the stacking domains. The spacing is established in the stock suspension during equilibration with the reservoir, before transfer into the measurement cell. The pressure-based analysis below should therefore be read as a reconstruction of the mechanical landscape consistent with the observed spacing, rather than as an independent prediction of it.

\subsection{Sedimentation sets the confinement-pressure scale}
\begin{figure}[t]
\centering
\includegraphics[width=0.92\linewidth]{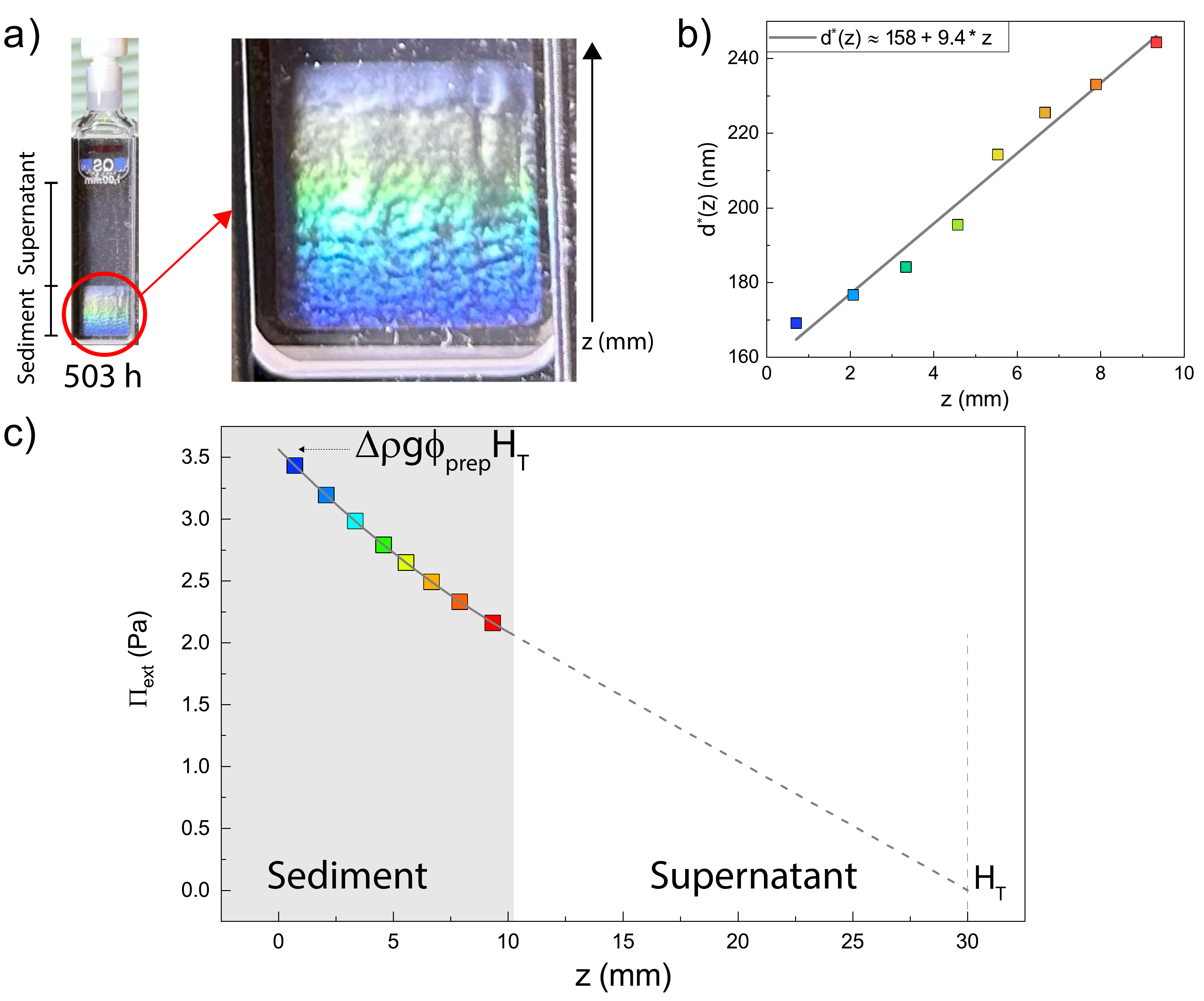}
\caption{\textbf{Sedimentation constrains the confinement-pressure scale.}
\textbf{a}, Photograph of a sedimented DBL suspension \SI{503}{\hour} after syringe injection, showing a vertical structural-colour gradient.
\textbf{b}, Reconstructed vertical profile of the pressure-selected spacing \(d^{\ast}(z)\). The linear fit shown in the panel is \(d^{\ast}(z) \approx \SI{158}{\nano\meter} + (\SI{9.4}{\nano\meter\per\milli\meter})\,z\), with \(z\) the height in the sediment.
\textbf{c}, Confinement-pressure profile inferred from the full vertical concentration profile, including the dilute supernatant above the sediment, of order a few pascal. The intercept \(\Delta\rho\,g\,\phi_{\mathrm{prep}}\,H_{\mathrm T}\) marked in the panel is the buoyancy pressure at the cell bottom. Here \(\Delta\rho\) is the nanosheet--medium density contrast, \(g\) the gravitational acceleration, \(\phi_{\mathrm{prep}}\) the prepared volume fraction and \(H_{\mathrm T}\) the total sample height. Reconstruction details are in Supplementary Fig.~S13 and Section~S4.}
\label{fig:fig4}
\end{figure}

To estimate the mechanical confinement acting on the swollen lamellae, we performed sedimentation experiments on DBL suspensions, using the structural colour that develops in stratified columns as a local probe of spacing\cite{Fossum2005,Ringdal2010}. The result (Fig.~\ref{fig:fig4}) places the relevant confinement-pressure scale at only a few pascal. From the colour gradient we extract a local reflected-wavelength profile by a simple RGB analysis of the imaged column (Supplementary Section~S4). We then convert it to a local spacing through the Bragg--Snell relation, infer the local clay volume fraction from the established spacing--concentration relation\cite{TrigueiroNeto2025}, and obtain the confinement-pressure profile from the buoyancy relation (Methods, Eq.~\eqref{eq:buoyancy}). Because suspended nanosheets remain in the supernatant above the visible sediment, this profile includes the full vertical concentration distribution. Sedimentation also fractionates larger multilayer stacks from singlets, so the local spacing--concentration relation is an effective one. The colour gradient should not be read as arising from confinement pressure alone: fractionation, the broad spacing distribution, and the finite-domain structure of the swollen state all shape the optical response (Supplementary Sections~S4 and~S11).

\subsection{A sub-pascal restoring slope}
At the spacings studied here, the classical Derjaguin--Landau--Verwey--Overbeek (DLVO) contributions are negligible. Van der Waals attraction, electrostatic double-layer repulsion and fluctuation-induced Helfrich repulsion all decay rapidly with separation, and lie several orders of magnitude below the osmotic and confinement scales (Supplementary Sections~S5--S7 and Table~S1)\cite{Derjaguin1941,VerweyOverbeek1948,Israelachvili,Helfrich1978}. We therefore model the swollen state as Donnan osmotic swelling under weak external confinement\cite{ElRifaii2022}. The spacing dependence of the osmotic pressure enters through the effective osmotic charge \(\sigma_{\mathrm{eff}}\) and the prepared volume fraction \(\phi_{\mathrm{prep}}\) (Methods, Eq.~\eqref{eq:donnan}). At the inter-platelet separations probed here, \(d\approx150\)--300\,nm with Debye length \(\lambda_{\mathrm D}\approx 8\)--18\,nm and \(\kappa d\gg 1\), the system is deep in the asymptotic regime. There the effective charge saturates to a value set by the boundary-matching condition rather than by the bare structural charge\cite{Alexander1984,BocquetTrizacAubouy2002,TrizacBocquet2004}, and is, to good approximation, \(d\)-independent (at given salinity) over the swollen interval (Supplementary Section~S8).

\begin{figure}[t]
\centering
\includegraphics[width=0.92\linewidth]{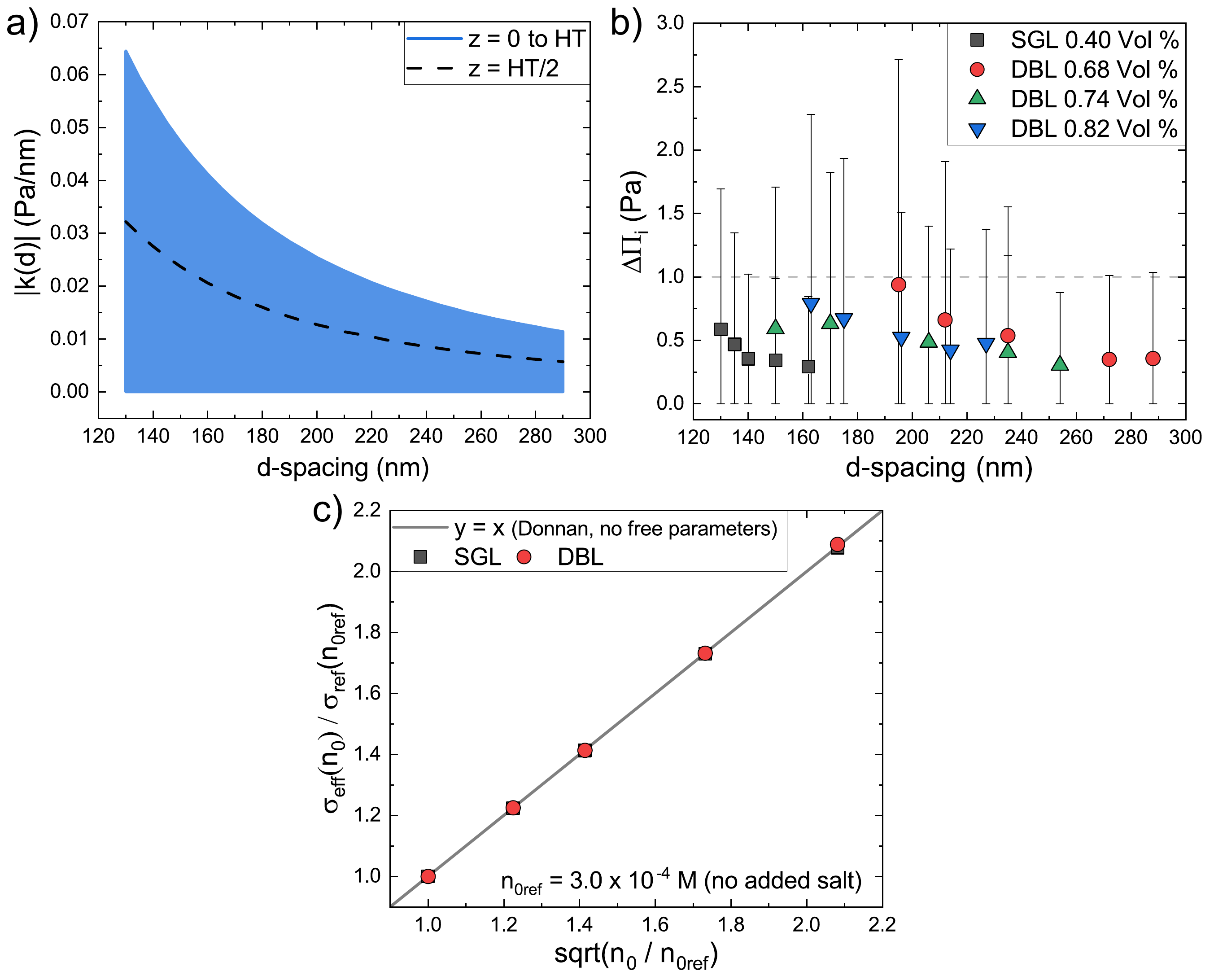}
\caption{\textbf{A sub-pascal restoring slope.}
\textbf{a}, Restoring slope at mechanical balance, \(|k(d,z)|\), from Eq.~\eqref{eq:kcompact_main} with the swelling relation of Sec.~S2 substituted for \(\phi_{\mathrm{prep}}(d)\). Each point is the curvature at the bottom of the well for a stack equilibrated at that spacing (the family of minima sketched in Fig.~\ref{fig:fig1}e), not the profile of a single well. Solid curve: mid-cell value (\(z=H_{\mathrm T}/2\)) computed for the DBL sedimentation column (\(\alpha=1.16\), \(A=176.1\,\mathrm{vol\%\,nm^{0.86}}\), taken as a dimensionless fraction in the pressure expressions; see Supplementary Section~S2, \(H_{\mathrm T}=\SI{30}{\milli\meter}\)); it is taken to represent all samples, since the mid-cell value \(H_{\mathrm T}/2\) varies little between them. Shaded band: buoyancy-controlled in-cell range, from the cell bottom (\(z=0\), upper edge) towards the meniscus (\(z\to H_{\mathrm T}\), lower edge).
\textbf{b}, Mid-cell restoring-pressure amplitude per sample, \(\Delta\Pi_i = \tfrac{1}{2}\alpha\,\Pi_{\mathrm{max}}(d_i^{\ast})\,(\Delta d/d)_i\) (Methods, Eq.~\eqref{eq:dPi}). Central markers: mid-cell value (\(z=H_{\mathrm T}/2\)). Error bars: buoyancy-controlled in-cell range, from near the meniscus (lower) to \(2\Delta\Pi_i\) at the cell bottom (upper).
\textbf{c}, Normalised effective osmotic charge \(\sigma_{\mathrm{eff}}(n_0)/\sigma_{\mathrm{eff}}(n_{\mathrm{ref}})\) against \(\sqrt{n_0/n_{\mathrm{ref}}}\), with \(n_{\mathrm{ref}}=3\times 10^{-4}~\mathrm{M}\) (the lowest reservoir salinity in the series, zero added salt). Grey line: \(y=x\). Solid markers: DBL cluster averages with within-cluster spread as error bars. Open markers: SGL (one stock per salinity). Absolute \(\sigma_{\mathrm{eff}}\) range across the series: \SIrange{0.003}{0.010}{\elementarycharge\per\nano\meter\squared} (Supplementary Fig.~S15 and Section~S8).}
\label{fig:fig5}
\end{figure}

The quantity that controls stacking stability is the restoring slope \(k(d)\), the spacing derivative of the net pressure \(\Pi_{\mathrm{osm}}-\Pi_{\mathrm{ext}}\) (Methods, Eq.~\eqref{eq:kslope}). At the equilibrium spacing \(d^{\ast}\) this net pressure vanishes (\(\Pi_{\mathrm{osm}}=\Pi_{\mathrm{ext}}\)). So \(k(d^{\ast})\) is not a force exerted at the minimum but the curvature of the free-energy well there (Fig.~\ref{fig:fig1}e), the rate at which a restoring pressure builds as the spacing is displaced \emph{away} from \(d^{\ast}\). Because SAXS and SANS measure Bragg peaks from the platelet packing, the Bragg spacing \(d^{\ast}\) is a local nanostructural quantity, the lamellar spacing of the platelet stack averaged over the scattering volume. \(k(d^{\ast})\) is correspondingly a local restoring slope, with \(\sigma_{\mathrm{eff}}\) held at the value set by the global Donnan balance with the reservoir (Supplementary Section~S8). Both \(\Pi_{\mathrm{ext}}\propto\phi_{\mathrm{prep}}\propto d^{-\alpha}\) and the quadratic Donnan form \(\Pi_{\mathrm{osm}}\propto(\sigma_{\mathrm{eff}}\phi_{\mathrm{prep}})^{2}\propto d^{-2\alpha}\) inherit their \(d\)-dependence from the empirical swelling relation, with \(\alpha\approx1.16\) the swelling exponent (the reciprocal of the measured \(d\)-versus-\(\phi\) exponent \(x\approx0.86\); Supplementary Section~S2). Differentiating each at fixed height and imposing mechanical balance \(\Pi_{\mathrm{osm}}=\Pi_{\mathrm{ext}}\equiv\Pi\), the two terms partially cancel and \(\sigma_{\mathrm{eff}}\) drops out, giving
\begin{equation}\label{eq:kcompact_main}
|k(d,z)| \;\approx\; \frac{\alpha\,\Pi}{d}.
\end{equation}
Equation~\eqref{eq:kcompact_main} contains no adjustable parameters. \(\sigma_{\mathrm{eff}}\) does not enter at all, and the remaining inputs are the swelling exponent \(\alpha\) measured independently by Trigueiro Neto et al.\cite{TrigueiroNeto2025}, the Bragg spacing \(d^{\ast}\), and \(\Pi\), itself fixed by sample geometry through the buoyancy term \(\Pi=\Delta\rho\,g\,\phi_{\mathrm{prep}}(H_{\mathrm T}-z)\) (Methods). The power-law form of the swelling relation is the general feature. The exponent \(\alpha\) is not a universal exponent but an effective, system-specific one, set by sheet stiffness, packing and container geometry. The small-\(|k|\) conclusion is insensitive to its precise value, because \(|k|\) scales linearly with \(\alpha\) and \(\alpha=1\) is the lower bound (the ideal dilution case; Supplementary Section~S2). Substituting the empirical swelling relation \(\phi_{\mathrm{prep}}(d)\propto d^{-\alpha}\) makes the full \(d\)-dependence explicit, \(|k(d,z)|\propto d^{-(\alpha+1)}\), or \(d^{-2.16}\) at the measured \(\alpha\) (Methods). This is steeper than the bare \(1/d\) of Eq.~\eqref{eq:kcompact_main}, because \(\Pi\) itself decreases with \(d\) through the volume-fraction term. This decrease of \(|k|\) with spacing is the quantitative counterpart of the schematic well in Fig.~\ref{fig:fig1}e. Weaker confinement at larger spacing gives a smaller curvature at the well bottom, the broad, shallow minimum of the weak-coupling regime, while stronger confinement at smaller spacing gives the deeper, narrower minimum. Near any one minimum the energy is harmonic to leading order, \(F\approx\tfrac{1}{2}\,k(d^{\ast})\,(d-d^{\ast})^{2}\), with \(d^{\ast}\) the location of the minimum and \(\Delta d = d-d^{\ast}\) the displacement within the well. Fig.~\ref{fig:fig5}a accordingly plots the bottom curvature \(k(d^{\ast})\) as a function of where the minimum sits, across the family of equilibria, not the profile of a single well. The quantity \(|k|\) is thus the local curvature at the bottom of each minimum, and a single harmonic slope captures only the response close to \(d^{\ast}\). The full well around a given minimum is markedly asymmetric, because \(\Pi_{\mathrm{osm}}\propto d^{-2\alpha}\) rises steeply against compression while \(\Pi_{\mathrm{ext}}\propto d^{-\alpha}\) and the net restoring pressure both decay toward larger spacings. The classical DLVO double-layer repulsion adds to the compression wall at small spacings, while remaining negligible near \(d^{\ast}\) (Supplementary Section~S8 and Supplementary Fig.~S16).

Equation~\eqref{eq:kcompact_main} carries a linear \(z\)-dependence through the buoyancy term \((H_{\mathrm T}-z)\). In the pre-sedimentation homogeneous suspension \(d\) is the same everywhere, but \(|k|\) is maximal at the cell bottom (\(z=0\), where \(\Pi=\Pi_{\mathrm{max}}=\Delta\rho\,g\,\phi_{\mathrm{prep}}H_{\mathrm T}\)) and the buoyancy contribution vanishes towards the meniscus (\(z\to H_{\mathrm T}\)). The buoyancy framework describes the bulk of the sample column. The meniscus boundary layer itself carries additional capillary, anchoring and surface-tension contributions, characterised previously in this family of charged-platelet suspensions\cite{MichelsBrito2021AirWater,Hansen2013Gradient,Pacakova2026GeometricDomain,Hemmen2009INinterface,Ringdal2010}, that lie outside the gravitational picture and are not modelled here (Supplementary Section~S8). The shaded band in Fig.~\ref{fig:fig5}a accordingly spans the buoyancy-controlled in-cell range, with the meniscus end marking where the buoyancy contribution becomes small rather than where \(|k|\) literally vanishes. We report \(\Pi\) at the representative mid-cell value (\(z=H_{\mathrm T}/2\), \(\Pi=\Pi_{\mathrm{max}}/2\)). The smallness of \(|k(d,z)|\) follows from the smallness of the gravitational prefactor \(\alpha\,\Delta\rho\,g\,A\,(H_{\mathrm T}-z)\) itself. For the dilute, large-spacing colloidal suspensions studied here, \(\Pi_{\mathrm{max}}\) is at most a few pascal (Fig.~\ref{fig:fig4}c) and \(d\) is in the hundred-nanometre range, so \(|k|\) is sub-pascal-per-nanometre across the accessible bulk range. With \(\Pi\approx\SI{2}{\pascal}\) and \(d^{\ast}\approx\SI{200}{\nano\meter}\), Eq.~\eqref{eq:kcompact_main} gives \(|k|\approx\SI{0.012}{\pascal\per\nano\meter}\), rising only to \(\approx\SI{0.025}{\pascal\per\nano\meter}\) at the cell-bottom maximum (Fig.~\ref{fig:fig5}a; full derivation in Supplementary Section~S8).

The reconstructed \(\sigma_{\mathrm{eff}}\) sits in the range \SIrange{0.003}{0.010}{\elementarycharge\per\nano\meter\squared} across the SGL and DBL samples and the salinity series. This is about two orders of magnitude below the bare structural charge of fluorohectorite (\(\sim\)\SI{1}{\elementarycharge\per\nano\meter\squared}, from the Mg-substitution stoichiometry), consistent with the strongly renormalised effective charge expected in the asymptotic-saturation regime\cite{Alexander1984,BocquetTrizacAubouy2002,TrizacBocquet2004}, where the screening length \(\lambda_{\mathrm D}\) sets the effective-charge scale rather than the bare charge itself. The Trizac--Bocquet--Aubouy prediction for an isolated planar surface, \(\sigma_{\mathrm{eff}}^{\mathrm{sat}}\approx 4\varepsilon\varepsilon_0 k_{\mathrm B}T\kappa/e\), evaluates to \SIrange{0.025}{0.053}{\elementarycharge\per\nano\meter\squared} across our reservoir salinities, within a factor of 3--5 of the reconstructed values. That offset is consistent with the two being different operational charges (an osmotic Donnan charge here, a far-field charge for an isolated plate), together with reconstruction-prefactor latitude, rather than with double-layer overlap, which is small at our \(\kappa d\) (Supplementary Section~S8). Colloidal-probe measurements on related fluorohectorites likewise show that the diffuse-layer potential does not scale with the stoichiometric charge\cite{Kuznetsov2020DL}. Under the parameter-free normalisation \(\sigma_{\mathrm{eff}}(n_0)/\sigma_{\mathrm{eff}}(n_{\mathrm{ref}}) = \sqrt{n_0/n_{\mathrm{ref}}}\), SGL and DBL collapse onto the same \(\sqrt{n_0}\) line (Fig.~\ref{fig:fig5}c), as the Donnan balance predicts when the product \(\sigma_{\mathrm{eff}}\,\phi_{\mathrm{prep}}\) is pinned to the reservoir. The modest reduction (\(\sim\)\SI{10}{\percent}) in DBL diffuse-layer potential relative to SGL, from Cs-intercalation of the non-swelling inner interlayer (zeta-potential values in Methods), sits well inside the regime where both architectures saturate to similar effective charge densities. The compact result Eq.~\eqref{eq:kcompact_main} is independent of this magnitude, because \(\sigma_{\mathrm{eff}}\) drops out at mechanical balance. Fig.~\ref{fig:fig5}c is therefore a self-consistency visualisation of the Donnan picture, not an input to \(k(d)\) (Supplementary Section~S8).

The same compact form gives the restoring pressure generated by the observed distortions. For each sample, evaluated at mid-cell (\(z=H_{\mathrm T}/2\)), the restoring-pressure amplitude is \(\Delta\Pi_i \approx |k(d_i^{\ast},H_{\mathrm T}/2)|\,\Delta d_i = \tfrac{1}{2}\alpha\,\Pi_{\mathrm{max}}(d_i^{\ast})\,(\Delta d/d)_i\) (Methods, Eq.~\eqref{eq:dPi}). This stays of order \SI{1}{\pascal} or below across the series (Fig.~\ref{fig:fig5}b), even though the relative spacing distortions are large, \(\Delta d/d \approx 0.2\)--\num{0.3} (up to \num{0.36} in the most compressed sample). Even substantial spacing distortions therefore generate only of-order-pascal restoring pressures.

\subsection{Weak coupling and slow rearrangement}
The mean spacing is first set in the stock suspension during gentle equilibration, where the weakly coupled lamellae already organise into finite domains. Transfer into the measurement cell by syringe injection adds transient shear and further stacking faults (domain walls) on top of this state. Because the restoring response is small, these mechanically introduced faults relax only slowly. The system therefore retains a fragmented, finite-domain configuration rather than annealing toward extended stacking order. A dimensionless comparison of the restoring-pressure amplitude with the mechanical loading stress, \(\Lambda \equiv \Delta\Pi/\tau\) with \(\Delta\Pi \lesssim \SI{1}{\pascal}\) and \(\tau \approx 10\)--\SI{100}{\pascal} (Supplementary Section~S9), gives \(\Lambda\) of order \numrange{e-2}{e-1}, that is \(\Lambda \ll 1\). This ratio is an operational loading diagnostic. The intrinsic softness of the landscape is set independently by the smallness of the bound on \(k(d)\) relative to the osmotic pressure (Supplementary Section~S8).

Consistent with slow relaxation, the sediment continues to brighten after its height has stabilised (Supplementary Fig.~S12 and Section~S4). This points to ongoing internal reorganisation on timescales longer than sedimentation itself. We use ``athermal'' in the sense established for mechanically driven disordered systems such as granular and soft glassy matter\cite{Bi2015AthermalMaterials,Baule2018Edwards}: thermal energy is negligible compared with the mechanical energies governing rearrangement. The configuration is therefore set by mechanical history rather than by thermal equilibration.

\section{Discussion}\label{sec:discussion}

Taken together, the measurements separate two properties that are usually locked together in a layered phase. One is where the layers sit. The other is how far their stacking order extends. In these suspensions the first is sharp and salinity-tunable. The second spans only two to three layers by scattering-weighted measure, while the number distribution is dominated by gallery-free singlets (Supplementary Section~S10). The slope that would restore stacking order is bounded at sub-pascal values. The osmotic balance selects a sharp mean spacing but does not enforce registration, so the stacking breaks into short, few-layer domains at faults that, once introduced, heal only slowly.

The retention of so many galleries at \(d^{\ast}\) needs no appeal to restoring dynamics: \(d^{\ast}\) is an equilibrium. A gallery at the pressure-selected spacing experiences no net force, so occupying \(d^{\ast}\) costs nothing, and nothing evicts it: the sheets are athermal (F1 in Supplementary Section~S12), and the only agitation they feel, Brownian kicks from the solvent bath, is bounded by equipartition within the well at \((k_{\mathrm B}T/|k|A)^{1/2} \approx 1\)--\SI{2}{\nano\meter} for the area-weighted stiffness \(K = |k|A\) of a \SIrange{10}{20}{\micro\meter} gallery, one per cent of \(d^{\ast}\). This is not a thermal width the system exhibits but the ceiling on what temperature could ever do to the spacing, the quantitative licence for treating it as athermal; temperature enters the system only through the thermalised counterions that power \(\Pi_{\mathrm{osm}}\) itself. The dimers, the minimal coloured objects, therefore persist at \(d^{\ast}\) not because they are held stiffly but because their single gallery occupies the equilibrium of the osmotic balance, and with thermal energy irrelevant and no drive, occupation is permanent until a mechanical event displaces it. The balance is moreover orientation-blind: the Donnan condition is a scalar relation between each gallery and the reservoir, and neither the mosaicity \(\beta\) nor the fate of neighbouring galleries enters \(\Pi_{\mathrm{osm}} = \Pi_{\mathrm{ext}}\), so domains at every orientation answer to the same reservoir, and the salinity-set spacing scale, and hence the reflected wavelength, is common to the orientationally disordered ensemble. The two-dimensional SAXS patterns show this directly: the lamellar scattering appears as sharp, tilted lobes at a common radial \(|q^{\ast}|\) (Supplementary Fig.~S4), the tilt registering the domain misalignment \(\beta\) while the shared radius fixes one spacing across orientations. The measured \(\Delta d/d\) sits inside this picture rather than against it: it is principally the quenched spread of equilibrium spacings between structurally inequivalent stacks (Supplementary Section~S12), each stack occupying its own equilibrium, with a smaller loading-induced component whose healing is priced by the sub-pascal slope. Maintaining and restoring are thus different things: holding an undisturbed gallery at its equilibrium spacing costs no restoring force at all; only the healing of displaced galleries is slow. What is destroyed and stays destroyed is the registry, which has no reservoir-anchored equilibrium value.

This decoupling is the behaviour expected for a soft colloidal smectic. Its layer-compression modulus is orders of magnitude softer than in thermotropic smectics. Even classical layered systems then support only quasi-long-range positional order along the stacking direction\cite{ColloidalSmectic2023,Petukhov2015}. Once mechanically introduced, quenched stacking faults are not healed by such a soft restoring landscape. The recovery time is correspondingly long, a relaxation time \(\tau\sim\zeta/|k|\). Here \(|k|\) is the sub-pascal restoring slope of Eq.~\eqref{eq:kcompact_main}, and \(\zeta\) is an effective hydrodynamic drag per unit area (the force per unit area per unit relative platelet velocity) for motion across the hundred-nanometre interlayer gaps. Because it is the restoring slope, not the mobility, that is small, this recovery is slow even for platelets freely suspended in water, where motion is otherwise unhindered. The standard statements that low-dimensional fluctuations suppress order are thermal, equilibrium results. This holds for the Mermin--Wagner theorem for continuous symmetries in two dimensions\cite{Mermin1966,Hohenberg1967,Illing2017}. It also holds for the layered Landau--Peierls counterpart, which renders smectic stacking only quasi-long-range even in three dimensions\cite{AlsNielsen1980,deGennesProst1993}. We invoke them only as analogies. The apt analogue here is instead the Imry--Ma picture of \emph{quenched} disorder destroying long-range order while preserving local periodicity\cite{ImryMa1975}. The present mechanism is athermal and history-dependent, and we make the distinction between these results explicit in Supplementary Section~S12. The slow rearrangement that continues after sedimentation has stopped is consistent with physical ageing. We interpret it as relaxation within a shallow, rugged landscape that would reshape as the stacks heal, lowering some barriers and raising others. The response would then be non-stationary and slow with the system's age, as in glassy and granular matter\cite{Bouchaud1990,Fisher1994}. Here ``ageing'' means that the relaxation slows as the system grows older, as it explores an increasingly rugged landscape. Such ageing has been studied in two contrasting settings. In low-dimensional ordered systems, for example two-dimensional crystals and smectics, it is driven by thermal fluctuations. In amorphous jammed matter, such as granular packings and soft glasses, it is athermal, governed by mechanical history rather than by temperature. The ageing we propose here would be entirely of the second, athermal kind: temperature plays no role, and the configuration is set by mechanical history alone. What is unusual is only its subject: it would act on a low-dimensional \emph{periodic} order, whereas athermal ageing has so far been reported only in amorphous matter. To our knowledge this combination, the athermal ageing of a low-dimensional \emph{periodic} order, has not previously been realised in a colloidal lamellar system of this kind. In Supplementary Section~S12 we set out a minimal one-dimensional elastic description of this regime and consider two candidate ageing scenarios that the present data cannot yet distinguish. In one, the stacking-order dynamics map onto trap-model ageing, with an athermal, self-generated noise in place of temperature, giving parameter-free power-law predictions such as two-time correlations that collapse onto an age-rescaled form rather than a single stationary decay. In the other, the stacks compact logarithmically, as in parking-lot and granular (Tetris) models\cite{TalbotTarjusViot2000,Caglioti1997Tetris}, so that an internal structural observable would instead relax as \(1/\ln t\). We work the trap-model mapping out in detail because the measured disorder reflects a rugged landscape of shallow metastable states, of the kind the trap model describes; this makes it a concrete, falsifiable example rather than a favoured outcome, and a long-baseline structural or two-time measurement would decide between the two. That this state relaxes rather than persisting indefinitely makes it no less well defined. Like a structural glass or an arrested colloidal gel, it is a long-lived non-equilibrium state, characterised on the timescale over which it is prepared and observed.

Our contribution therefore goes beyond the observation of finite stacking domains\cite{TrigueiroNeto2025,Hotton2025}. It is a quantitative reconstruction of the restoring landscape. It is also a clean realisation that connects spacing selection to its mechanical and osmotic origin. The restoring-slope expression Eq.~\eqref{eq:kcompact_main} follows from independently anchored inputs and contains no adjustable parameters. The effective charge \(\sigma_{\mathrm{eff}}\) drops out at mechanical balance. The conclusion that \(|k|\) is small therefore does not depend on the Donnan reconstruction (Supplementary Section~S8). The reconstructed \(\sigma_{\mathrm{eff}}\) magnitudes are nonetheless in the range expected from asymptotic-saturation theory at our screening length, about two orders of magnitude below the bare structural charge of fluorohectorite (Supplementary Section~S8). The broader control space of repulsive nanosheet suspensions\cite{vanderKooij2000} is spanned by platelet loading and reservoir salinity. It is this decoupling that defines the swollen state studied here as a distinct regime within that plane (Fig.~\ref{fig:fig1}f): a regime of soft layered matter that is periodically ordered yet mechanically out of register. A sharply defined, salinity-tunable spacing coexists with stacking confined to a few layers (scattering-weighted), because the restoring slope that would extend stacking order is sub-pascal. This sector sits apart from the crystalline- and Wigner-swelling regimes in which layered nanosheet systems are usually described. It is held together by weak osmotic and buoyant forces, not by the strong electrostatic or steric coupling those regimes assume. A separate claim of long-range periodic alignment in titanate-nanosheet suspensions at comparable spacings rests on visible-light Bragg reflection rather than on a scattering structure factor\cite{Kikuchi2025}. As shown here, optical interference is set by the local spacing and persists even when extended stacking order is absent. The two observations need not be in conflict, and may simply reflect their being different materials.

This also bears on a long-standing question in clay science: what sets the size of a tactoid, the number of consecutively aligned layers. That size is conventionally attributed to a balance of electrostatic, van der Waals and osmotic forces. It is known to depend on sample history, salinity and applied stress\cite{Segad2012,Kleijn1982,Faisal2021}. In the strongly swollen state studied here those interactions are negligible (Supplementary Sections~S5--S6). Yet the stacks still terminate after a few layers. Their size is therefore limited not by attractive cohesion but by the weakness of the response that restores stacking order. Synthetic fluorohectorite is structurally uniform, which also enabled ordered interstratification in this material\cite{Loch2020}. That uniformity makes it an unusually clean realisation of this weak-coupling limit, which it reveals rather than creates. Independent support comes from confined Cs--fluorohectorite of the same family. There, wall anchoring orders the stack over a finite penetration length before it crosses over to a multidomain state\cite{Pacakova2026GeometricDomain}. The orientational alignment of the platelets thus propagates over mesoscopic distances, favoured by their very high aspect ratio, even though the stacking registry itself remains short-ranged.

Several independent measurements would test the picture developed above. The ageing interpretation in particular is not established by the present, largely structural data. A waiting-time-dependent, two-time measurement, such as X-ray photon correlation spectroscopy, would test it directly (Supplementary Section~S12). Spatially resolved microbeam SAXS under controlled shear or injection could likewise track how the stacks fragment and recover, testing the weak-coupling quench picture directly. The rearrangement seen in the sediment after the height stabilises leaves an endpoint open. The stacking-domain size may saturate at a few layers, or it may slowly grow toward the positional coherence length \(d^{2}/\Delta d\) set by the spacing disorder. Whether that endpoint is a permanently finite-domain state or eventual full healing is not resolved by the present data. Depth-resolved microfocus SAXS through the sediment is the measurement suited to resolving it. A single vertical scan maps the local spacing and domain size against the height-dependent confinement. Repeating the scan over time tracks their evolution. In such a rugged, ageing landscape the relaxation is predicted to be intermittent rather than smooth. Long-baseline sedimentation experiments that follow an internal structural observable (the spacing, the peak intensity or the relative disorder), after the sediment height has stabilised, would test this. Step-like rearrangements with broadly distributed waiting times, rather than smooth decay, would be a direct signature of the ageing dynamics. Separately, a direct structure-factor measurement on the titanate-nanosheet system\cite{Kikuchi2025} would establish whether its apparent long-range periodic alignment, inferred from optical Bragg reflection, reflects true extended stacking order or, as here, only a well-defined local spacing.

That separation has a practical consequence. Properties governed by the local interlayer spacing need not require long-range stacking order. Structural colour is the example we observe directly. This colour was reported for the same fluorohectorite system by Michels-Brito et al.\cite{MichelsBrito2022}; here it appears as the most visible instance of a more general situation, in which a precisely selected spacing, rather than long-range stacking, sets the response. Because that decoupling follows from weak osmotic coupling and not from anything specific to fluorohectorite or to optics, the same principle is expected across the class and for spacing-controlled functions well beyond colour. The bright, spectrally defined colour is an incoherent sum over a large number of co-aligned domains, each only a few layers deep (Supplementary Section~S3). The two decoupled quantities then control distinct optical attributes: the mean spacing sets the reflected wavelength, while its spread ($\Delta d/d$), together with the short stacking coherence, broadens the reflection and suppresses the sharp, registry-dependent Bragg signatures. A spacing-set optical response can therefore be retained while the angular signatures that require extended stacking order are absent. This decoupling is also what would make solid optical composites practical, because the colour is set by the local spacing and not by the scattering-weighted registry, which extends over only two to three galleries. A solid pigment therefore need not reproduce an extended Bragg stack. It is enough to pin pairwise spacings at the visible-interference distance and embed them in an arbitrary transparent matrix, reducing the fabrication problem from maintaining long-range stacking order to fixing a single spacing. The suspension could therefore be dispersed in a transparent, low-viscosity precursor and then set, by UV curing or chemical or thermal gelation. It would set into a stiff, ideally sustainable matrix, such as a hydrogel or biopolymer, that preserves the swollen spacing and provides sufficient refractive-index contrast to the sheets. The slow recovery of long-range stacking order has a consequence for processing. The metastable two-to-three-layer domains outlive the time needed to solidify the composite. A function that needs only a few aligned layers is thus compatible with arresting the structure in a rigid matrix. One that demands long, ordered stacks would instead be defeated by the very act of stiffening, which suppresses any further ordering (Supplementary Section~S13). Other properties depend on the extent of ordered, co-facial stacking, such as mechanical reinforcement\cite{Liang2025} or gas-barrier tortuosity\cite{Uhlig2026}. These should behave differently. Because that order recovers only slowly after processing, the achievable composite performance would depend on the waiting time between dispersion and setting. The slow relaxation is therefore an asset in both regimes. For colour it is harmless, and indeed welcome, because it simply buys processing time. For the stacking-extent-dependent functions it is a built-in processing knob, since performance can be tuned through the waiting time before setting (Supplementary Section~S13). These consequences are not specific to fluorohectorite. Because membership in this class requires only stiff, high-aspect-ratio sheets swollen to large spacings, the same design rule holds across it, for vermiculite, beidellite and large delaminated oxide nanosheets alike: a functional interlayer spacing can be selected and tuned, by salinity or loading, and then locked into a solid, while stacking registry stays short and history-set. Spacing-controlled functions, such as structural colour and photonics, spacing-selective ion and molecular transport, as demonstrated in graphene-oxide membranes\cite{Abraham2017}, barrier films\cite{Dudko2022,Uhlig2026}, and dielectric spacing in composites, are thus generically accessible in this family, whereas functions that demand extended stacking order are generically hard. The same softness that frustrates the stacking is what leaves the functional length freely tunable and, once set, mechanically robust. Realising any such composite means moving beyond water. The spacing and its restoring slope are set by the aqueous osmotic and electrostatic balance. Carrying the swollen architecture into a curable matrix of different dielectric and ionic character, without collapsing it, is the central challenge.

In summary, the charge and structural perfection of synthetic fluorohectorite enables the clean realisation of a case that is generic to stiff, large-spacing layered matter but is ordinarily hidden behind defects and charge heterogeneity. The state realised is a well-defined, salinity-controlled interlayer spacing that coexists with stacking domains limited to a few (scattering-weighted) layers, sustained because the restoring slope opposing spacing mismatch is bounded at sub-pascal values and mechanically introduced stacking faults relax slowly. The spacing thus exhibits stability without stiffness: it survives not because its restoring landscape is stiff, but because it is an occupied equilibrium in a system with nothing to evict it. The weak coupling responsible follows simply from a small osmotic pressure acting across a large spacing, a condition met whenever stiff, large-spacing layered systems are swollen far enough. The same spacing-without-stacking state is therefore expected to be a general feature of such systems, including oxide nanosheets\cite{Ma2010,Osada2012}, and is one that the structural homogeneity of synthetic fluorohectorite allows to be isolated and quantified cleanly here. Swollen nanosheet suspensions thus provide a model system for finite-domain organisation and disorder-limited ordering in low-dimensional soft matter.

\section{Methods}\label{sec:methods}

\subsection*{Clay material and suspension preparation}
The starting material, \([\mathrm{Na}_{0.5}]^{\mathrm{inter}}[\mathrm{Mg}_{2.5}\mathrm{Li}_{0.5}]^{\mathrm{oct}}[\mathrm{Si}_4]^{\mathrm{tet}}\mathrm{O}_{10}\mathrm{F}_2\) (NaHec), was prepared by melt synthesis of fluorohectorite following Stöter et al.\cite{Stoter2013}. Ordered interstratification of alternating Na- and Cs-interlayers, used for double-layer production, followed Stöter et al.\cite{Stoter2016}, and the cation exchange to Cs\(^{+}\) followed Michels-Brito et al.\cite{MichelsBrito2022}. The hydrophilic Na-interlayers swell without limit in water, whereas the Cs-interlayers have a lower hydration enthalpy and do not swell, so Na/CsHec delaminates into double-layer building blocks. Synthesis and interstratification were carried out at the University of Bayreuth; suspensions were prepared at NTNU. The negative zeta potential of the delaminated single sheets, measured by electrophoresis, is \(-90.3\pm1.1\)~mV (Na form) and \(-70.9\pm2.8\)~mV (Li form); for a double-stack (NH\(_4\)/Na) unit, the only DBL form available for this measurement, it is reduced in magnitude to \(-63.3\pm3.2\)~mV, consistent with the cation-rich non-swelling interlayer lowering the charge presented to the electrolyte.

The stock suspension had a base ionic strength of \SI{3e-4}{\mole\per\liter} (NaCl-equivalent), determined by conductivity (Metrohm 712 Conductometer). This baseline is set by the nanosheets' own counterions and so depends on loading. Final suspensions were prepared by adding sodium chloride (NaCl; EMSURE ACS, ISO, Reag.\ Ph.\ Eur., \(\geq\)\SI{99.5}{\percent}) to reach reservoir ionic strengths between \num{3e-4} and \SI{1.3e-3}{\mole\per\liter}, chosen to place the first-order structural colour in the visible range. Suspensions were kept on overhead shakers to ensure homogeneity, following the equilibration procedure of Michels-Brito et al.\cite{MichelsBrito2022}; the mean spacing is established in the stock suspension during this equilibration. Immediately before measurement, samples were gently shaken and injected by syringe into quartz cuvettes (Hellma, \SI{1.00}{\milli\meter} path length) or borosilicate capillaries (Hilgenberg, \SI{2.00}{\milli\meter} diameter). Syringe injection introduces transient shear and stacking faults; because the restoring slope is small in the swollen regime, these relax slowly after loading. Optical and sedimentation experiments used a clay concentration of \SI{0.7}{\percent} by volume and five salt concentrations (Supplementary Table~S3); SAXS and SANS used similar salinity ranges with additional concentrations chosen for scattering signal.

\subsection*{Small-angle X-ray scattering}
SAXS was performed at the CoSAXS beamline, MAX IV Laboratory\cite{Plivelic2019}, with a sample-to-detector distance of \SI{9.95}{\meter}, wavelength \SI{0.99}{\angstrom} and beam size at the sample of about \SI{150}{\micro\meter\squared}. Samples in borosilicate capillaries were measured at \SI{25}{\celsius}. The accessible range was \(1.4\times10^{-2} < q < 9.4\times10^{-1}\,\mathrm{nm}^{-1}\), with \(q = (4\pi/\lambda)\sin\theta\). Two-dimensional images were azimuthally integrated over \SI{360}{\degree} using standard beamline reduction. The monochromator bandwidth is \(\Delta E/E \approx 2\times10^{-4}\)\cite{Plivelic2019}, so the relative \(q\)-resolution at the lamellar peak is well below \SI{1}{\percent}, far narrower than the intrinsic peak width of a few-layer stack; the SAXS peak widths are therefore sample-limited and provide the primary evidence for finite stacking domains (Supplementary Section~S3).

\subsection*{Small-angle neutron scattering}
SANS was performed at the SANS2D beamline, ISIS Neutron and Muon Source (STFC; DOI \href{https://doi.org/10.5286/ISIS.E.RB2310488}{10.5286/ISIS.E.RB2310488}), with a sample-to-detector distance of \SI{12}{\meter} and a circular beam of about \SI{8}{\milli\meter} diameter. The accessible range was \(0.0015 < q < 0.8\,\mathrm{\AA}^{-1}\), obtained in time-of-flight mode with wavelengths from \num{1.75} to \SI{12.5}{\angstrom}. Samples in flat quartz cuvettes were held at \SI{20}{\celsius}; to limit shear, the needle was inserted to the cuvette corner and withdrawn slowly during injection. Background subtraction and absolute normalisation used a polystyrene standard, and reduction used Mantid\cite{Arnold2014}. The instrument resolution \(\Delta Q/Q\) from the Mantid pipeline for this configuration is approximately \num{12} to \SI{19}{\percent} over the lamellar-peak range\cite{Heenan2011}. This is not negligible against the intrinsic peak width, so SANS serves as cross-validation from a different contrast mechanism rather than as the primary domain-size probe; the agreement between the two indicates that the finite domain size is not a resolution artefact (Supplementary Section~S3).

\subsection*{Scattering analysis}
One-dimensional SAXS and SANS curves were analysed with the Förster lamellar model\cite{Forster2010} in the freely available \textsc{Scatter} package (the version distributed with Ref.~\cite{Forster2010}), previously applied to layered clays\cite{Rosenfeldt2016,Pacakova2026GeometricDomain}. The same stacking-domain sizes are obtained with independent SASView-based lamellar structure-factor modelling\cite{TrigueiroNeto2025}, so the result does not depend on the choice of lamellar model; the Förster model is retained only because it fits equally well with one fewer parameter (Supplementary Section~S3). The spacing follows from the principal peak position, \(d^{\ast} = 2\pi/q_{\mathrm{peak}}\); the peak width yields the relative spacing disorder \(\Delta d/d\) and the finite stacking-domain size. We report the stacking-domain size via its thickness \(L_{\mathrm{dom}}=(N-1)\,d^{\ast}\) returned by the fit, where \(N\) is the number of registered layers. The fit gives \(L_{\mathrm{dom}}/d^{\ast} = N-1 \approx 1.3\) to \(2.3\) interlayer spacings (Supplementary Table~S2), i.e.\ \(N \approx\) two to three registered layers. This finite stacking-domain size is distinct from, and shorter than, the positional coherence length that the F\"orster model associates with the spacing fluctuation \(\Delta d/d\), which is itself only a few layers here because \(\Delta d/d\) is large; the separate length \(d^{2}/\Delta d\) tabulated in Supplementary Table~S2 is that microstrain measure, not the stacking-domain size.

The SAXS measurements were repeated on three independently prepared suspensions (a triplicate);
the error bars on the SAXS points in Fig.~\ref{fig:fig3} give the spread across these
replicates. This sample-to-sample spread is below \SI{2}{\percent} on
\(d\) and on the number of registered layers; the corresponding spread on the displacement
disorder \(\Delta d/d\) (Supplementary Fig.~S6) is about \SI{4}{\percent}. The
numerical precision of a single Förster fit is about \SI{1}{\percent} of the
fitted parameter and is sub-dominant to this sample-to-sample variation. The SANS data were not replicated and therefore carry no corresponding error bars. The uncertainty on the restoring slope \(k(d)\) and the restoring-pressure amplitude \(\Delta\Pi\) (Fig.~\ref{fig:fig5}a,b) is dominated by the in-cell position at which mechanical balance is realised, propagated as a robustness envelope in Supplementary Section~S8 and shown as the shaded band in Fig.~\ref{fig:fig5}a and as vertical error bars in Fig.~\ref{fig:fig5}b. The within-cluster spread in \(\sigma_{\mathrm{eff}}\) (error bars in Fig.~\ref{fig:fig5}c) is the small residual finite-\(\kappa d\) variation across the stocks at a given salinity and reflects between-sample differences in \(\phi_{\mathrm{prep}}\); the per-sample values, grouped by reservoir salinity, are shown in Supplementary Fig.~S15.

The same finite-domain result is obtained whether the spacing is controlled by concentration or by salinity, and across independently prepared batches (this work), confirming that the limited stacking-domain size is a property of the swollen state.

\subsection*{Reflectance, sedimentation and pressure reconstruction}
Reflectance spectra were measured with an Avantes AvaSpec-ULS2048CL-EVO spectrometer (\SIrange{200}{1100}{\nano\meter}) and an AvaSphere integrating sphere, following the acquisition and calibration of our earlier work\cite{MichelsBrito2022,TrigueiroNeto2025}. The first-order peak position \(\lambda_{\mathrm{peak}}\) was converted to a spacing through the Bragg--Snell relation \(\lambda_{\mathrm{peak}} = 2 n_{\mathrm{eff}} d\). Sedimentation was monitored by imaging cuvettes at fixed intervals with an iPhone~14 camera on a fixed stand under diffuse front illumination, with the vertical scale calibrated against the known cuvette dimension. The clay concentration was \SI{0.7}{\percent} by volume and salt concentrations were \num{3.0e-4}, \num{4.5e-4}, \num{6.0e-4}, \num{9.0e-4} and \SI{1.3e-3}{\mole\per\liter}; a long-term run used \SI{3.0e-4}{\mole\per\liter} imaged over about one month.

The local spacing \(d(z)\) in the sediment follows from the reflected-wavelength profile through Bragg--Snell with \(n_{\mathrm{eff}} \approx 1.33\) (water; the perturbation by clay is negligible below \SI{1}{\percent} by volume). The local volume fraction \(\phi(z)\) follows from the empirical spacing--concentration relation\cite{TrigueiroNeto2025}, and the confinement pressure from
\begin{equation}\label{eq:buoyancy}
\Pi(z) = g\,\Delta\rho \int_{z}^{H_{\mathrm T}} \phi(z')\,\mathrm{d}z',
\end{equation}
where \(\Delta\rho\) is the nanosheet--medium density contrast, \(g\) the gravitational acceleration and \(H_{\mathrm T}\) the total sample height. At the salt concentrations studied (\(n_0\le\SI{1.3e-3}{\mole\per\liter}\)), the dependence of the solvent density on salinity is below one part in \num{e4} and \(\Delta\rho\) is therefore salinity-independent to that level; \(\phi_{\mathrm{prep}}\) is set by sample preparation and is independent of salinity by construction, and the suspension is homogeneous on the measurement timescale (sedimentation develops over days, whereas scattering measurements are completed within hours), so the local \(\phi\) at the scattering measurement point coincides with \(\phi_{\mathrm{prep}}\). The full salinity dependence of \(d^{\ast}\) therefore enters through \(\Pi_{\mathrm{osm}}\) (Eq.~\eqref{eq:donnan}), not through \(\Pi_{\mathrm{ext}}\). The osmotic pressure of the confined electrolyte follows the Donnan expression of El Rifaii et al.\cite{ElRifaii2022} (Supplementary Section~S8),
\begin{equation}\label{eq:donnan}
\Pi_{\mathrm{osm}} \approx \frac{k_{\mathrm B}T}{n_0}\left(\frac{\sigma_{\mathrm{eff}}\,\phi_{\mathrm{prep}}}{2 e h}\right)^{2},
\end{equation}
with \(n_0\) the reservoir salt concentration, \(\sigma_{\mathrm{eff}}\) an effective osmotic charge, \(\phi_{\mathrm{prep}}\) the prepared volume fraction, \(e\) the elementary charge and \(h\) the nanosheet thickness. The restoring slope is
\begin{equation}\label{eq:kslope}
k(d) = \frac{\partial}{\partial d}\left[\Pi_{\mathrm{osm}}(d) - \Pi_{\mathrm{ext}}\right].
\end{equation}
With the empirical swelling relation \(\phi_{\mathrm{prep}}(d)=A\,d^{-\alpha}\) (Supplementary Section~S2), the saturation of \(\sigma_{\mathrm{eff}}\) at large \(d\) (Supplementary Section~S8 and Refs.~\cite{Alexander1984,BocquetTrizacAubouy2002,TrizacBocquet2004}), and the Donnan balance \(\Pi_{\mathrm{osm}}=\Pi_{\mathrm{ext}}\equiv\Pi\), Eq.~\eqref{eq:kslope} reduces at mechanical balance to the chained expression
\begin{equation}\label{eq:kfull}
|k(d,z)|\;\approx\;\frac{\alpha\,\Pi}{d}\;=\;\frac{\alpha\,\Delta\rho\,g\,A\,(H_{\mathrm T}-z)}{d^{\,\alpha+1}}\;\propto\;d^{-(\alpha+1)},
\end{equation}
where the second equality follows from substituting the buoyancy form for \(\Pi\), with \(z\) the in-cell height coordinate (\(z=0\) at the cell bottom, \(z=H_{\mathrm T}\) at the meniscus), and \(A\) the prefactor of the empirical swelling relation. The first equality is the form used as Eq.~\eqref{eq:kcompact_main} in the main text. For the swelling exponent we use \(\alpha\approx 1.16\) (equivalently \(x=1/\alpha\approx 0.86\)), the value measured for DBL nanosheets in cuvette geometry at zero added salt by Trigueiro Neto et al.\cite{TrigueiroNeto2025}, which corresponds to the sample architecture and cell geometry used throughout the present work. To estimate the restoring-pressure amplitude associated with the measured spacing distortion \(\Delta d\), we evaluate, for each sample \(i\), at mid-cell (\(z=H_{\mathrm T}/2\), where \(\Pi=\Pi_{\mathrm{max}}/2\)),
\begin{equation}\label{eq:dPi}
\Delta\Pi_i \;\approx\; |k(d_i^{\ast},H_{\mathrm T}/2)|\,\Delta d_i \;=\; \tfrac{1}{2}\alpha\,\Pi_{\mathrm{max}}(d_i^{\ast})\left(\frac{\Delta d}{d}\right)_i,
\end{equation}
where \(\Pi_{\mathrm{max}}(d_i^{\ast})=\Delta\rho\,g\,\phi_{\mathrm{prep},i}\,H_{\mathrm T}\) uses each sample's actual prepared volume fraction \(\phi_{\mathrm{prep},i}\) (not the swelling-relation value evaluated at \(d_i^{\ast}\)). Because the balance \(\Pi_{\mathrm{osm}}=\Pi_{\mathrm{ext}}\) underlying the compact slope holds only at the equilibrium spacing \(d^{\ast}\), and the measured distortions take the spacing a finite distance \(\Delta d/d \approx 0.2\)--\num{0.3} away from it (so \(\Pi_{\mathrm{osm}}\neq\Pi_{\mathrm{ext}}\) there), Eq.~\eqref{eq:dPi} is read as an order-of-magnitude estimate of the restoring-pressure scale opposing the measured distortion rather than as a strict linear-response calculation about \(d^{\ast}\) (Supplementary Section~S8). The dimensionless ratio \(\Lambda \equiv \Delta\Pi/\tau\) compares the restoring-pressure amplitude with the loading stress \(\tau \approx 10\)--\SI{100}{\pascal} (Supplementary Section~S9). An intrinsic dimensionless group is \(\Lambda^{\ast} \equiv |k(d)|\,d/\Pi_{\mathrm{osm}}\), which is small in this regime because \(k(d)\) is small relative to the osmotic scale and does not depend on the choice of \(\tau\).

\section*{Use of AI}
During the preparation of this manuscript the authors used large language models (ChatGPT, OpenAI; and Claude, Anthropic) to assist with language editing and with the wording and organisation of author-written text. These tools were not used to generate scientific ideas, study designs, data, analyses or interpretations. The authors reviewed and edited all output and take full responsibility for the content.

\section*{Data availability}
The scattering, reflectance and sedimentation datasets and the fitted parameters that support the findings of this study are available in the DataverseNO repository at \url{https://doi.org/10.18710/TDIVQY}\cite{TDIVQY_2026}.

\section*{Code availability}
The Förster-fitting and pressure-reconstruction code used in this study is available in the same DataverseNO repository at \url{https://doi.org/10.18710/TDIVQY}\cite{TDIVQY_2026}.

\section*{Author contributions}
O.T.N.\ performed the reflectance, sedimentation and part of the SAXS measurements, carried out the data reduction and analysis, and prepared the figures. P.H.M.B.\ prepared the samples and supported the SANS experiments. B.C.T.\ and N.J.E.\ supported the SANS data acquisition. L.P.C.\ supervised the SANS experiments. T.S.P.\ performed the SAXS experiments. J.B.\ provided the synthesised materials and contributed to interpretation. J.O.F.\ conceived and supervised the study, developed the pressure-based framework, and led the analysis and writing. All authors discussed the results and edited the manuscript.

\section*{ORCID}
O.\ Trigueiro Neto 0000-0001-5086-0859; P.\ H.\ Michels-Brito 0000-0002-6320-1171; B.\ Ceccato Telli 0000-0002-1414-8511; L.\ P.\ Cavalcanti 0000-0002-0408-0058; T.\ S.\ Plivelic 0000-0003-0193-3251; J.\ Breu 0000-0002-2547-3950; J.\ O.\ Fossum 0000-0002-8952-303X.

\section*{Competing interests}
J.O.F.\ owns U.S.\ patent application 18/577,609, ``Nano-functionalised clay minerals for structural colouration'', which concerns the clay system studied here, and is the owner of SCML International AS (Soft and Complex Matter Lab). The remaining authors declare no competing interests.

\section*{Acknowledgements}
We acknowledge MAX IV Laboratory for beamtime on CoSAXS under proposals 20210657 and 20220537. Research at MAX IV, a Swedish national user facility, is supported by the Swedish Research Council (2018-07152), Vinnova (2018-04969) and Formas (2019-02496). We thank Barbara Pacáková, Éric Clément and Françoise Brochard-Wyart for helpful discussions.

\section*{Funding}
This work was supported by the Research Council of Norway through the PETROMAKS2 programme (project 280643; O.T.N., J.O.F.) and the FRIPRO programme (project 315135; P.H.M.B., L.P.C., J.O.F.), by the EU Horizon 2020 Marie Skłodowska-Curie Actions (project 956248; B.C.T., J.O.F.), and by the German Research Foundation (DFG, project 492723217, CRC 1585, subproject A5; N.J.E., J.B.).

\bibliographystyle{naturemag}
\bibliography{references}

\clearpage


\section*{Supplementary material}

# Supplementary Information for: Layered matter that maintains spacing but loses stacking order

Osvaldo Trigueiro Neto[1], Paulo Henrique Michels-Brito[1],
Bruno Ceccato Telli[1,2], Natalie Jasmin Eichstaedt[3],
Leide P. Cavalcanti[4], Tomás S. Plivelic[5], Josef Breu[3], Jon Otto Fossum[1,3,6,7]

[1]Department of Physics, Norwegian University of Science and Technology (NTNU), Trondheim, Norway
[2]School of Chemical Engineering, University of Campinas (UNICAMP), Campinas, Brazil
[3]Department of Chemistry, University of Bayreuth, Bayreuth, Germany
[4]ISIS Neutron and Muon Source, STFC, Rutherford Appleton Laboratory, Didcot, United Kingdom
[5]MAX IV Laboratory, Lund University, Lund, Sweden
[6]SCML International, Oslo, Norway
[7]Laboratoire de Physique et Mécanique des Milieux Hétérogènes (PMMH), ESPCI Paris, Paris, France
Corresponding authors: Jon Otto Fossum (jo@scml.no); Osvaldo Trigueiro Neto (osvaldo.t.neto@ntnu.no)

## S1. Materials and experimental methods

Full materials and experimental methods (clay synthesis and suspension preparation, SAXS, SANS, scattering analysis, reflectance spectrophotometry, sedimentation and optical monitoring, and confinement-pressure reconstruction) are given in the Methods section of the main text.

### S2. Nanosheet architectures and crowded lamellar environment

The two figures in this section define the structural and mechanical context of the system studied in the main text. Fig. S1 specifies the nanosheet building units (SGL and DBL), while Fig. S2 illustrates the crowded lamellar environment in which these units are embedded. This environment is essential for understanding the origin of the weak but finite external confinement pressure $\Pi_{ext}$ that governs the swelling behaviour and restoring response of the lamellar stacks.

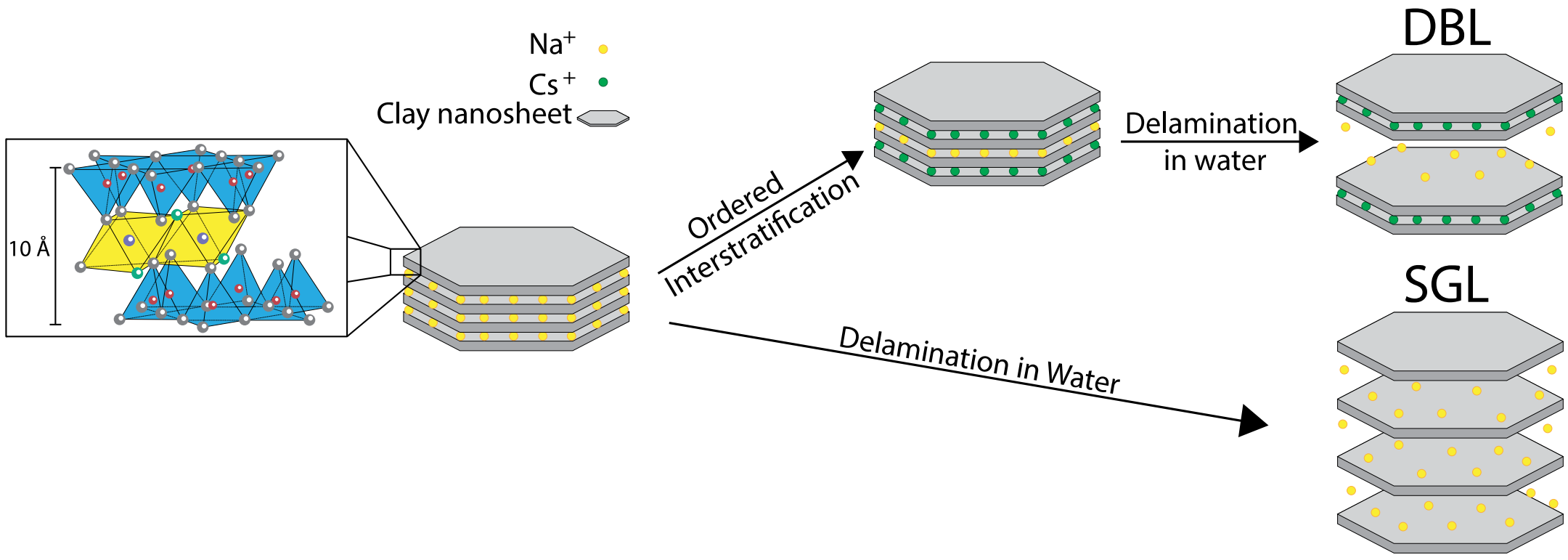


Figure S1: Detailed structural representation of single-layer (SGL) and double-layer (DBL) nanosheet building units derived from synthetic fluorohectorite. In the SGL system, individual silicate sheets are fully delaminated and separated by hydrated interlayers containing mobile counterions. In the DBL system, two adjacent layers remain associated after interstratification and subsequent delamination, forming composite nanosheet units with approximately twice the thickness and reduced effective surface charge density.[1] This structural distinction defines the effective lamellar building block in the swollen regime studied in the main text, while preserving identical chemical composition. The left inset shows the crystalline gallery (about 10 Å) of the parent fluorohectorite. In the crystalline (non-osmotic) regime this gallery swells only in discrete cation-hydration steps, the basal spacing stepping through about 9.6 Å, 12.2 Å and 15.1 Å for the zero-, one- and two-layer hydrates of Na-fluorohectorite, with adjacent layers held in near-perfect registry by interlayer hydrogen bonding[2]; the osmotically swollen state studied here lies far beyond these discrete states and loses that registry. The upper route, ordered Na/Cs interstratification followed by delamination in water, yields DBL units; the lower route, direct delamination in water, yields SGL units. Yellow markers denote $Na^{+}$ and green markers $Cs^{+}$.

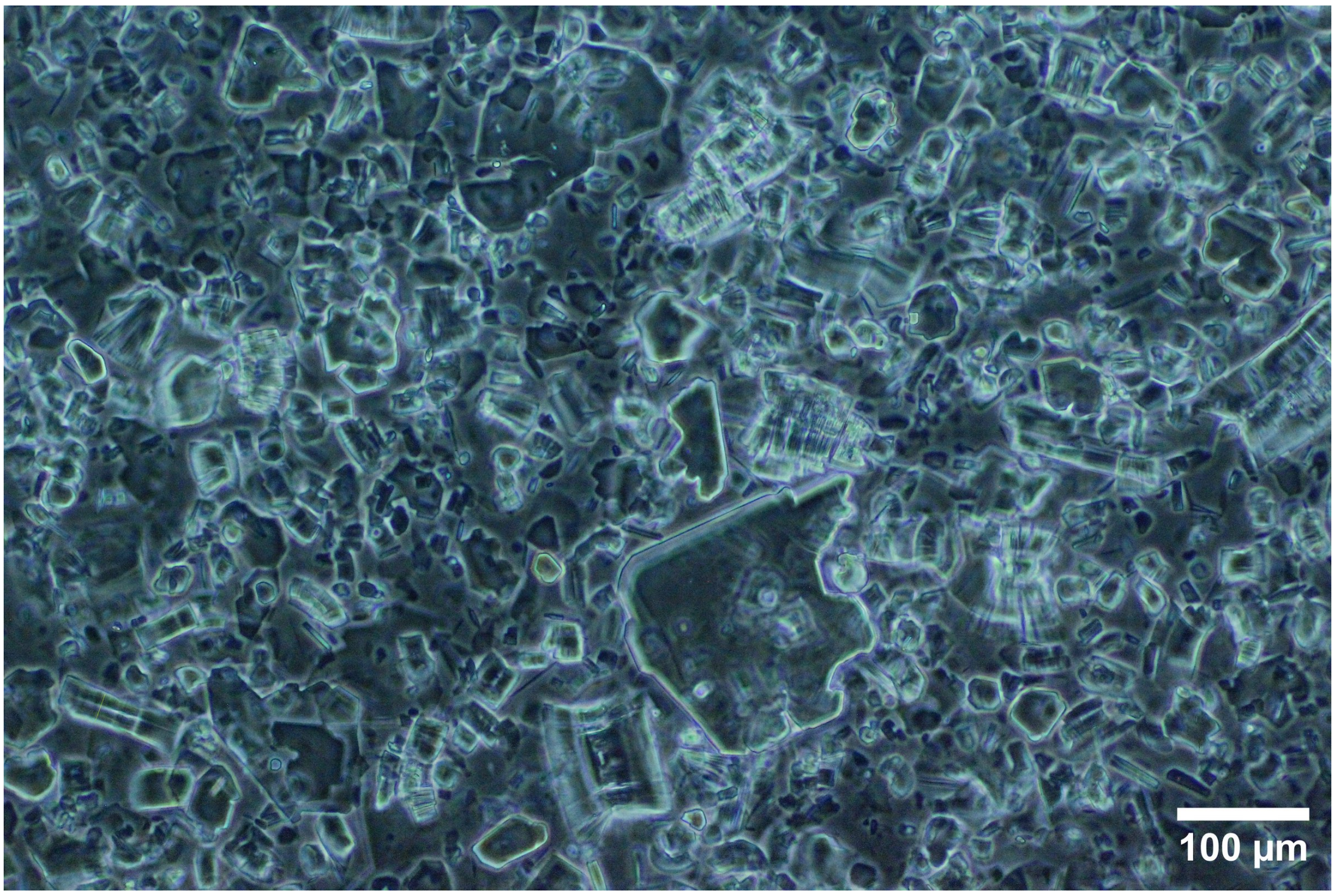


Figure S2: Optical micrograph (scale bar, 100 µm) of a swollen nanosheet suspension illustrating the crowded lamellar environment. The image shows that each lamellar stack is embedded in a surrounding population of neighbouring stacks rather than existing as an isolated object. This crowding has two important physical consequences. First, neighbouring stacks exert an effective confinement through osmotic and steric interactions, contributing to the external pressure $\Pi_{\mathrm{ext}}$ acting on a given stack. Second, the finite sample height and overall suspension geometry impose an additional global confinement, which together with the surrounding stack population sets the mechanical boundary conditions of the system. The relevant confinement pressure $\Pi_{\mathrm{ext}}$ discussed in the main text should therefore be understood as an emergent, collective quantity arising from both the crowded lamellar environment and the macroscopic confinement of the sample.

Fig. S2 suggests that even at low overall clay volume fraction, the large lateral extent of the nanosheets leads to an effectively confined interlayer geometry, in which neighbouring platelets impose constraints on the local swelling state.

In addition to this intrinsic crowding, the experimentally observed swelling behaviour is influenced by the container geometry. The effective confinement imposed by the sample cell affects both the local spatial organisation and wall anchoring of the nanosheets[3], as well as the relation between interlayer spacing and clay volume fraction.

In the main text we use the empirical power-law relation $\phi(d) \approx A\, d^{-\alpha}$. The parameters $A$ and $\alpha$ were determined for the zero-added-salt case and depend on both the container geometry and the nanosheet architecture (SGL or DBL)[4]:

$$
\begin{aligned}
d^* &\approx 176.1\,\phi^{-0.86} && \text{(DBLs in cuvette geometry)},\\
d^* &\approx 166.5\,\phi^{-0.67} && \text{(DBLs in capillary geometry)},\\
d^* &\approx 92.4\,\phi^{-0.77} && \text{(SGLs in cuvette geometry)},\\
d^* &\approx 94.0\,\phi^{-0.64} && \text{(SGLs in capillary geometry)}.
\end{aligned}
\tag{S1}
$$

Here $d^*$ is expressed in nm and $\phi$ in vol%, so the tabulated coefficients $A$ carry units of vol% nm$^x$. When these swelling relations are substituted into the osmotic and buoyancy pressure expressions of Sec. S8, $\phi$ must be taken as the *dimensionless* volume fraction (vol% divided by 100) and

all lengths in SI units, so that the pressures come out in pascal; equivalently, the coefficient $A$ used in the closed-form restoring slope Eq. (S23) is $A/100$ with lengths in metres. Note that the exponents listed above are the $d$-versus-$\phi$ exponents $x$ (that is, $d^* \propto \phi^{-x}$); the exponent $\alpha$ appearing in $\phi(d) \approx A\, d^{-\alpha}$, and hence in the restoring slope (Sec. S8), is the reciprocal, $\alpha = 1/x$. For the zero-added-salt DBL-in-cuvette case, $x = 0.86$ gives $\alpha \approx 1.16$. The cuvette geometry corresponds to the sedimentation, structural-colour and SANS measurements and sets the representative bound quoted in the main text, whereas the SAXS measurements were performed in capillaries, for which $x \approx 0.67$ gives $\alpha \approx 1.49$ (DBLs); both are of order unity, so this distinction does not affect the sub-pascal conclusion.

These results show that the empirical swelling relation is not universal, but depends on the confinement imposed by the container, the packing achieved during preparation (including the loading protocol), and the internal structure and stiffness of the nanosheet building blocks. This dependence is nevertheless sufficiently weak for the power-law form to provide a robust description of the swelling behaviour across the experimentally accessible visible regime. Although these exponents were determined at zero added salt, the restoring-slope bound (Sec. S8) depends on the swelling exponent only linearly, through $\alpha = 1/x$. The $d$-versus-$\phi$ exponent cannot exceed the ideal one-dimensional swelling value $x = 1$ (spacing grows no faster than $\phi^{-1}$ when added solvent enters the interlayer galleries) and lies empirically between 0.64 and 0.86 across the architectures and geometries above, so $\alpha$ remains of order unity (about 1 to 1.6); even $x = 0.5$ would raise $\alpha$ only to 2, changing the bound by about a factor of two, comparable to the confinement-pressure uncertainty. The sub-pascal restoring-slope conclusion therefore does not depend on the precise exponent or on its variation with salinity.

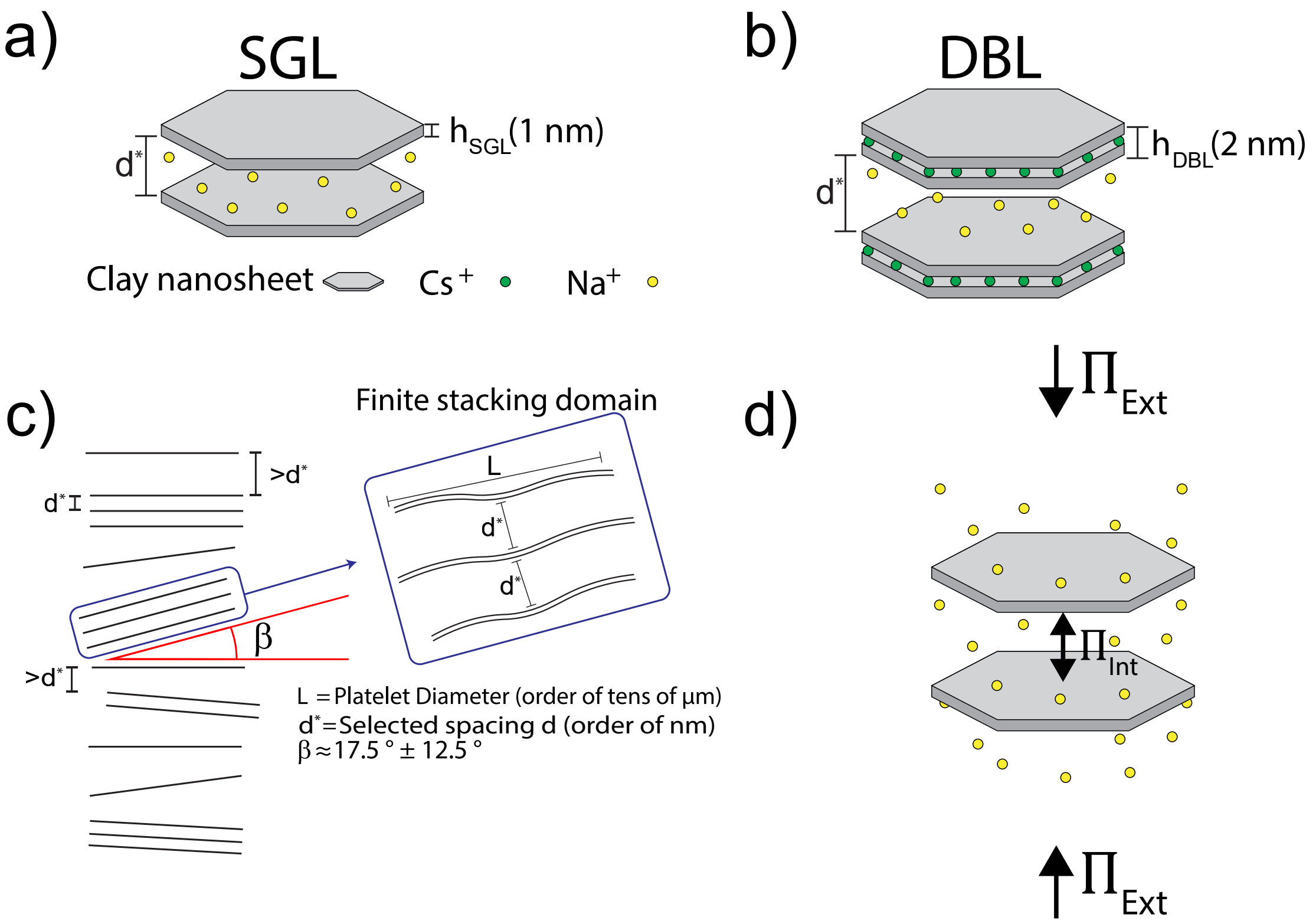


Figure S3: **Architecture of SGL and DBL nanosheet systems and schematic phenomenology of swollen lamellar stacks.** (a) Schematic structure of single-layer nanosheets (SGL) obtained by osmotic delamination of Na-fluorohectorite, in which adjacent sheets interact repulsively and the swelling is quasi-continuous, in contrast to the stepwise, cohesive galleries of the crystalline hydrate[5]. Individual silicate sheets are separated by hydrated interlayers containing mobile counterions. (b) Structure of double-layer nanosheets (DBL) produced through ordered interstratification followed by delamination. In this system two adjacent clay layers remain associated, forming a composite nanosheet with approximately twice the thickness of the SGL units. (c) Schematic representation of the swollen lamellar state formed by these nanosheets in aqueous suspension. Rather than forming extended ordered lamellar stacks, the swollen lamellar state is composed of finite stacking domains consisting of singlets, dimers, trimers and short multilayer stacks. The finite stacking domains can be misaligned relative to one another, represented by the angle $\beta$ (the mosaic spread, or mosaicity), typically ranging from 5° to 30°; such cofacial but orientationally disordered domains are characteristic of nematic suspensions of these high-aspect-ratio nanosheets, in which individual cofacial domains are randomly oriented[6]. See also Fig. S2. $\beta$ is exaggerated here for visualisation. (d) Schematic of the pressure balance on a lamellar stack. The external confinement pressure $\Pi_{\mathrm{Ext}}$, arising from the crowded surrounding stacks and the finite sample height (Sec. S2), acts on the whole accordion from outside, not within the individual interlayers; it is opposed, within each gallery, by the internal osmotic pressure $\Pi_{\mathrm{Int}}$ of the confined electrolyte, and the balance $\Pi_{\mathrm{Int}} = \Pi_{\mathrm{Ext}}$ selects the mean interlayer spacing $d^*$. Because the restoring slope about $d^*$ is extremely weak in the swollen regime, mechanically introduced perturbations relax only inefficiently after sample loading, so the observed state exhibits limited stacking order and finite stacking domains whose statistical distribution is discussed in Sec. S10.

In DBL systems, two adjacent clay layers remain associated after delamination and form paired composite units. This pairing preserves the lateral platelet dimensions while reducing the effective electrostatic layer charge seen by the surrounding electrolyte[7].

**Finite-domain lamellar organisation.** The swollen lamellar state formed by these nanosheets is illustrated schematically in Fig. S3c,d. Rather than forming extended ordered lamellar stacks, the system consists of finite domains composed of singlets, dimers, trimers and short multilayer stacks. The crowded lamellar environment that provides the weak but finite confinement

background is illustrated in Fig. S2 and discussed in Sec. S2.

## S3. SAXS and SANS data and model fits

The main fit parameters extracted from the SAXS and SANS analysis, including spacing disorder and stacking-domain sizes, are summarised in Table S2. An overview of all sample conditions and the corresponding experimental techniques is given in Table S3.

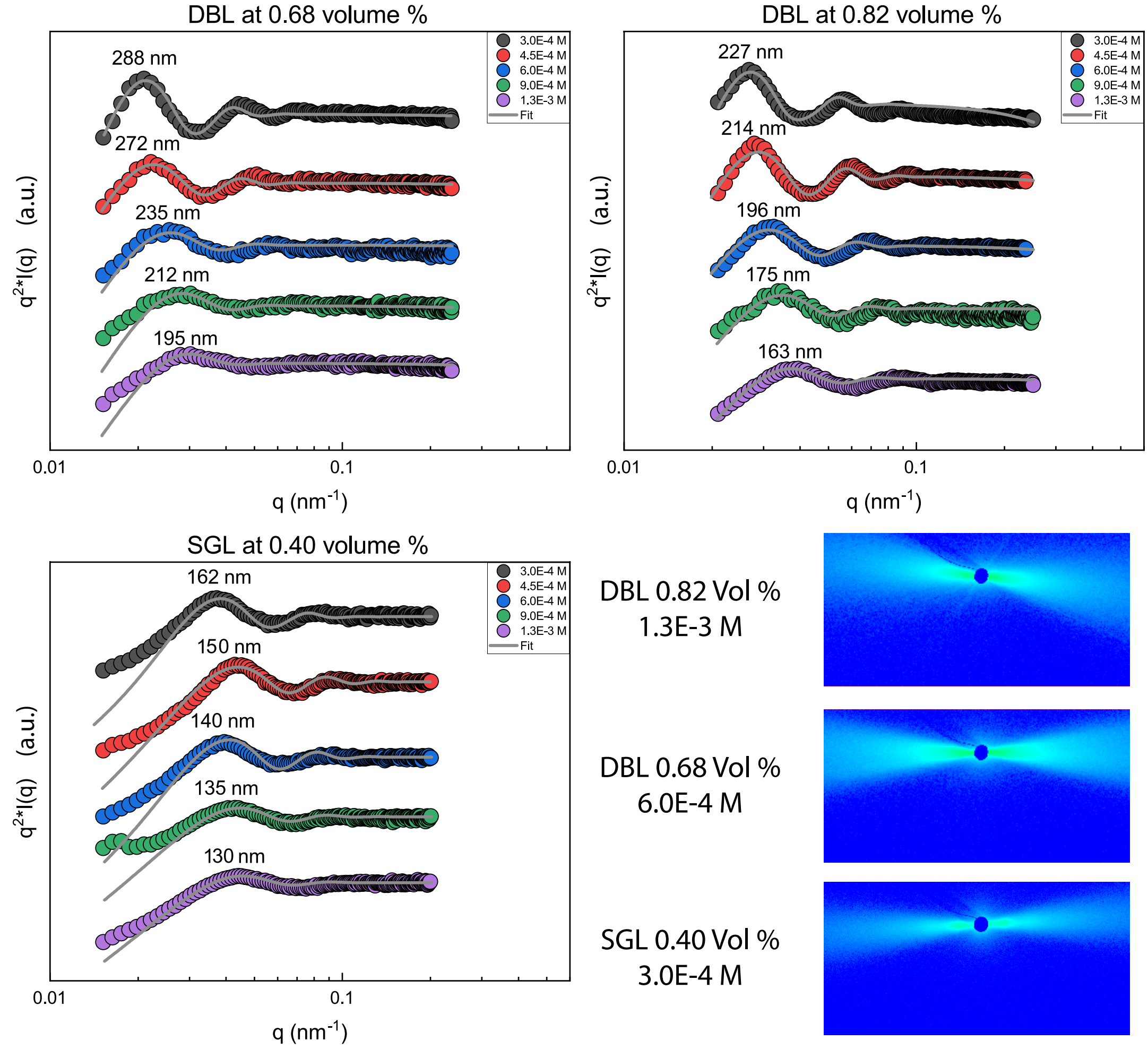


Figure S4: Small-angle X-ray scattering (SAXS) intensity profiles for SGL and DBL nanosheet suspensions at different reservoir salinities, together with fits using the Förster model and three selected two-dimensional diffractograms illustrating the relative domain misalignment, or mosaicity ($\beta$) represented schematically in Fig. S3c. In these patterns the lamellar scattering appears as sharp, azimuthally tilted lobes rather than a uniform ring: the tilt registers the domain misalignment $\beta$, while the lobes lie at a common radial $|q^*|$ to within the peak width, so the interlayer spacing is shared across domain orientations. This is the direct scattering signature of the orientation-blind Donnan balance discussed in the main text and Sec. S8: mosaicity varies the direction of the layering, not the spacing. The data are presented in the same form as the SANS results shown in Fig. 2 of the main text, with the dominant single-particle form factor removed in order to emphasise the lamellar structure factor. The positions of the principal peaks determine the interlayer spacing through $d^* = 2\pi/q_{\text{peak}}$, while the peak widths provide a measure of the finite stacking domains. The SAXS profiles are dominated by a first-order lamellar peak, with higher-order harmonics weak and broad. Fits using the Förster model[8] yield stacking-domain sizes corresponding to only a few layers together with substantial interlayer displacement disorder, consistent with the finite-domain interpretation discussed in the main text.

**Interpretation of lamellar scattering features.** The scattering profiles are dominated by a first-order peak corresponding to the lamellar spacing, with higher-order peaks weak and broad. This shows that although a well-defined mean interlayer spacing exists, the stacking domains remain limited. The stacking-domain size extracted from the peak width corresponds to only a few layers. This stacking-domain size, the stacking-domain thickness $L_{\text{dom}} = (N-1)\,d^*$, with $N \approx 2$–3 the number of registered layers, is distinct from, and shorter than, the positional coherence length that the Förster model associates with the spacing fluctuation $\Delta d/d$, which is itself only a few layers here because $\Delta d/d$ is large.

The weakness and breadth of the higher-order peaks reflect substantial interlayer displacement disorder together with limited stacking order. The measured scattering signal therefore represents a scattering-weighted average over a distribution of finite stacking domains rather than extended lamellar order.

**Robustness of the finite-domain interpretation.** The finite stacking-domain size extracted here is not an artefact of the particular fitting model. Equivalent stacking-domain sizes and the same finite-domain phenomenology were obtained for this material system using both the Förster formalism and independent SASView-based lamellar structure-factor modelling[4]; we retain the Förster model here because it describes the lamellar structure factor equally well with one fewer fitting parameter. Both descriptions encode the standard powder-diffraction decomposition of Bragg-peak broadening into interlayer spacing disorder (microstrain), $\Delta d/d$, and finite stacking-domain thickness, the same separation of strain and finite-size broadening expressed by the Scherrer equation[9] and a Williamson–Hall analysis[10]. With the measured $\Delta d/d \sim 0.2$–0.3, a substantially larger disorder would suppress the first-order peak entirely and eliminate the observed structural coloration; the persistence of a well-defined first-order peak therefore requires a finite finite-domain contribution, so the extracted stacking-domain size of $\sim 2$–3 registered layers is physically meaningful rather than model-generated. We note, however, that with a single resolved lamellar order the finite-size and disorder contributions to the peak width are not fully separable, so the precise layer count (two versus three) should be read together with the fit uncertainties; the sample-to-sample reproducibility of the extracted parameters is quantified below. The same limited domain size is observed by both SAXS and SANS, which rely on fundamentally different contrast mechanisms (electron-density versus neutron scattering-length-density contrast); consistency across the two techniques argues strongly against a method-specific origin for the finite-domain signal.

**Bounded joint inference of $N$ and $\Delta d/d$.** The two width contributions extracted from the Förster fits, the relative spacing disorder $\Delta d/d$ and the stacking-domain layer count $N$, enter the first-order lamellar peak width additively: a finite-size (Scherrer-like) contribution that scales as $1/N$, and a paracrystalline disorder contribution that scales with $\Delta d/d$. For a given observed peak width, the two are therefore positively correlated in the fit: a larger $N$ requires a larger $\Delta d/d$ to reproduce the same width, and a smaller $N$ requires a smaller $\Delta d/d$. Both parameters are also bounded by hard physical interference constraints, not merely by Gaussian fit statistics. The disorder is bounded above, $\Delta d/d \lesssim 0.3$, because above this value the first-order Bragg peak is destroyed (paracrystalline broadening exceeds the peak position) and the observed structural coloration would not exist. The layer count is bounded below by $N \geq 2$: a single isolated platelet ($N = 1$) has no neighbour to interfere with, so it produces no Bragg peak at all.

The fit therefore lands at the unique corner consistent with the observed peak width and both bounds simultaneously: $\Delta d/d$ is close to its upper bound ($\sim 0.2$–0.3, just below the peak-destruction limit) and $N$ is close to its lower bound ($\sim 2$–3, just above the interference

minimum). Moving along the fit-degenerate direction would immediately violate one of the two hard bounds. Were the fit to return a smaller $\Delta d/d$, the peak-width constraint would require a smaller $N$ to compensate, but $N$ is already at the interference floor. Were the fit to return a larger $\Delta d/d$, the Bragg peak would vanish. The joint $(N, \Delta d/d)$ region admissible to the data is in this sense not a degenerate ridge extending to infinity, as a Gaussian fit-statistic reading might suggest, but a single corner pinned by hard interference physics (Fig. S6b). This rules out the two alternatives a critical reader might invoke, a fully ordered lamellar phase with very large $\Delta d/d$, or an isolated single-platelet population with vanishing $N$, without recourse to fit statistics. The second-order behaviour is independently informative. A fully ordered, many-layer alternative ($N \gg 1$) would produce a sharp, strong second-order Bragg peak at $2q^*$; what is observed instead, in both SAXS (Fig. S4) and SANS (Fig. S5), is a weak, broad second order together with finite-size interference fringes, the signature of short, internally ordered stacks of $N \approx 2$–$3$ layers rather than of extended order. The survival of any resolved second order is itself informative about the disorder: a purely cumulative (paracrystalline) $\Delta d/d$ of 0.2–0.3 would broaden the second order by a factor $\sim 4$ and wash it out, which is not seen. The measured $\Delta d/d$ is therefore dominated by static, first-kind displacement disorder, principally the spread of equilibrium spacings between structurally inequivalent stacks (Sec. S12), which broadens the peak while leaving the within-stack registry intact; only a small cumulative component limits the coherence.

**Scattering weighting and the real-space layer count.** The scattering-weighted average reported by the Förster fit is biased towards the larger domains in the ensemble, because the coherent scattering intensity scales as $N^2$ (the square of the number of platelets in the coherent stack). Even at unchanged real-space domain population, this weighting shifts the inferred $N$ upward relative to a number-averaged value. The number-averaged real-space layer count is therefore expected to be *below* the 2–3 reported here, plausibly close to $N \sim 1$–$2$ in number-average terms, that is, fewer than one gallery per object on number average. This same weighting bias sharpens what the second order tells us: a sub-population of well-ordered, many-layer stacks would dominate the coherent signal and produce a sharp, strong $2q^*$ peak. The second order actually observed is weak and broad, consistent with the coherent signal being dominated by short stacks of only a few layers rather than by any well-ordered, many-layer minority. The scattering-weighted $N \approx 2$–$3$ is thus the upper end of a real-space distribution that includes a substantial fraction of singlets and pairs.

**Instrumental resolution.** The extracted stacking-domain sizes are sample-limited, as established by the following analysis. For the SAXS measurements (CoSAXS, MAX IV), the monochromator bandwidth is $\Delta E/E \approx 2 \times 10^{-4}$[11], so the total $\Delta q/q$ at the lamellar peak is well below 1%, negligible compared with the intrinsic relative peak width of $\sim 1/N \approx 0.3$–$0.5$ for stacking domains of $N \approx 2$–$3$ layers. The SAXS peak widths are therefore unambiguously sample-limited and provide the primary evidence for finite stacking domains.

For the time-of-flight SANS measurements (SANS2D, ISIS), the situation is more involved. The instrument-calculated resolution $\Delta Q/Q$ provided by the Mantid data-reduction pipeline for the actual experimental configuration (run 80504; $\lambda = 1.75$–$12.5$ Å, sample–detector distance 12 m, beam diameter 8 mm) ranges from approximately 12% at the higher-$q$ end of the lamellar peak range ($q \approx 0.004$ Å$^{-1}$) to approximately 19% at the lowest $q$ values ($q \approx 0.002$ Å$^{-1}$). This SANS instrumental broadening is not negligible compared with the intrinsic peak width of $\sim 40\%$, and could in principle broaden the observed peaks beyond the true sample width, leading to an apparent stacking-domain size somewhat shorter than the true value. However, the SAXS

resolution ($< 1\%$) is more than an order of magnitude narrower, and SAXS and SANS yield the same extracted stacking-domain sizes and spacing-disorder parameters within uncertainty. If the SANS domain size were substantially shortened by instrumental broadening, the SAXS measurements, performed at negligible resolution broadening, would return systematically longer stacking-domain sizes. They do not. The agreement between SAXS and SANS therefore demonstrates that the finite stacking-domain size is a genuine sample property: the SAXS result establishes it at negligible resolution, and the SANS result independently confirms it from a fundamentally different contrast mechanism. The same few-layer stacking-domain size is also reproducible rather than batch- or preparation-specific. It is obtained across different fluorohectorite synthesis batches (and, for repeat aliquots, within the same batch), for stock suspensions prepared mainly by one of us (P.H.M.-B.) and for cuvette and capillary samples filled by several others, and from both SAXS and SANS despite their different contrast mechanisms and instrumental resolutions. It is also consistent with the peak widths reported independently by Trigueiro Neto *et al.*[4], obtained on separately prepared samples. Taken together, these checks indicate that the few-layer stacking-domain size is an intrinsic property of the osmotically swollen state.

The finite-domain lamellar state inferred from the scattering data remains compatible with strong optical functionality because optical interference depends primarily on locally preserved interlayer spacing rather than extended stacking order. The experimentally observed spacing distortions therefore broaden the reflected spectral distribution while preserving the local spacing periodicity scales associated with the pressure-selected spacing $d^*$. This decoupling of structural colour from long-range order parallels disordered colloidal photonic materials, in which short-range positional order alone, without long-range periodicity, suffices to produce bright structural colour[12].

**Why few-layer domains still produce bright structural colour.** That stacking domains of only 2–3 layers can nonetheless generate vivid structural colour is not paradoxical, and follows from three observations. First, bright interference colour requires a well-defined reflecting spacing, not a tall stack: the most familiar structural colours, soap films, oil slicks and Newton's rings, arise from only one or two interfaces, where a single fluid layer between two surfaces is sufficient to produce saturated, angle-dependent colour. Adding layers narrows and intensifies the reflection but is not a prerequisite for it. Second, the spectral *position* of the reflection is set by the shared pressure-selected spacing $d^*$, whereas the stacking-domain height controls only the spectral *width* of each domain's contribution; a population of short domains that share a common $d^*$ therefore reflects at a common wavelength. Third, and most importantly, the macroscopically observed colour is an *incoherent* sum over an enormous ensemble of such domains. Along a 1 mm optical path with $d^* \approx 150$–$300\,\text{nm}$, the beam traverses $\sim 3 \times 10^3$–$7 \times 10^3$ layers, i.e. $\sim 10^3$–$3 \times 10^3$ short stacking domains in depth. This is multiplied by the number of domains illuminated in parallel across the lateral footprint of the probe: for the integrating-sphere spectrometer used here (illuminated spot $\sim 8\,\text{mm}$ in diameter, area $\sim 50\ \text{mm}^2$) and platelets of average lateral size $\sim 10\ \mu\text{m}$ (area $\sim 100\ \mu\text{m}^2$), a single layer already spans $\sim 5 \times 10^5$ platelets. The total number of finite domains sampled by the measurement is therefore of order $10^8$–$10^9$, and their incoherent superposition, all centred on the same $d^*$, yields a bright, spectrally well-defined colour even though each individual domain is only a few layers thick. In this picture the stacking-domain height governs the *sharpness* of the reflection, while the *brightness* is set by the number of co-aligned domains sampled, which remains enormous in a millimetre-scale suspension. Although singlets are numerically the most abundant species (Sec. S10) and are optically inactive, removing them leaves of order half the ensemble, still $\sim 10^8$ optically active multilayer domains across the

millimetre path and 8 mm footprint, which is far more than sufficient to produce a bright colour; sedimentation further enriches this active fraction and intensifies the colour.

The figure below displays the full SANS dataset.

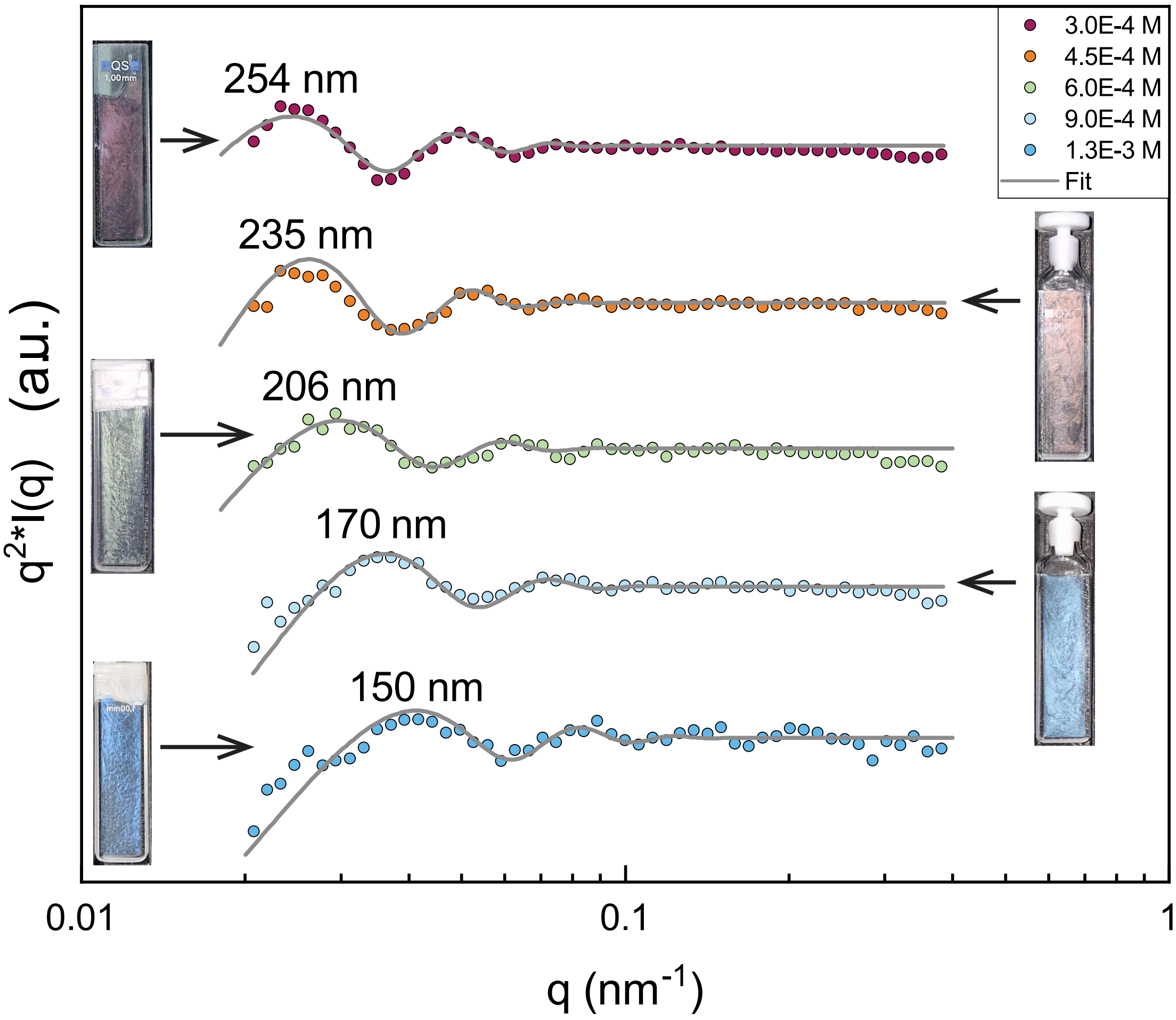


Figure S5: **SANS intensity profiles of DBL suspensions at different reservoir salinities.** Small-angle neutron scattering intensity curves $I(q)$ measured for DBL suspensions equilibrated with five reservoir salinities. The data are shown together with model fits used to extract the lamellar spacing and stacking-domain size, and are plotted after removal of the dominant single-particle form factor (after scaling by $q^2$) to emphasise the lamellar structure factor. The positions of the principal peaks correspond to the interlayer spacing through $d^* = 2\pi/q_{\text{peak}}$. The higher-order lamellar peaks are weak and broad, indicating substantial interlayer displacement disorder and limited stacking order. Analysis of the peak widths yields stacking-domain sizes corresponding to only a few layers (typically 2–3), reflecting a scattering-weighted average over a distribution of finite stacking domains. Photographs of the samples in the SANS cuvettes illustrate the structural colours produced by optical interference from the swollen lamellar stacks.

**Salinity dependence of structural parameters.** The interlayer spacing decreases systematically with increasing salinity, while the number of registered layers remains approximately constant (two to three); the stacking-domain size $L_{\text{dom}}$ itself decreases with salinity, like the spacing. The relative displacement disorder likewise remains large throughout the explored salinity range. The largest value in the series, $\Delta d/d = 35.9\%$ (DBL, $\phi_{\text{prep}} = 0.0068$, $1.3 \times 10^{-3}$ M, Table S2), occurs for the most compressed, highest-salinity capillary sample, where the smallest spacing and the correspondingly weakest restoring response make the lamellar stack most susceptible to mechanically introduced registry distortion; it is consistent with the overall trend rather than an analysis artefact.

These trends indicate that the scattering signal reflects a lamellar system in which the pressure-selected spacing $d^*$ is controlled by osmotic conditions, whereas the stacking domains remain limited by the weak restoring response of the swollen lamellar structure.

The figure below preserves the full structural-parameter summary. A simplified version emphasising the central trends is shown in the main text, while the complete parameter overview is retained here.

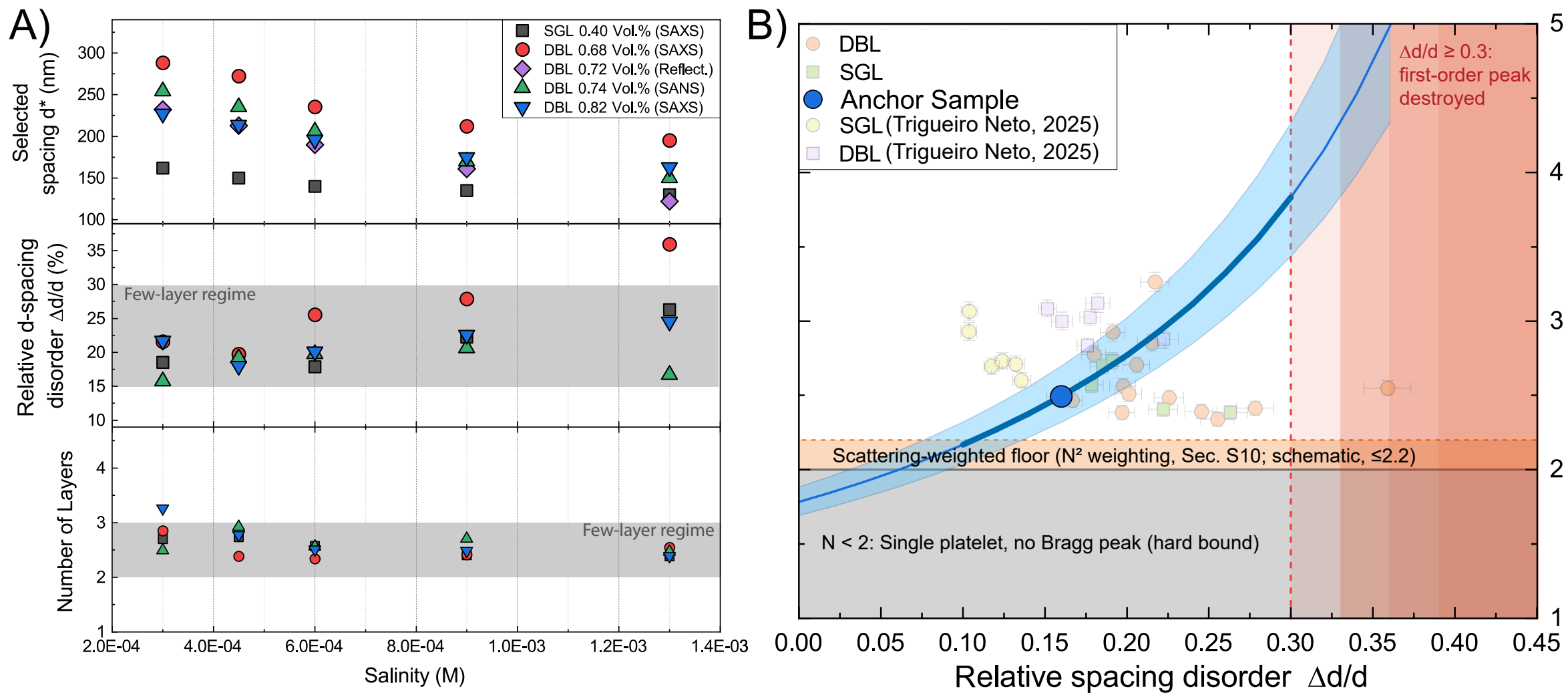


Figure S6: **Structural parameters and their joint confinement.** **(a)** The interlayer spacing $d^*$, the number of registered layers $N = L_{\mathrm{dom}}/d^* + 1$, and the relative interlayer displacement disorder $\Delta d/d$, as functions of reservoir salinity $n_0$, from Förster-model fits to SAXS, SANS and reflectivity (five series; see legend). Increasing salinity reduces the spacing, while $N$ and $\Delta d/d$ remain approximately constant; $N$ lies between two and three across the range, reaching about 3.3 at the lowest salinity. Error bars (SAXS series) give the spread across three independently prepared replicates: below 2 % on $d$ and $N$ and about 4 % on $\Delta d/d$; SANS and reflectivity points are single measurements. **(b)** The same data in the $N$–$\Delta d/d$ plane. The two quantities are degenerate in the peak-width fit yet confined to a small corner by three bounds: the hard interference floor $N \geq 2$ (a single platelet gives no Bragg peak); a scattering-weighted floor (the $N^2$ weighting of Sec. S10 lifts the reported $N$ above the number average; schematic, $\lesssim 2.2$); and the disorder wall ($\Delta d/d \approx 0.3$–$0.4$, where the first-order peak is progressively destroyed). All 20 samples of this work (Table S2) lie in the corner; the highlighted anchor sample is the one whose free-energy well is reconstructed in Sec. S8. The earlier data of Trigueiro Neto *et al.*[4] (zero added salt; DBL and SGL) are overlaid as individual faint points and fall in the same corner at comparable or lower disorder, so the few-layer, bounded-disorder state is common to both datasets. The confinement is set by interference physics, not by fitting statistics.

**Extracted structural parameters from scattering fits.** The parameters shown in Fig. S6 are obtained from fits to the SAXS and SANS data using the Förster model, as described above. The interlayer spacing $d^*$ is determined from the position of the principal lamellar peak, while the displacement disorder and stacking-domain size are extracted from the peak shape and width.

The results show that the interlayer spacing decreases systematically with increasing salinity, reflecting the salinity dependence of the osmotic swelling balance. In contrast, the number of registered layers remains limited to a few (two to three), while the relative displacement disorder remains approximately constant over the explored range. The reproducibility of these parameters was checked with a SAXS triplicate on independently prepared samples: the sample-to-sample spread is below 2 % on the interlayer spacing and on the number of registered layers and about 4 % on the relative displacement disorder $\Delta d/d$, while the precision of a single fit ($\approx 1$ %) is smaller.

These observations are consistent with the picture that the system forms a lamellar structure

with a well-defined spacing but without the development of extended stacking order. The complete parameter analysis presented here provides the experimental basis for the decoupling between spacing periodicity and stacking order discussed in the main text.

## S4. RGB analysis and optical interpretation of the sediment colour gradient

**Sedimentation and emergence of structural colour.** In the main text (Fig. 4), we analyse the optical response of a fully developed, vertically stratified sediment. Before reaching this final state, however, the system undergoes a slow sedimentation and structural reorganisation process. This evolution is illustrated in Fig. S7, which shows the gradual formation of the sediment together with the emergence of structural colour.

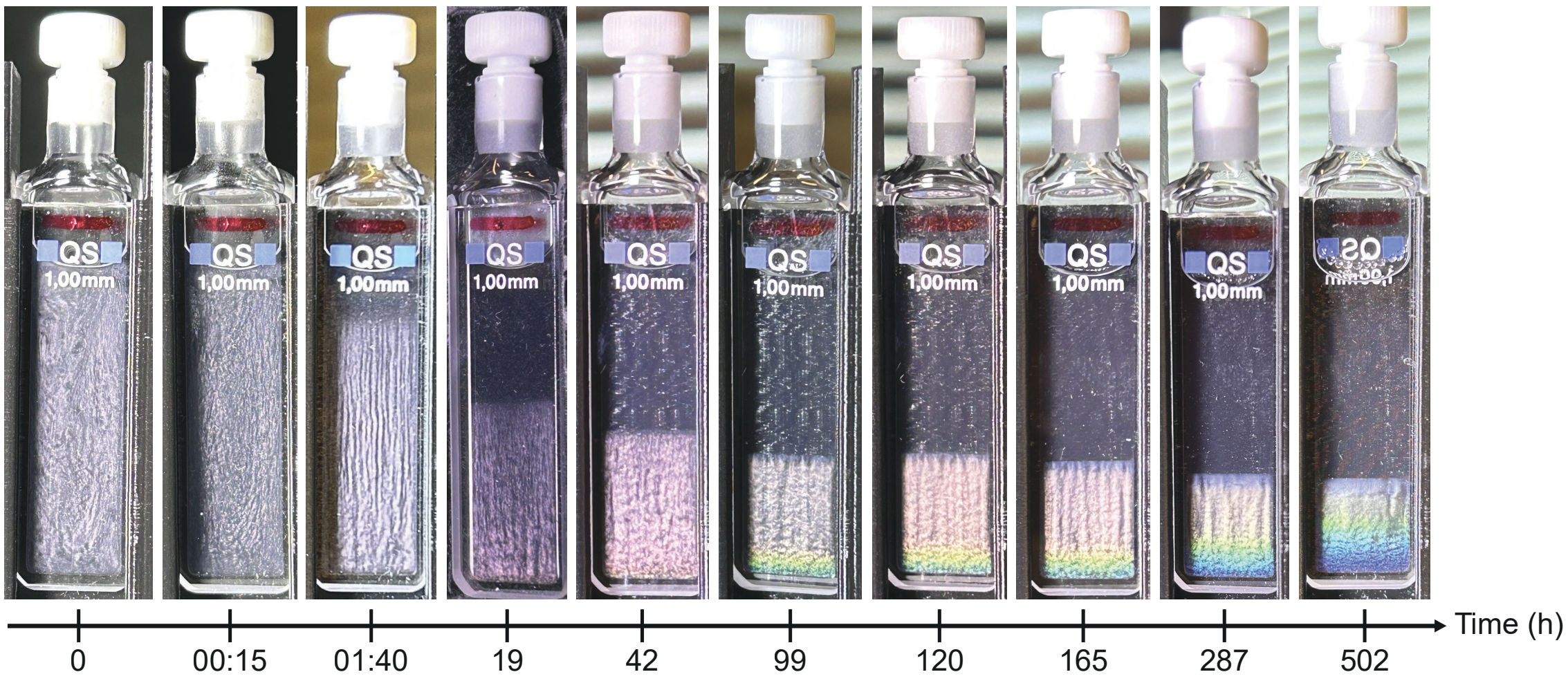


Figure S7: Time evolution of sedimentation and structural colour development in the nanosheet suspension. The sequence shows the gradual formation of a vertically stratified sediment with increasing optical contrast and colour separation over time. The emergence of a colour gradient reflects the development of both a pressure gradient and a redistribution of lamellar stack populations during sedimentation.

After sedimentation, the system reaches the stratified state analysed in the main text (Fig. 4). The procedure used to extract the RGB signal from this sediment is shown in Fig. S8.

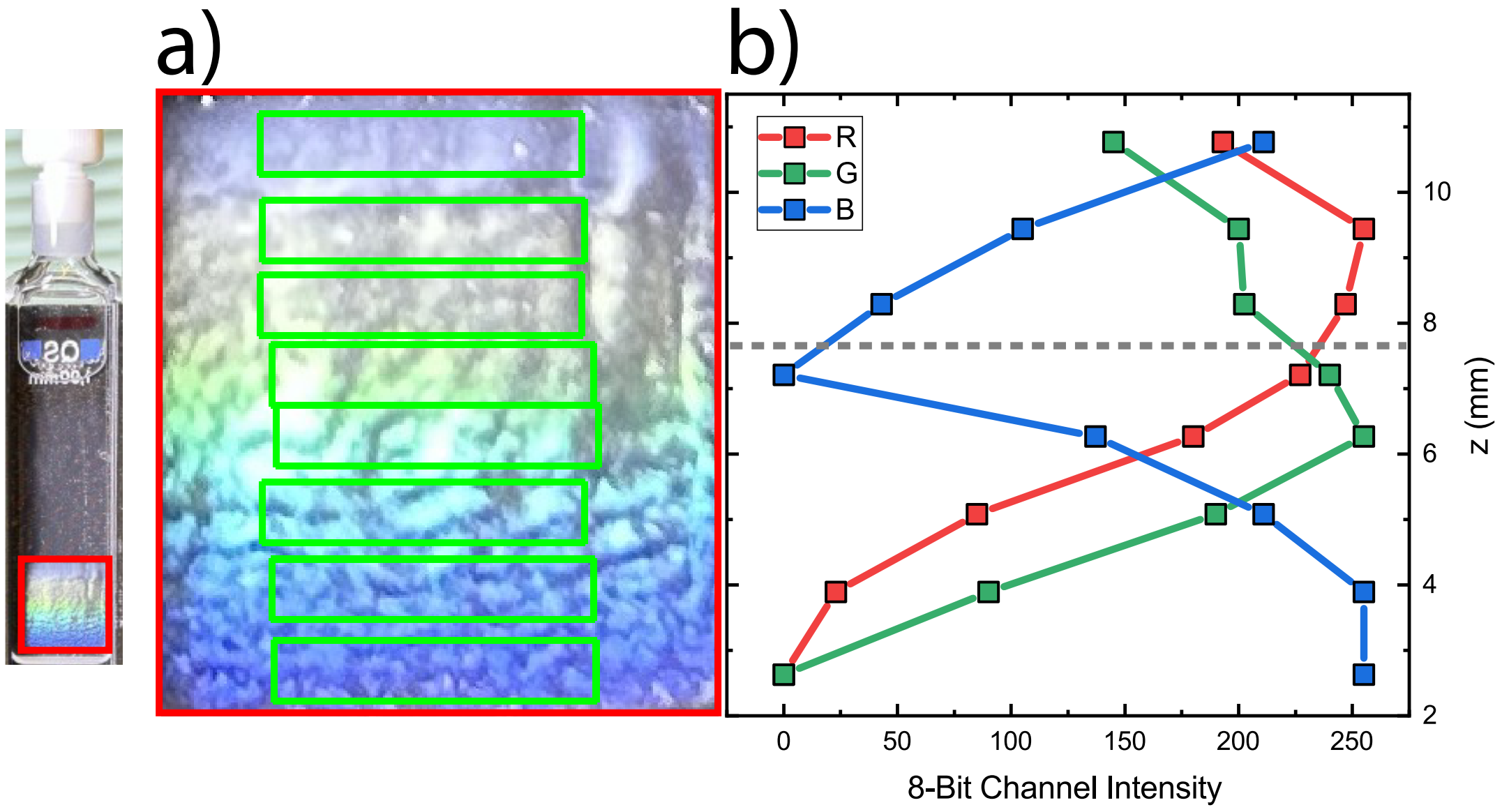


Figure S8: Sediment photograph and corresponding RGB analysis. (a) Photograph of the sediment column, with marked regions used for RGB extraction along the vertical direction. (b) Red, green, and blue channel intensities extracted from these regions as a function of height.

**RGB analysis of the stratified sediment.** The RGB analysis reveals a clear vertical colour gradient. The bottom of the sediment exhibits a bright, strongly blue-dominated response, while the upper part appears increasingly red-shifted. However, the upper sediment does not form a pure red band: a substantial blue contribution remains, resulting in a mixed and comparatively desaturated colour.

This behaviour reflects the combined effects of sedimentation, confinement-pressure variation, and stack-size selection.

**Population gradients generated during sedimentation.** During sedimentation, larger and heavier lamellar stacks sediment more rapidly and accumulate towards the bottom of the sample, while smaller stacks and singlets remain suspended for longer times. As a result, the sediment develops a vertical population gradient: the bottom region becomes enriched in larger multilayer stacks, whereas the upper sediment contains an increasing fraction of smaller stacks and residual singlets.

Stack-size selection may influence not only the coherence of the optical response, but also the local spacing selected at a given height. If smaller lamellar objects remain suspended for longer times or are advected upward during sedimentation, while larger multilayer stacks sediment downward more efficiently, then the local population of nanosheet assemblies becomes height-dependent. In that case, the empirical relation between spacing and concentration used to reconstruct $\phi(z)$ should be interpreted as an effective average over a changing population of lamellar objects rather than as a strictly universal equation of state.

This possibility is qualitatively consistent with previous studies of nanosheet colloids showing strong particle-size effects on liquid-crystalline organisation,[13] as well as with binary niobate–clay nanosheet systems in which mesoscopic composition-dependent ordering and phase coexistence were observed.[14]

**Pressure gradients and spacing selection.** At the same time, the external confinement pressure increases with depth due to the weight of the overlying material. This pressure gradient

selects different mean interlayer spacings along the vertical direction: smaller spacings at the bottom and larger spacings towards the top.

These two gradients have distinct optical consequences. First, enrichment of multilayer stacks at the bottom enhances coherent interference, leading to a bright and relatively pure structural colour. This is consistent with previous work in which removal of singlets by centrifugation produced highly saturated structural colours.[1] In the present case, sedimentation provides a natural, spatially resolved analogue of this purification process. The corresponding conditional redistribution of stack populations is discussed quantitatively in Sec. S11 and illustrated in Fig. S18.

Second, the increase in interlayer spacing towards the top shifts the reflected wavelength to longer values.

**Weak-coupling broadening of the optical response.** Importantly, optical interference depends primarily on local spacing periodicity rather than extended stacking order. Fragmentation of the swollen lamellar structure into finite domains therefore does not eliminate its spacing-selective functionality, provided the local spacing remains centred around the pressure-selected spacing $d^*$.

Because the restoring slope $k(d) = \partial\Pi/\partial d$ is exceptionally small in the swollen regime, the system supports a broad distribution of local spacings rather than a single sharply defined value. Local spacing distortions ($d < d^*$ or $d > d^*$) therefore broaden the reflected spectral distribution, so that even in the red-shifted upper sediment a significant blue contribution persists.

In addition, if the upper part of the sediment is enriched in smaller stacks and residual singlets, as suggested by the sedimentation-driven population gradient, then the local optical response may reflect not only larger mean spacing but also a different local population composition. The residual blue contribution at the top should therefore be understood as arising from both broad spacing distributions and incomplete compositional purification of the lamellar stack population.

The increasing fraction of singlets and small stacks towards the top further reduces optical coherence and intensity, contributing to the more diffuse and less saturated appearance of this region.

Taken together, the observations indicate coexistence of:

- a pressure gradient, selecting the mean spacing and hence the central reflected wavelength,
- a population-composition gradient, controlling the degree of coherent optical reflection.

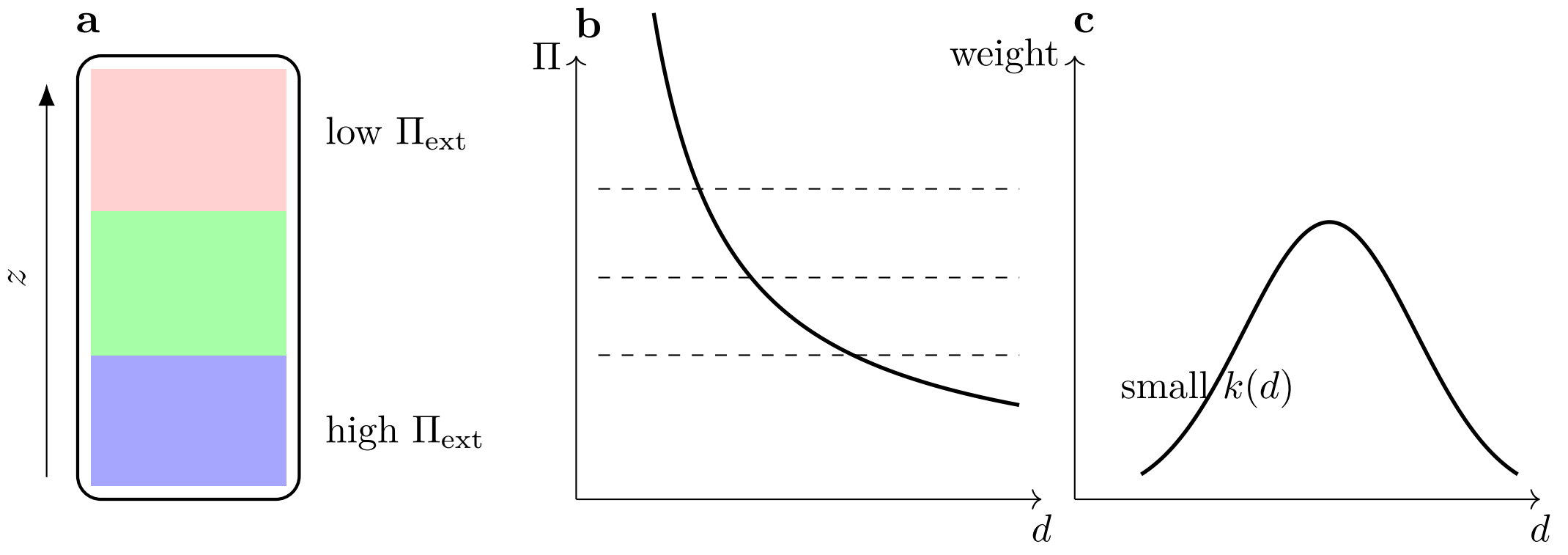


Figure S9: Conceptual interpretation of the sediment colour gradient. (a) A vertical confinement-pressure gradient develops during sedimentation. (b) The local interlayer spacing is selected by the balance between osmotic and external pressures. (c) Weak restoring slope allows a distribution of spacings, leading to a broadened optical response.

**Time evolution and stripe formation during sedimentation.** To further resolve how the optical response develops during sedimentation, we examine the time evolution of both the reflectance spectra and the spatial structure of the sample. These measurements provide insight into the dynamical processes that lead to the final stratified state.

In particular, the sedimentation process is not spatially uniform. Instead, the system exhibits transient vertical heterogeneities that reflect ongoing material transport and rearrangement during sediment formation.

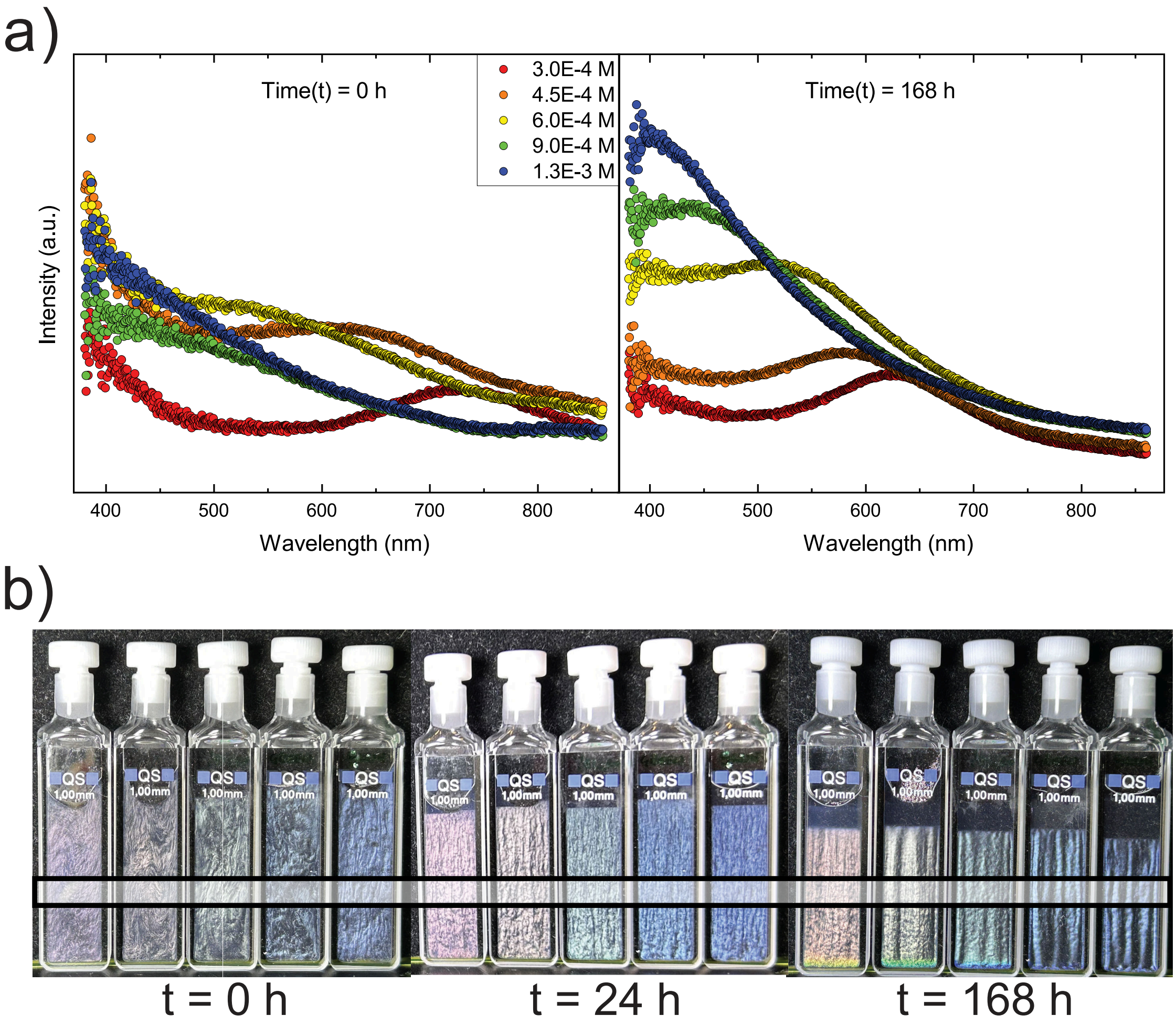


Figure S10: Time evolution of the optical response during sedimentation. (a) Reflectance spectra recorded during sedimentation. (b) Time-dependent images showing the evolving optical response and the emergence of a vertical stripe pattern associated with spatially heterogeneous material transport and structural rearrangement. The horizontal marked region indicates the location of the integrating-sphere reflectometer used for the spectra shown in panel (a).

A striking feature of the sedimentation process is the emergence of a vertical stripe pattern, visible in Fig. S10. These stripes indicate that sedimentation proceeds through spatially heterogeneous dynamics rather than uniform settling.

One possible interpretation is that these bands correspond to regions of differential material transport, with some regions dominated by downward sedimentation of larger stacks, while others accommodate upward displacement of fluid together with smaller particles or shorter lamellar domains. The stripe pattern may therefore represent transient fractionation pathways superimposed on the overall settling process, leading to mesoscopic heterogeneity in both local population composition and optical response.

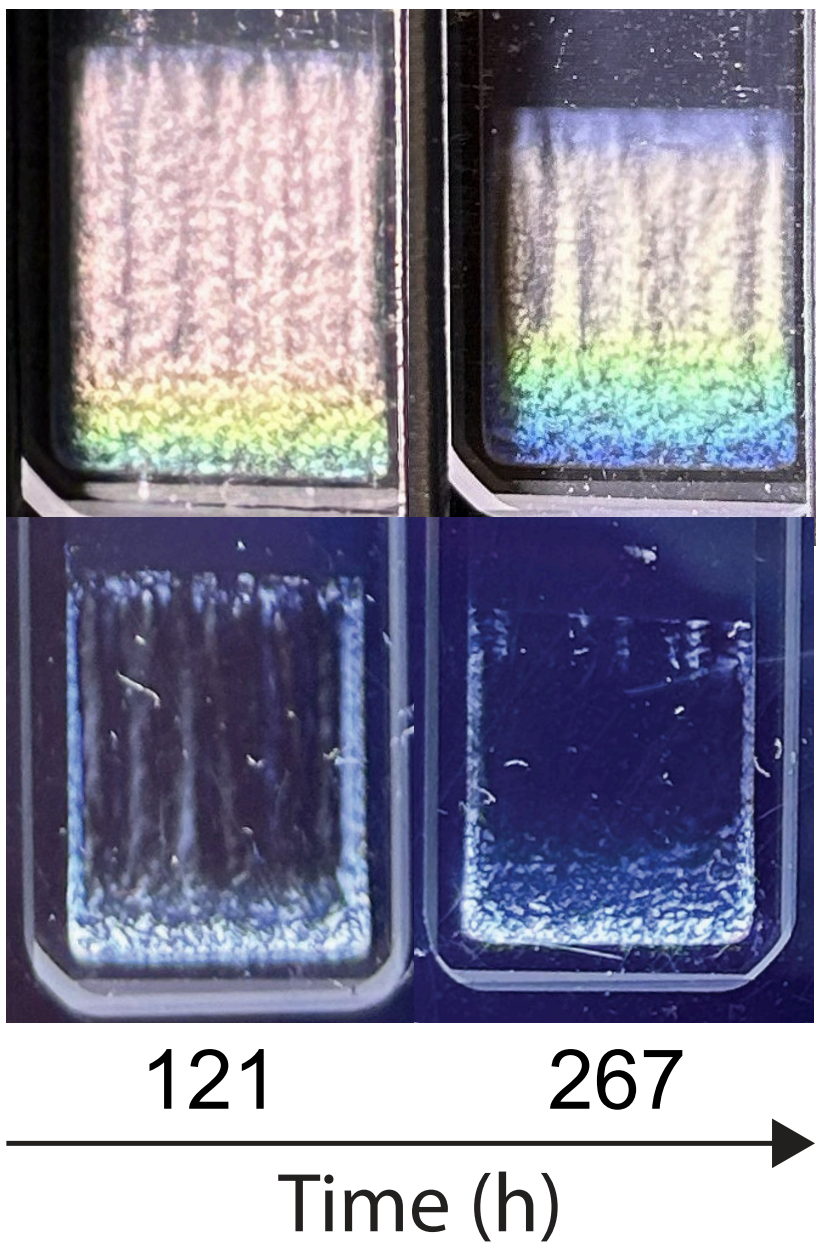


Figure S11: Bright-field (top row) and cross-polarised (bottom row) reflection images of the same sediment region at two times during sedimentation (121 and 267 h). Under crossed polarisers the dark vertical bands still contain birefringent material, demonstrating that the stripe pattern does not correspond to empty regions but rather to regions of differing internal structure and orientation.

**Cross-polarised imaging of the stripe structure.** To clarify the nature of these stripe patterns, we examine the sediment under crossed polarisers (Fig. S11). The dark vertical bands observed in polarised light still contain material, demonstrating that they are not voids or regions depleted of nanosheets. Instead, they correspond to regions with different internal organisation or orientation of the lamellar structures.

This indicates that the stripe pattern reflects internal structural heterogeneity and flow-driven rearrangement rather than macroscopic phase separation.

**Late-stage structural evolution after sedimentation.** At later times, after the macroscopic sediment height has nearly stabilised, the system continues to evolve. This late-stage behaviour is illustrated in Fig. S12.

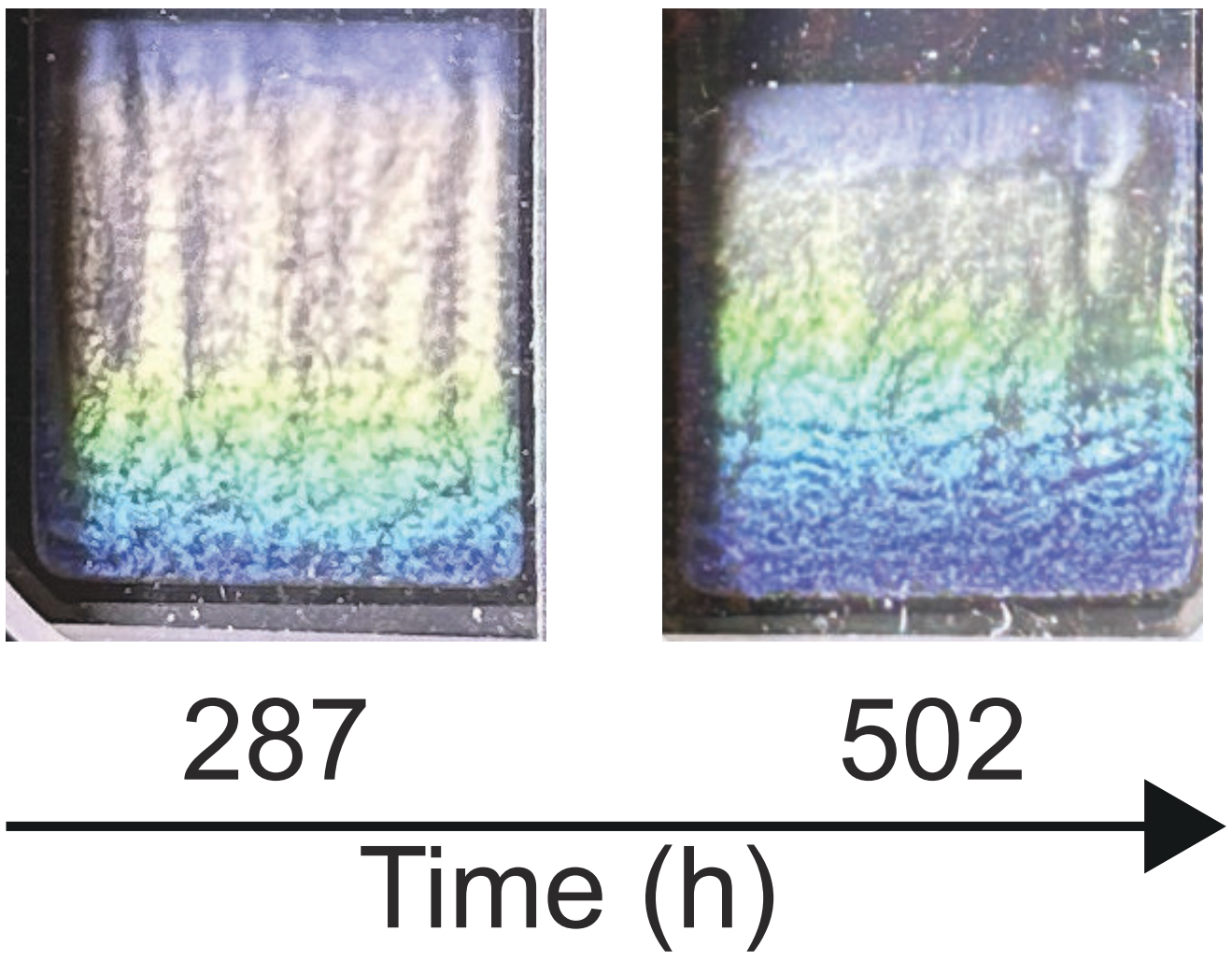


Figure S12: Near-final sediment imaged at 287 and 502 h, showing a continued increase in colour brightness over time. Even after the sediment height has stabilised, the optical response continues to evolve, indicating ongoing internal structural rearrangements.

Beyond sediment formation, the system continues to evolve. Even after the sediment height has reached its near-final value, a further increase in colour brightness is observed. This indicates ongoing internal rearrangements within the sediment on timescales much longer than the sedimentation process itself.

Such behaviour points to slow relaxation of stacking configurations, likely involving partial restacking of short domains together with gradual reduction of interlayer registry disorder. The vertical stripe pattern observed during sedimentation suggests that material is continuously redistributed, with regions of downward sedimentation coexisting with upward flow.

Sedimentation-induced fractionation may help prepare the lower region by enriching it in larger multilayer objects, but the continued post-sedimentation brightening is more naturally attributed to increasing stacking order and reduced registry disorder within that enriched population than to a continuing monotonic shift of the mean spacing alone.

These slow dynamics are a direct consequence of the weak-coupling regime identified in the main text. Because the restoring response is exceptionally weak, spacing distortions and registry mismatches relax only inefficiently and therefore persist over long timescales. As a result, structural reorganisation continues even after macroscopic sedimentation is complete; what this slow, history-dependent relaxation implies, and its analogy to ageing in disordered systems, is developed in Sec. S12.

The sediment therefore provides a direct, spatially resolved manifestation of a mechanically soft lamellar system in which weak restoring response allows a broad distribution of local spacing configurations to persist rather than efficiently restoring a unique local spacing.

**Pressure reconstruction and weak-confinement interpretation.** The figure below preserves the full sedimentation and pressure-reconstruction figure. In the main manuscript, a simplified sediment figure is used, while the full reconstruction is retained here.

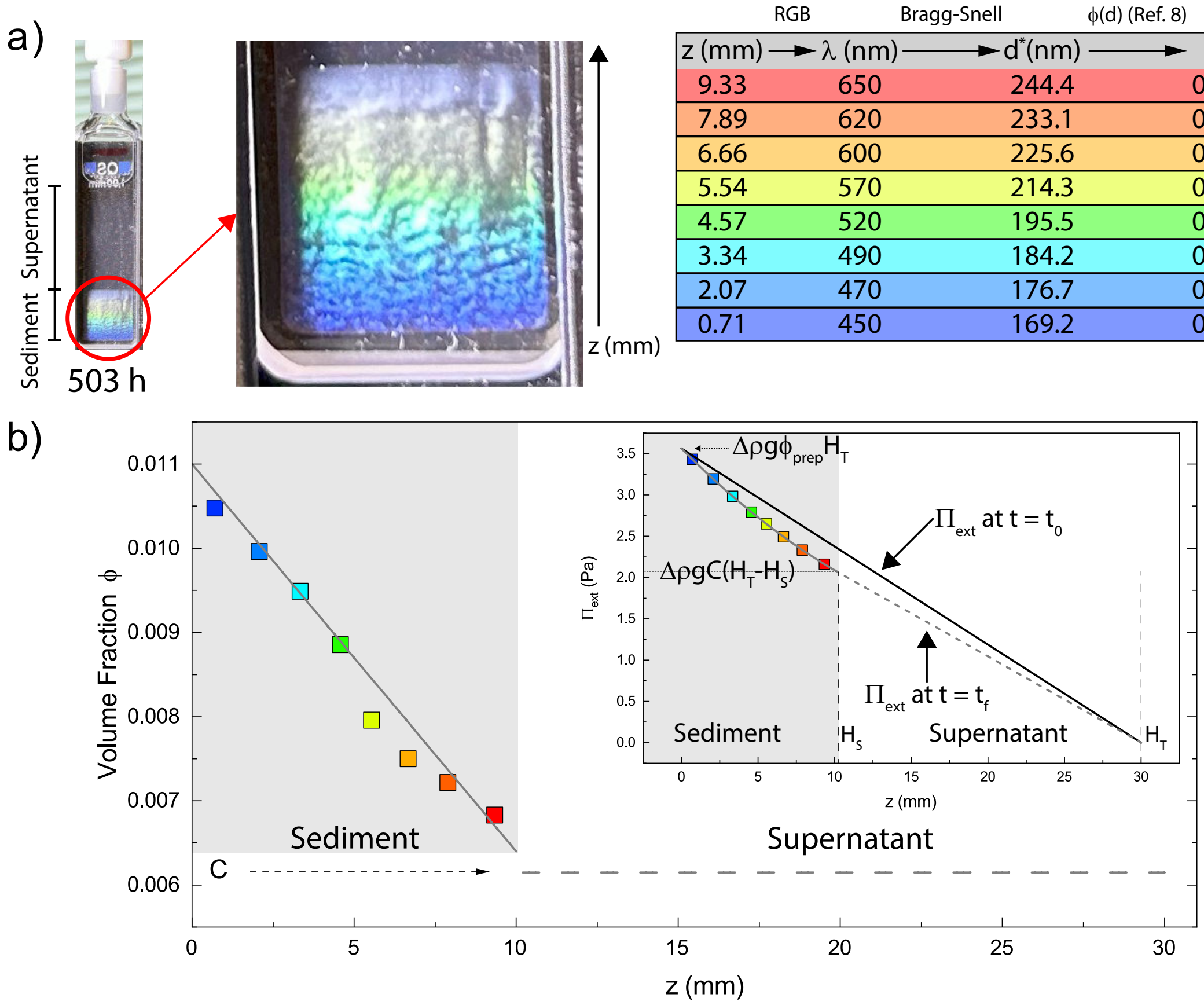


| z (mm) | λ (nm) | d* (nm) | ϕ |
|---|---|---|---|
| 9.33 | 650 | 244.4 | 0.0068 |
| 7.89 | 620 | 233.1 | 0.0072 |
| 6.66 | 600 | 225.6 | 0.0075 |
| 5.54 | 570 | 214.3 | 0.0080 |
| 4.57 | 520 | 195.5 | 0.0088 |
| 3.34 | 490 | 184.2 | 0.0095 |
| 2.07 | 470 | 176.7 | 0.0100 |
| 0.71 | 450 | 169.2 | 0.0105 |

Figure S13: **Sedimentation experiment used to estimate the confinement-pressure scale.** (a) Photograph of a DBL suspension after 503 hours of sedimentation. The vertical colour gradient reflects a spatial variation of the interlayer spacing associated with the increasing clay concentration towards the bottom of the sample. RGB colour values extracted from the image are converted to reflected wavelength and subsequently to interlayer spacing using the Bragg–Snell relation at normal incidence, $\lambda = 2n_{\rm eff}d$, and subsequently into clay volume fraction using an empirical power-law relation.[4] (b) Main panel: reconstructed clay volume-fraction profile $\phi(z)$ along the column (markers), with a fit (grey) that approaches the residual supernatant volume fraction $C$ at large height; the grey shaded area marks the sediment. Inset: external confinement-pressure profile $\Pi_{\rm ext}(z) = \Delta\rho\, g \int_z^{H_T} \phi(z')\, dz'$, shown for the initial homogeneous state ($t = t_0$, solid) and the final stratified state ($t = t_f$, dashed). The intercepts $\Delta\rho\, g\, \phi_{\rm prep}\, H_T$ and $\Delta\rho\, g\, C\, (H_T - H_S)$ give the pressures at the cell bottom for the two states, where $H_T$ is the total sample height, $H_S$ marks the top of the sediment, $\phi_{\rm prep}$ is the prepared volume fraction and $\Delta\rho$ the nanosheet–medium density contrast. The resulting pressure scale is of order only a few pascal, indicating that swollen lamellar stacks experience extremely weak external confinement.

**Pressure profile extracted from sedimentation.** The external confinement-pressure profile is obtained by integrating the concentration profile over height using the density difference between clay and solvent. The resulting pressure varies smoothly with depth and remains on the order of only a few Pascal across the sediment.

The inset in Fig. S13 shows the reconstructed confinement pressure as a function of height. The nearly linear variation reflects the weak gravitational loading and the low overall clay volume fraction.

These results indicate that the lamellar stacks are subject to extremely weak confinement, providing the physical basis for the weak-coupling regime discussed in the main text.

**Connection to the weak-coupling picture.** The broadened optical response observed in the sediment is consistent with the weak restoring slope identified in the main text. Because

the restoring slope $k(d) = \partial\Pi/\partial d$ is exceptionally small over the relevant spacing range, the system supports a broad distribution of local spacings rather than a single sharply defined value. As a result, even at fixed height the optical response reflects a superposition of local spacing configurations rather than a unique spacing.

## S5. Estimate of van der Waals attraction

For two nanosheets in the colloidal accordion separated by a distance $d$, the attractive van der Waals pressure is given by [15]

$$\Pi_{\rm vdW} = 20B\frac{h^2}{d^6}, \tag{S2}$$

where $B$ is the Hamaker constant, $h$ is the nanosheet thickness (1 nm for SGL and 2 nm for DBL), and $d$ is the interlayer spacing. Using $B \sim 10^{-28}$ J m and the experimentally measured spacing range $d = 150$–300 nm, one obtains van der Waals pressures of order

$$\Pi_{\rm vdW} \sim 10^{-7}\text{–}10^{-5}\ \text{Pa}. \tag{S3}$$

The corresponding sample-by-sample values are listed in Table S1.

The van der Waals attraction is therefore negligible compared with both the Donnan osmotic pressure and the confinement-pressure scale inferred from sedimentation.

## S6. Estimate of classical electrostatic double-layer repulsion

At the large interlayer spacings investigated in the present work, classical DLVO electrostatic double-layer interactions are negligible compared with the Donnan osmotic pressure associated with ion partitioning between the galleries and the external reservoir.

For a monovalent electrolyte, the Debye screening length in water at room temperature is well approximated by [16]

$$\lambda_D(\text{nm}) \approx \frac{0.304}{\sqrt{n_0(\text{M})}}, \tag{S4}$$

where $n_0$ is the ionic strength in molar units. For the salinity range used in this work, the corresponding Debye lengths are

$$\begin{aligned} n_0 &= 3.0\times10^{-4}\,\text{M} \Rightarrow \lambda_D \approx 18\ \text{nm},\\ n_0 &= 4.5\times10^{-4}\,\text{M} \Rightarrow \lambda_D \approx 14\ \text{nm},\\ n_0 &= 6.0\times10^{-4}\,\text{M} \Rightarrow \lambda_D \approx 12\ \text{nm},\\ n_0 &= 9.0\times10^{-4}\,\text{M} \Rightarrow \lambda_D \approx 10\ \text{nm},\\ n_0 &= 1.3\times10^{-3}\,\text{M} \Rightarrow \lambda_D \approx 8\ \text{nm}. \end{aligned} \tag{S5}$$

These screening lengths are much smaller than the experimentally observed interlayer spacings,

$$d^* \approx 150\text{–}288\ \text{nm}, \tag{S6}$$

as compiled for all samples in Table S1.

Following Ref. [15], the electrostatic double-layer repulsion between two charged lamellae in the colloidal accordion can be written as

$$\Pi_{\rm Elec}(d) = \frac{8k_BT\exp\left(-\frac{d}{\lambda_D}\right)}{\pi l_B\lambda_D^2}, \tag{S7}$$

where $k_B$ is the Boltzmann constant, $T$ the temperature, $l_B$ the Bjerrum length, and $\lambda_D$ the Debye length.

For the present system (Table S1), this expression yields

$$\Pi_{\mathrm{Elec}} \sim 10^{-5}\text{–}10^{-1}\,\mathrm{Pa}. \tag{S8}$$

Classical double-layer repulsion is therefore negligible compared with both the confinement-pressure scale (0–4 Pa) and the Donnan osmotic pressure relevant to the swollen lamellar regime.

The figure below preserves the interaction-scale comparison from the original main text.

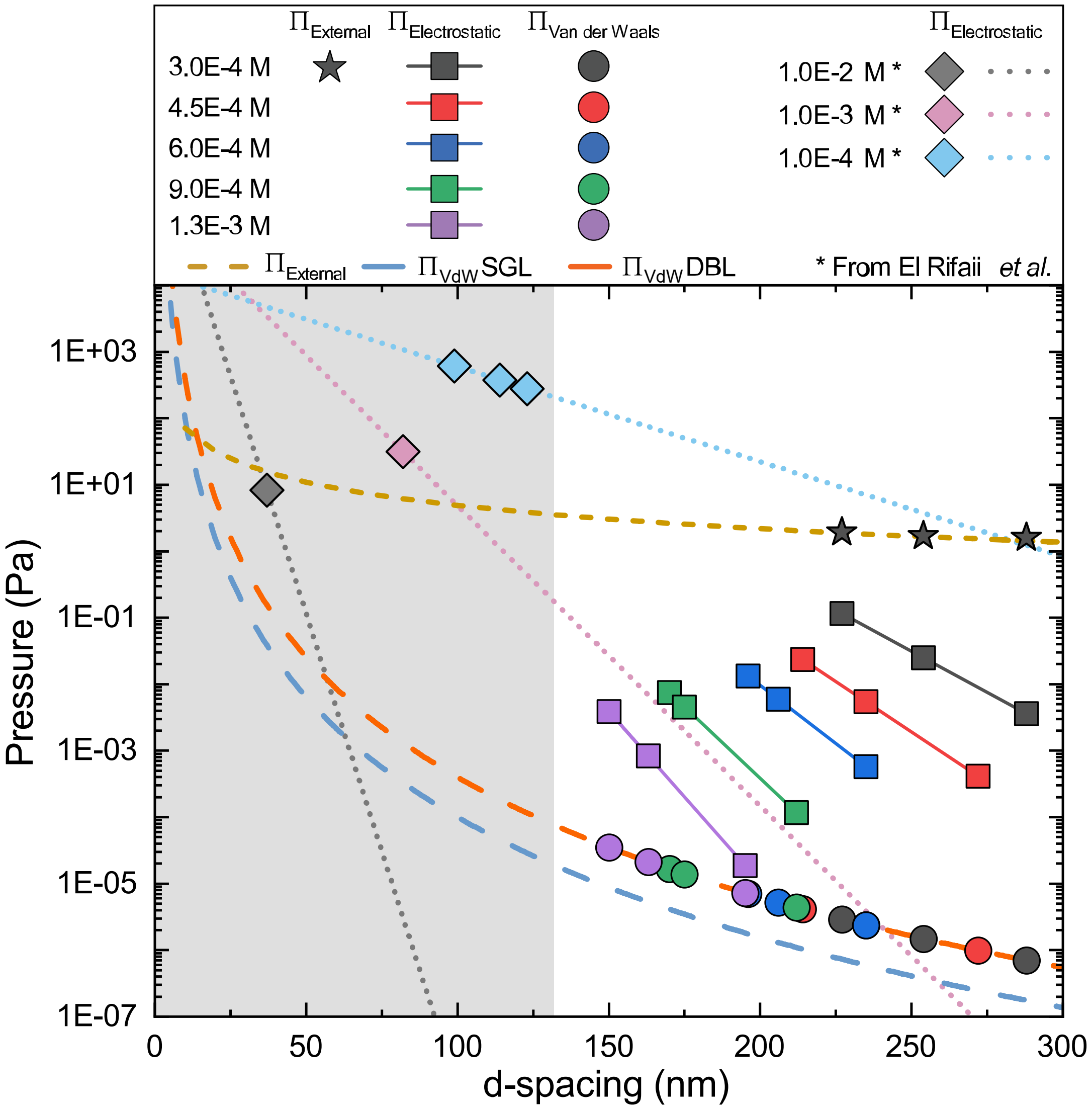


Figure S14: **Interaction pressure scales in the swollen lamellar regime.** Comparison of the confinement-pressure range inferred from sedimentation ($\Pi_{\mathrm{Ext}}$, labelled $\Pi_{\mathrm{External}}$ in the figure) with estimates of the van der Waals ($\Pi_{\mathrm{vdW}}$) and electrostatic double-layer ($\Pi_{\mathrm{Elec}}$, DLVO) interactions; these three quantities are the curves plotted. The shaded region indicates the experimental range investigated by El Rifaii *et al.*[15]. At the large interlayer spacings studied here ($d^* = 150$–$300$ nm), both the van der Waals and the classical double-layer interactions are negligible on the scale of the experimentally constrained confinement pressure; the fluctuation-induced Helfrich repulsion, estimated separately in Sec. S7, is smaller still and is not plotted here. The swollen lamellar regime is therefore governed primarily by Donnan osmotic swelling under extremely weak mechanical confinement.

**Comparison of interaction contributions.** The contributions from van der Waals attraction and electrostatic double-layer interaction are shown in Fig. S14 over the experimentally relevant range of interlayer spacings.

All three interaction contributions remain small compared with the confinement pressures inferred from sedimentation. In particular, the magnitudes of these interactions remain well below the Pascal scale throughout the experimentally accessible spacing range.

These results show that classical DLVO and fluctuation-induced interactions do not dominate the restoring response of the swollen lamellar state in the regime investigated here. Instead, the dominant contribution arises from Donnan osmotic swelling, as discussed in Sec. S8.

## S7. Estimate of fluctuation-induced Helfrich repulsion

The fluctuation-induced Helfrich repulsion between flexible sheets can be written as [16, 17]

$$\Pi_{\mathrm{H}} = \frac{3\pi^2}{128}\frac{(k_BT)^2}{\kappa d^3}, \tag{S9}$$

where $\kappa$ is the bending modulus of the nanosheets.

For silicate nanosheets with nanometer-scale thickness, the bending modulus may be estimated from thin-plate elasticity,

$$\kappa = \frac{Et^3}{12(1-\nu^2)}, \tag{S10}$$

where $E$ is the in-plane Young modulus and $t$ the sheet thickness. Using typical values $E \sim 10^{11}$ Pa for silicate nanosheets [18] and $t \sim 1\,\mathrm{nm}$ gives

$$\kappa \sim 10^{-17}\,\mathrm{J} \sim 10^3\text{–}10^4\,k_BT. \tag{S11}$$

For the spacing range relevant to the present work,

$$d^* = 150\text{–}300\ \mathrm{nm}, \tag{S12}$$

this yields

$$\Pi_{\mathrm{H}} \sim 10^{-4}\text{–}10^{-3}\ \mathrm{Pa}. \tag{S13}$$

The strong $d^{-3}$ dependence together with the large bending modulus places the system deep in a regime where fluctuation-induced repulsion is strongly suppressed.

The Helfrich repulsion is therefore several orders of magnitude smaller than both the Donnan osmotic pressure and the experimentally inferred confinement-pressure scale, and can safely be neglected.

## S8. Donnan osmotic pressure model following El Rifaii *et al.*

The osmotic pressure arises from Donnan partitioning between the electrolyte confined in the interlayer galleries and the external salt reservoir. The model assumes a monovalent electrolyte and a uniform effective charge distribution associated with the nanosheets. Following the formulation of El Rifaii *et al.*[15], the osmotic pressure is written as

$$\Pi_{\mathrm{osm}} = 2n_0k_BT\left[\sqrt{1+\left(\frac{Z_{\mathrm{eff}}n_S}{2n_0}\right)^2}-1\right], \tag{S14}$$

where $n_0$ is the salt concentration in the external reservoir, $k_B$ is the Boltzmann constant, $T$ the temperature, $Z_{\text{eff}}$ is the effective charge per sheet, and $n_S$ is the sheet number density.

The product $Z_{\text{eff}} n_S$ can be expressed in terms of an effective charge parameter $\sigma_{\text{eff}}$ and the sheet packing fraction $\phi_{\text{prep}}$, where $\phi_{\text{prep}}$ denotes the prepared volume fraction. The effective charge parameter is defined as

$$\sigma_{\text{eff}} = \frac{Z_{\text{eff}} e}{A}, \tag{S15}$$

where $A$ is the lateral area of a nanosheet. The packing fraction is given by

$$\phi_{\text{prep}} = n_S V_{\text{sheet}}, \tag{S16}$$

where $V_{\text{sheet}} = Ah$ is the sheet volume and $h$ is the sheet thickness (1 nm for SGL and 2 nm for DBL). Combining these relations gives

$$Z_{\text{eff}} n_S = \frac{\sigma_{\text{eff}} \phi_{\text{prep}}}{eh}. \tag{S17}$$

Substituting into the osmotic pressure expression yields

$$\Pi_{\text{osm}} = 2n_0 k_B T \left[ \sqrt{1 + \left( \frac{\sigma_{\text{eff}} \phi_{\text{prep}}}{2n_0 eh} \right)^2} - 1 \right]. \tag{S18}$$

In the present regime, the dimensionless parameter

$$\left( \frac{\sigma_{\text{eff}} \phi_{\text{prep}}}{2n_0 eh} \right)^2 \ll 1, \tag{S19}$$

so that the osmotic pressure reduces to the quadratic form

$$\Pi_{\text{osm}} \approx \frac{k_B T}{n_0} \left( \frac{\sigma_{\text{eff}} \phi_{\text{prep}}}{2eh} \right)^2. \tag{S20}$$

Here $\sigma_{\text{eff}}$ should be understood as an effective osmotic parameter, far below the bare structural charge, that enters the Donnan description of the swollen state. In the present regime the inter-platelet separations $d \approx 150$–$300\,\text{nm}$ are very large compared with the Debye length $\lambda_{\text{D}} \approx 8$–$18\,\text{nm}$ (Sec. S6), so $\kappa d \gg 1$ and the system is deep in the asymptotic regime of effective-charge renormalisation, where the effective charge tends to a saturated, weakly varying value set by the boundary-matching condition rather than by the bare structural charge[19, 20, 21]. We accordingly take $\sigma_{\text{eff}}$ as approximately independent of $d$ in what follows; this is the only assumption needed to evaluate the spacing derivative of $\Pi_{\text{osm}}$ in closed form, and $\sigma_{\text{eff}}$ itself drops out of the resulting expression for the restoring slope at mechanical balance (see below). El Rifaii *et al.*[15] characterised the same family of clays at smaller spacings, where DLVO double-layer and osmotic contributions both enter; the present, larger-$d$ regime is the asymptotic pure-osmotic limit where DLVO interactions have decayed below the pascal scale (Secs. S5–S7).

This small-argument approximation is well satisfied for the present data range, although it is not generally valid for the higher-concentration systems considered in Ref. [15]. The corresponding sample conditions, measured spacings, and experimental techniques used throughout this analysis are summarised in Table S3.

The figure below shows the per-sample $\sigma_{\text{eff}}(d)$ reconstruction.

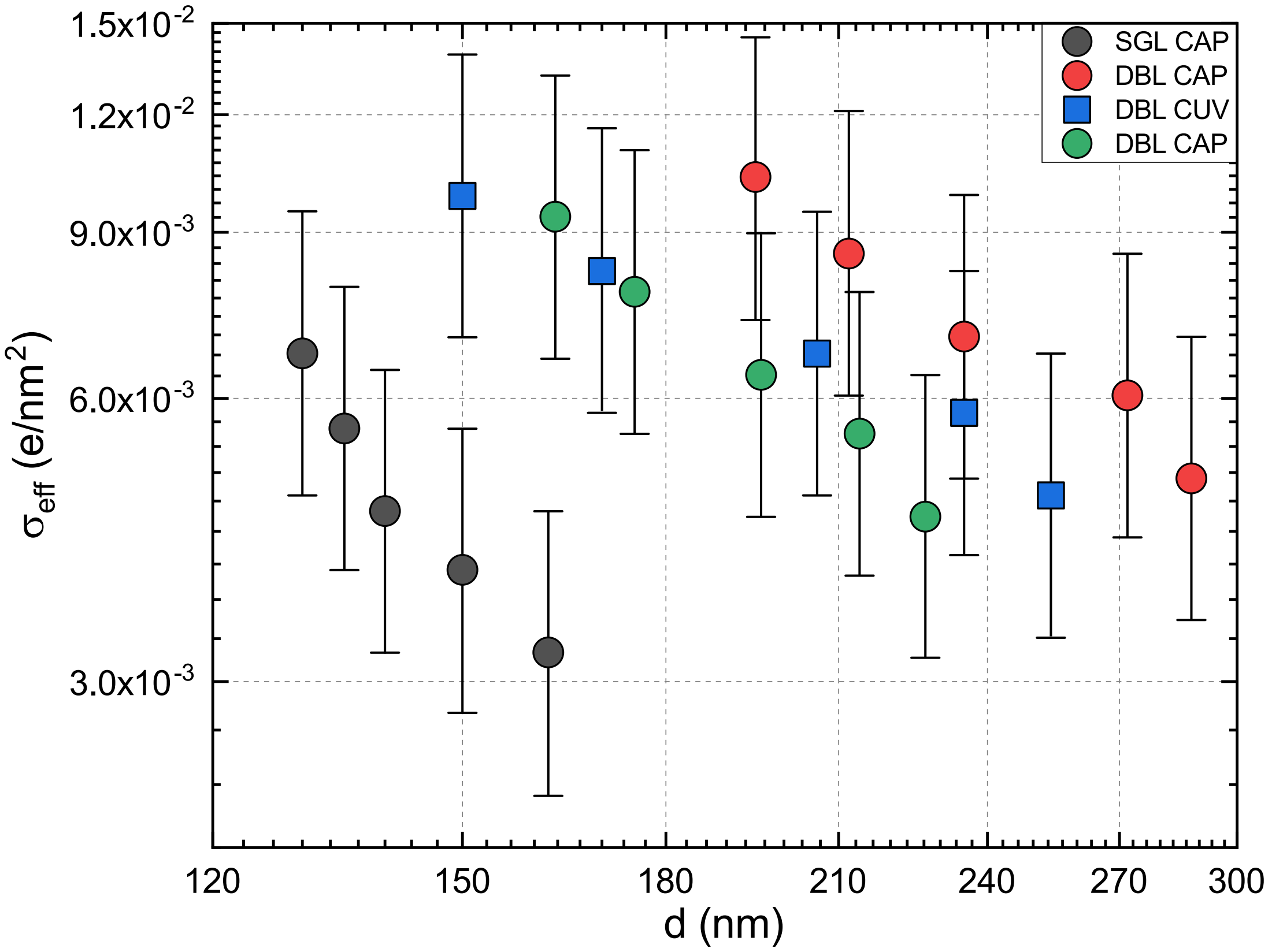


Figure S15: **Reconstructed effective charge parameter.** Per-sample effective charge parameter $\sigma_{\text{eff}}$ obtained from the Donnan inversion using the sample-specific mid-cell loading $\Pi_{\text{ext}} = \frac{1}{2}\Delta\rho\, g\, \phi_{\text{prep}} H_{\text{T}}$ (Eq. (S24), with $H_{\text{T}} = 30\,\text{mm}$, the sedimentation-column height used as representative of all samples), plotted against the measured spacing $d^*$ for all 20 samples (Table S1). Marker shape indicates the container (squares: cuvette; circles: capillary) and colour denotes the sample series (grey: SGL capillary; red and green: DBL capillary; blue: DBL cuvette). Across the set $\sigma_{\text{eff}}$ follows the Donnan $\sqrt{n_0}$ scaling, made parameter-free in main-text Fig. 5c. The scatter at a given spacing reflects the small residual finite-$\kappa d$ deviation from strict effective-charge saturation across stocks; it is not a single-sample $\sigma_{\text{eff}}(d)$ response (Section S8 "Local versus global"). The systematic vertical offset between SGL and DBL is set by the reconstruction prefactor $h\sqrt{n_0/\phi_{\text{prep}}}$ (stacking-unit thickness $h$ and prepared volume fraction), not by a physical charge difference between architectures; it is removed by the normalisation of main-text Fig. 5c (Section S8).

**Restoring slope at mechanical balance.** The restoring slope is the spacing derivative of the net pressure,

$$k(d) = \frac{\partial \Pi_{\text{osm}}}{\partial d} - \frac{\partial \Pi_{\text{ext}}}{\partial d}, \tag{S21}$$

At the equilibrium spacing $d^*$ the net pressure $\Pi_{\text{osm}} - \Pi_{\text{ext}}$ vanishes by the balance condition, so the first derivative of the net free energy is zero there and no net force acts at the minimum; $k(d^*)$ is the next-order quantity, the curvature of the free-energy well at the minimum, which measures the restoring pressure that develops as the spacing is displaced away from $d^*$, not anything evaluated at the force-free minimum itself. Both terms carry an explicit $d$-dependence that is set entirely by the empirical swelling relation $\phi_{\text{prep}}(d) \approx A\, d^{-\alpha}$ (Sec. S2), with $\alpha$ the swelling exponent reported independently by Trigueiro Neto *et al.*[4]. The buoyancy expression $\Pi_{\text{ext}} = \Delta\rho\, g\, \phi_{\text{prep}}(d)(H_{\text{T}} - z)$ is linear in $\phi_{\text{prep}}$ and so scales as $\Pi_{\text{ext}} \propto d^{-\alpha}$; the quadratic Donnan form $\Pi_{\text{osm}} \propto (\sigma_{\text{eff}}\, \phi_{\text{prep}})^2$, with $\sigma_{\text{eff}}$ approximately $d$-independent in the asymptotic

large-$d$ regime, scales as $\Pi_{\rm osm} \propto d^{-2\alpha}$. Differentiating each at fixed $z$ gives

$$\frac{\partial \Pi_{\rm ext}}{\partial d} = -\frac{\alpha\,\Pi_{\rm ext}}{d}, \qquad \frac{\partial \Pi_{\rm osm}}{\partial d} = -\frac{2\alpha\,\Pi_{\rm osm}}{d}. \tag{S22}$$

At mechanical balance $\Pi_{\rm osm} = \Pi_{\rm ext} \equiv \Pi$, the two terms in Eq. (S22) combine and $\sigma_{\rm eff}$ drops out of the result. Substituting in turn the explicit buoyancy form $\Pi = \Delta\rho\, g\, \phi_{\rm prep}(d)\,(H_{\rm T} - z)$ with $\phi_{\rm prep}(d) = A\, d^{-\alpha}$ gives the restoring slope in two equivalent forms,

$$|k(d,z)| \;\approx\; \frac{\alpha\,\Pi}{d} \;=\; \frac{\alpha\,\Delta\rho\, g\, A\,(H_{\rm T} - z)}{d^{\,\alpha+1}} \;\propto\; d^{-(\alpha+1)}. \tag{S23}$$

Equation (S23) is the central result of the section and contains no adjustable parameters: the value of $\sigma_{\rm eff}$ does not enter; the only inputs are $\alpha$ from the independently measured swelling relation, the Bragg spacing $d^*$, the in-cell height $z$, and the buoyancy prefactor set by sample geometry and the nanosheet–medium density contrast. The relation is written as an approximate equality because it assumes $\sigma_{\rm eff}$ exactly $d$-independent: a residual $d$-dependence of $\sigma_{\rm eff}$ would add a term $2\Pi\,\partial(\ln\sigma_{\rm eff})/\partial d$ to the derivative, small in the asymptotic-saturation regime but not strictly zero, and the swelling relation $\phi_{\rm prep} \approx A\,d^{-\alpha}$ is itself an empirical fit. (In the explicit second form, $A$ is the dimensionless-fraction coefficient, i.e. the vol% coefficient of Sec. S2 divided by 100, with $d$, $z$ and $H_{\rm T}$ in metres, so that $|k|$ comes out in Pa m$^{-1}$; dividing by $10^9$ expresses it per nanometre.) The first equality emphasises the structural form, an explicit $1/d$ factor multiplying the local osmotic pressure; the second equality makes the full $d$-dependence explicit: because $\Pi$ itself decreases with $d$ through $\phi_{\rm prep} \propto d^{-\alpha}$, the explicit $1/d$ factor combines with this implicit dependence to give $|k(d,z)| \propto d^{-(\alpha+1)}$. With $\alpha \approx 1.16$ (cuvette geometry; Sec. S2) the predicted scaling is $|k(d,z)| \propto d^{-2.16}$, steeper than the bare $1/d$. Equivalently, substituting $\Pi = \Pi_{\rm osm}$ at balance gives the same scaling, since the two pressures are equal there.

Equation (S23) is also $z$-dependent through the buoyancy term $(H_{\rm T} - z)$: in the pre-sedimentation, homogeneous suspension $d$ is the same everywhere, but $\Pi_{\rm ext}$ varies linearly with height from $\Pi_{\rm max} = \Delta\rho\, g\, \phi_{\rm prep}\, H_{\rm T}$ at the cell bottom to a vanishing buoyancy contribution at the meniscus, so the restoring slope inherits the same linear profile. We emphasise that this is the buoyancy contribution to the local pressure balance; the meniscus boundary layer itself (region of order the capillary length below $z = H_{\rm T}$) carries additional pressure terms from surface tension, capillary forces and platelet anchoring at the air–water interface that are not captured by the gravitational picture. These boundary contributions are not modelled here; their existence and qualitative magnitude in the same family of charged-platelet suspensions are documented in earlier work[22, 23, 3, 24, 25]. The analysis applies to the bulk of the sample column where the buoyancy term sets the local pressure; the meniscus end of the in-cell range in main-text Fig. 5a should be read as the buoyancy contribution becoming small near $z \to H_{\rm T}$, not as a claim that $|k|$ literally vanishes at the air–water interface. The maximum buoyancy value is $|k|_{\rm max} = \alpha\,\Delta\rho\, g\, A\, H_{\rm T}/d^{\alpha+1} = \alpha\,\Pi_{\rm max}/d$, realised at $z = 0$; a representative mid-cell value uses $\Pi \approx \Pi_{\rm max}/2$. The smallness of $|k(d,z)|$ is then traceable to the smallness of the gravitational prefactor $\alpha\,\Delta\rho\, g\, A\,(H_{\rm T} - z)$ itself: with the dilute platelet loading characteristic of the osmotically swollen regime ($\phi_{\rm prep}$ at the sub-percent level), the modest cell height ($H_{\rm T}$ of order centimetres) and the moderate density contrast of hydrated silicate nanosheets in water, $\Pi_{\rm max}$ is at most a few pascal across the experimental range (Fig. 4c) and the prefactor is correspondingly small in absolute terms. Combined with $d$ in the hundred-nanometre range in the denominator, this fixes $|k(d,z)|$ at sub-pascal-per-nanometre values: with $\Pi \approx 2$ Pa and $d^* \approx 200$ nm, Eq. (S23) gives $|k| \approx 0.012$ Pa nm$^{-1}$, and even at the cell-bottom maximum $\Pi_{\rm max} \approx 4$ Pa it remains below 0.025 Pa nm$^{-1}$. The weak-coupling regime of the swollen lamellar state is in this sense an intrinsic

consequence of large-spacing colloidal lamellar matter held by weak buoyancy in a finite cell, not a finely tuned property.

The buoyancy term $\Delta\rho\, g\, \phi_{\rm prep}$ entering Eq. (S23) is not a reconstructed quantity. It is set directly by the sample geometry and composition: $\Delta\rho$ is the nanosheet–medium density contrast, $g$ the gravitational acceleration and $\phi_{\rm prep}$ the prepared volume fraction of each stock, all independently known. Its value is also directly visible in the sedimentation data: the slope of $\Pi_{\rm ext}(z)$ in the supernatant region of Fig. 4c (and Fig. S13) equals $\Delta\rho\, g\, \phi(z)$, which in the homogeneous pre-sedimentation state reduces to $\Delta\rho\, g\, \phi_{\rm prep}$, and the intercept at the cell bottom is the buoyancy pressure $\Pi_{\rm max} = \Delta\rho\, g\, \phi_{\rm prep} H_{\rm T}$ marked in that panel. The few-pascal pressure scale, and the resulting sub-pascal-per-nanometre value of $|k(d, z)|$, therefore rest on independently anchored inputs, namely $\alpha$ from Ref. [4], the buoyancy term from sample preparation and the sedimentation gradient, and the Bragg spacing from scattering. They do not rest on the reconstruction of any internal model quantity. In particular, $\sigma_{\rm eff}$ drops out of Eq. (S23) at mechanical balance, so the conclusion that $|k|$ is small does not depend on the effective-charge reconstruction discussed below.

**Why a slope reconstructed from the Donnan balance is uninformative.** It is tempting to try to extract $k(d)$ by differentiating the reconstructed osmotic-pressure curve directly. This route is uninformative, because the Donnan balance pins only the *product* $\sigma_{\rm eff}\, \phi_{\rm prep}$ to the reservoir: from $\Pi_{\rm osm} \propto (\sigma_{\rm eff}\, \phi_{\rm prep})^2$ and $\Pi_{\rm osm} = \Pi_{\rm ext}$, one obtains $\sigma_{\rm eff}\, \phi_{\rm prep} \propto \sqrt{n_0\, \Pi_{\rm ext}}$. The split of the spacing dependence between $\sigma_{\rm eff}$ and $\phi_{\rm prep}$ is not constrained by the data, and a slope obtained by differentiating the pinned product is identically zero by construction. Eq. (S23) circumvents this redundancy by inserting the externally known $\phi_{\rm prep}(d)$ relation from Ref. [4] and treating $\sigma_{\rm eff}$ as the saturated, $d$-independent quantity expected in the asymptotic regime[19, 20, 21]. Both inputs are independently grounded, and only their combination is needed to evaluate Eq. (S23).

**Reconstruction of the effective charge.** The effective charge parameter $\sigma_{\rm eff}$ is reconstructed from the Donnan inversion using the sample-specific gravitational loading: each stock sits at its own $\Pi_{\rm ext} = \Delta\rho\, g\, \phi_{\rm prep}(H_{\rm T} - z)$, set by its own $\phi_{\rm prep}$ and cell geometry, which we evaluate at mid-cell ($z = H_{\rm T}/2$). Substituting this into the Donnan inversion gives

$$\sigma_{\rm eff} \;=\; \frac{2eh}{\sqrt{k_{\rm B}T}}\,\frac{\sqrt{n_0\, \Pi_{\rm ext}}}{\phi_{\rm prep}} \;=\; \frac{2eh}{\sqrt{k_{\rm B}T}}\sqrt{\frac{n_0\, \Delta\rho\, g\, H_{\rm T}}{2}} \cdot \phi_{\rm prep}^{-1/2}, \tag{S24}$$

with all quantities sample-specific. The key prediction at fixed sample preparation is the $\sqrt{n_0}$ trend between salinity clusters,

$$\sigma_{\rm eff} \;\propto\; \sqrt{n_0}, \tag{S25}$$

which is what we observe and what is rendered parameter-free in main-text Fig. 5c. The reconstructed values shown in Fig. S15 are strongly grouped by reservoir salinity $n_0$; the small within-cluster spread visible in the figure reflects the modest variation of $\phi_{\rm prep}$ across stocks at the same salinity, with $\kappa d$ of order 5–30 placing the system well inside, but not infinitely deep in, the asymptotic effective-charge saturation regime[19, 20, 21]. The within-cluster spread is not a meaningful $\sigma_{\rm eff}(d)$ relation for a single sample; it reflects the dispersion of effective charges across distinct global preparations at the same reservoir salinity (see “Local versus global” below).

**The restoring slope is a within-sample, off-equilibrium response.** The restoring slope k(d) that controls the free-energy curvature of Fig. 1e is the response of the net pressure to a spacing perturbation $\Delta d$ *away from the equilibrium* of a single sample. The sample is prepared

at some $(\phi_{\rm prep}, n_0)$ and sits at its equilibrium spacing $d^*$; a transient excursion to $d^* + \Delta d$ takes it off the curve $\Pi_{\rm osm} = \Pi_{\rm ext}$. The reservoir-set quantities ($n_0$, and through it $\sigma_{\rm eff}$ via the bulk-equilibrium Donnan balance) do not have time to follow such a fluctuation: their re-equilibration is set by ion diffusion to the reservoir across the full cell, which is slow compared with the platelet motion that defines the perturbation. $\sigma_{\rm eff}$ is therefore correctly held at its bulk-equilibrium saturation value when evaluating $\partial(\Pi_{\rm osm} - \Pi_{\rm ext})/\partial d$ at $d^*$, and Eq. (S23), $|k(d^*, z)| \approx \alpha\Pi/d^*$, is the within-sample restoring slope. It is the curvature of the free-energy well of Fig. 1e, not a slope along a curve connecting different equilibrium states.

The free-energy well is, moreover, strongly asymmetric, and $|k|$ captures only its harmonic part. The well is obtained by integrating the net pressure outward from the equilibrium spacing,

$$F(d) = -\int_{d^*}^{d} [\Pi_{\rm rep}(d') - \Pi_{\rm ext}(d')]\, {\rm d}d',$$

where $\Pi_{\rm ext} \propto d^{-\alpha}$ is the buoyancy (confinement) term and $\Pi_{\rm rep}$ the inter-platelet repulsion; $F$ has its minimum at $d^*$ with curvature $F''(d^*) = |k| = \alpha\Pi/d^*$, the quantity plotted in main-text Fig. 5a. Figure S16 shows $F$ as a function of the interlayer spacing $d$ for the two real DBL-cuvette samples that bracket our salinity range, each at its own equilibrium spacing $d^*$ (Table S1).

With the repulsion taken as the Donnan osmotic term alone, $\Pi_{\rm rep} = \Pi_{\rm osm} \propto d^{-2\alpha}$, the well is already asymmetric: against compression $\Pi_{\rm osm}$ rises faster than $\Pi_{\rm ext}$, giving a hard osmotic wall, whereas against expansion both terms decay and the well is far softer. For the zero-added-salt sample ($\alpha = 1.16$, $d^* = 254\,{\rm nm}$, $\Pi = 1.74\,{\rm Pa}$, $|k| \approx 0.008\,{\rm Pa\,nm^{-1}}$) a compression to $d^* - 0.3\,d^*$ costs about 2.5 times the equal-magnitude expansion (the osmotic curves in Fig. S16).

The classical DLVO double-layer repulsion adds to the compression side. We take it in the same form as Sec. S6, $\Pi_{\rm Elec}(d) = 8k_{\rm B}T\, e^{-d/\lambda_{\rm D}}/(\pi l_{\rm B}\lambda_{\rm D}^2)$; this is the short-ranged double-layer term, not a second independent charge contribution added to $\Pi_{\rm osm}$ (which already carries the renormalised effective charge), but the part of the interaction that the long-range Donnan balance does not capture. At the equilibrium spacing it is negligible, $\Pi_{\rm Elec}(d^*) \approx 0.03\,{\rm Pa} \ll \Pi$ (consistent with Fig. S14 and Table S1), but it grows exponentially under compression and overtakes the osmotic term at spacings well below $d^*$, in the small-spacing regime probed directly by El Rifaii *et al.*[15]. This margin rests on a measured screening length rather than an assumed one: the lowest, zero-added-salt reservoir ionic strength of the present series, $n_0 = 3 \times 10^{-4}$ M, is fixed by conductivity (Methods), so the Debye length $\lambda_{\rm D} \approx 18\,{\rm nm}$ entering $\Pi_{\rm Elec}$ is set by experiment and the negligibility of the double-layer term at $d^*$ does not depend on the value of $n_0$ being assumed. Including it (the osmotic + DLVO curves) steepens the compression wall and raises the asymmetry over the observed $|\Delta d| \lesssim 0.3\,d^*$ from 2.5 to about 5 at zero added salt. The two facts, DLVO negligible at $d^*$ and DLVO reviving under compression, are consistent because they concern different spacings. The asymmetry is in any case not a small perturbation across the observed $\Delta d/d \approx 0.2$–$0.3$. This is a second reason, in addition to $\Delta d/d$ not being small, why $\Delta\Pi \approx |k|\,\Delta d$ (Eq. (5) of the main text) is read as an order-of-magnitude estimate rather than a strict linear-response result: compression costs more, and expansion less, than the single harmonic slope $|k|$ would imply. The small value of $|k|$ reflects the weak curvature at the bottom of the well, not proximity to unbinding: the free-energy cost of separating the layers to large spacing remains well above the depth sampled by the observed distortions. The sub-pascal magnitude of $|k|$ bounds the restoring response near $d^*$ in either direction, which is what governs the slow recovery of stacking order.

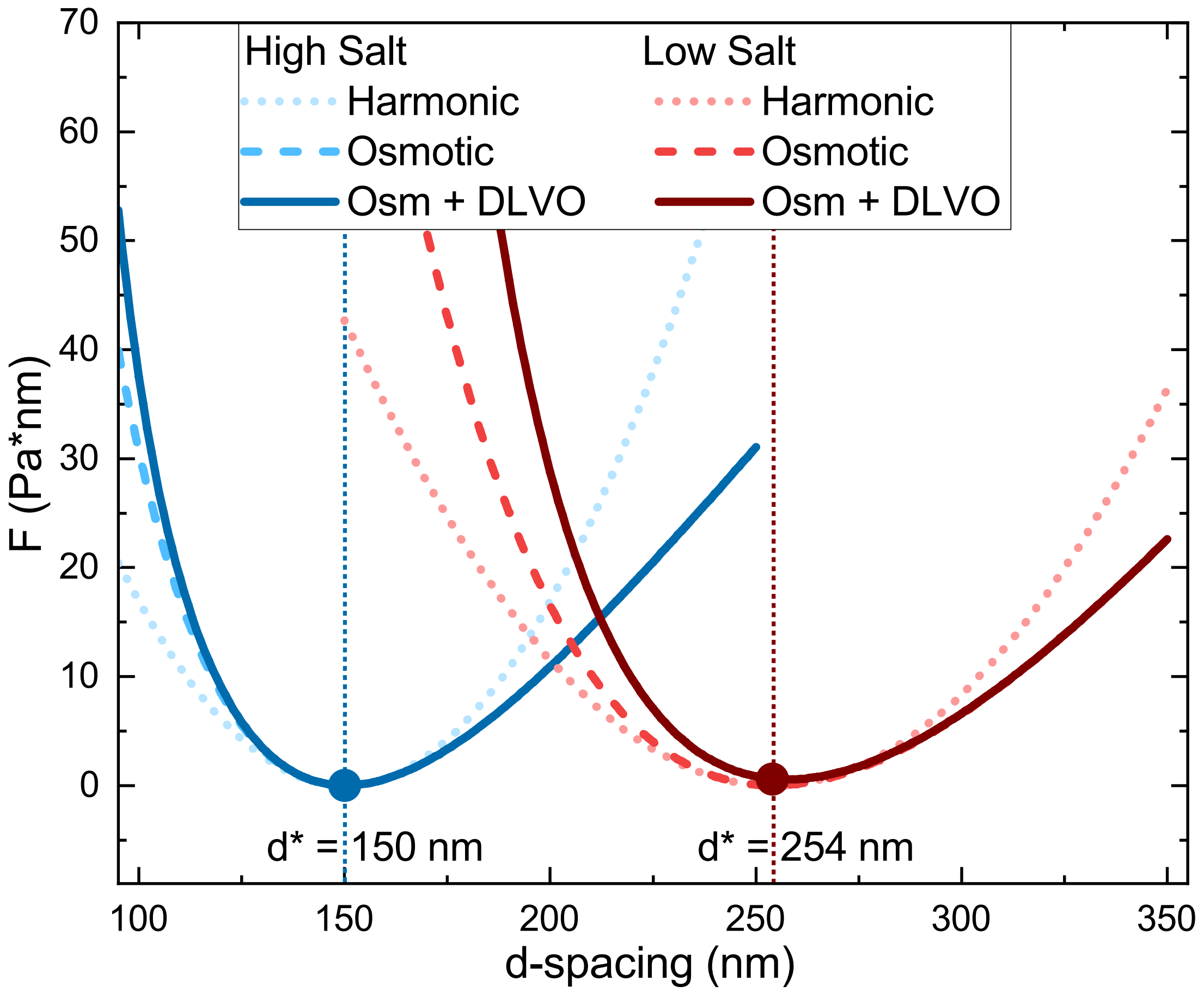


Figure S16: Free-energy well $F(d) = \int_{d^*}^{d} (\Pi_{\text{ext}} - \Pi_{\text{rep}})\, \mathrm{d}d'$ as a function of interlayer spacing $d$, for the two real DBL-cuvette samples bracketing the salinity range, each shown at its own equilibrium spacing (Table S1; $\alpha = 1.16$, $\Pi = 1.74\,\text{Pa}$). The high-salt well (left, $d^* = 150\,\text{nm}$) is shifted to smaller spacing and is intrinsically narrower than the no-salt well (right, $d^* = 254\,\text{nm}$), because $|k| = \alpha\Pi/d^*$ grows as $d^*$ falls. Dotted: the harmonic approximation $\frac{1}{2}|k|(d - d^*)^2$ for each well (red, no salt; blue, high salt), symmetric by construction; the high-salt harmonic is narrower because $|k| = \alpha\Pi/d^*$ is larger at the smaller $d^*$ (0.0135 versus $0.0079\,\text{Pa}\,\text{nm}^{-1}$). Dashed curves: the Donnan osmotic well alone ($\Pi_{\text{rep}} = \Pi_{\text{osm}} \propto d^{-2\alpha}$); solid curves: the same with the DLVO double-layer term of Sec. S6 added ($\lambda_{\text{D}} = 18\,\text{nm}$ no salt, $8\,\text{nm}$ high salt). The gap between the solid and dashed curve of each colour is the DLVO contribution: at no salt it opens on the compression side (asymmetry rising from about 2.5 to 5), while at high salt the shorter Debye length screens it and the two curves nearly coincide. The DLVO term is negligible at $d^*$ for both samples ($\Pi_{\text{Elec}}(d^*) \approx 0.03$ and $0.002\,\text{Pa}$; Fig. S14), reviving only under compression toward the small-spacing regime of El Rifaii *et al.*[15].

**Effect of added salt.** Adding salt to the reservoir shortens the Debye length ($\lambda_{\text{D}} \propto n_0^{-1/2}$, from $\approx 18\,\text{nm}$ at the lowest, zero-added-salt level $n_0 = 3 \times 10^{-4}$ M to $\approx 8\,\text{nm}$ at $1.3 \times 10^{-3}$ M) and acts on the well in two ways. First, the equilibrium spacing decreases ($d^*$ from 254 to 150 nm across the DBL-cuvette series, Table S1), so the well moves inward and, since $|k| = \alpha\Pi/d^*$, its bottom becomes intrinsically narrower at higher salt. Second, the DLVO term $\Pi_{\text{Elec}} \propto \lambda_{\text{D}}^{-2} e^{-d/\lambda_{\text{D}}}$ is screened to shorter range, so its revival under compression is pushed to smaller $d$: in Fig. S16 the osmotic and osmotic + DLVO curves nearly coincide for the high-salt sample over the displayed range, the double-layer enhancement of the compression wall having been screened away. The asymmetry of the high-salt well is therefore close to the Donnan value set by the $d^{-2\alpha}/d^{-\alpha}$ contrast alone; the electrostatic steepening of the compression wall is largest at the lowest salt, where $\lambda_{\text{D}}$ is longest. The location of these samples in the broader $(\phi_{\text{prep}}, n_0)$ control plane, and

the salinity traverse linking the two reconstructed wells, are shown in main-text Fig. 1f.

**Local versus global: nanostructure and reservoir.** What $k(d)$ acts on is the *local* interlayer spacing $d$, the gallery-to-gallery distance between adjacent platelets, which varies across the sample. SAXS and SANS do not return this single-gallery value directly: the Bragg peak *position* fixes the mean spacing averaged over the scattering volume (of order mm$^3$ for SANS and tens of $\mu$m$^3$ for SAXS, each spanning a very large number of galleries), which we report as $d^*$, while the peak *width* encodes the spread of the local $d$ about that mean, the relative distortion $\Delta d/d$ analysed above. The measured $d^*$ is therefore a volume-averaged *structural* spacing at the platelet length scale, and the local $d$ on which $k(d)$ acts fluctuates around it. The empirical $d$–versus–$\phi$ power laws of Sec. S2, also obtained from SAXS[4], are nanostructural relations of the same kind, linking the prepared volume fraction to the lamellar spacing.

The restoring slope $k(d)$ is correspondingly a local response: evaluated at the mean spacing $d^*$, it describes how stiffly the local platelet packing resists departures from its equilibrium spacing. Such departures can be set in motion by local mechanical perturbations (transient shear and registry distortions during sample loading) or by a globally imposed shift in the preparation parameters ($\phi_{\text{prep}}$, $n_0$, or mechanical history) that moves the local equilibrium $d^*$ itself; in either case the same $k(d^*)$ sets the local restoring response, and a small $|k|$ implies the local nanostructure heals softly toward its (possibly shifted) equilibrium.

The effective osmotic charge $\sigma_{\text{eff}}$, by contrast, is a global thermodynamic property set by the bulk Donnan balance with the reservoir: its value is fixed by the externally imposed $n_0$ and the prepared clay loading, not by the local Bragg spacing reached at any nanostructural location in the cell. This is the precise sense in which $\sigma_{\text{eff}}$ is correctly held at its bulk-equilibrium value when evaluating the local spacing derivative in Eq. (S23): the local nanoscale $d$ that $k(d)$ describes can fluctuate around $d^*$, but $\sigma_{\text{eff}}$ is set globally and does not track those fluctuations. The dominant observable in the reconstruction (Fig. S15) is accordingly the salinity clustering, with $\sigma_{\text{eff}} \propto \sqrt{n_0}$ between groups; the modest within-cluster spread is residual variation across different global preparations at the same reservoir salinity and is not a single-sample $\sigma_{\text{eff}}(d)$ response.

**Retention at $d^*$ is an equilibrium statement, not a dynamical one.** The population of galleries found at the pressure-selected spacing requires no restoring dynamics to explain: $d^*$ is the equilibrium of the balance $\Pi_{\text{osm}} = \Pi_{\text{ext}}$, and a gallery at equilibrium experiences no net force, so its occupation is free. Occupation is also permanent, and here the athermal condition (F1) acquires a number for the spacing coordinate. Temperature is present in the system only in its thermodynamic role, powering $\Pi_{\text{osm}}$ through the fully thermalised counterions ("What replaces temperature" below); its residual dynamical role, Brownian agitation of the sheets by the solvent bath, is bounded by equipartition within the well. The relevant stiffness is area-weighted, $K = |k|A$: with $|k| \approx 0.008$–$0.0135\,\text{Pa}\,\text{nm}^{-1}$ and gallery areas set by lateral sizes of $10\,\mu\text{m}$ to $20\,\mu\text{m}$, the bath can displace the spacing by at most $(k_{\text{B}}T/|k|A)^{1/2} \approx 1$–$2\,\text{nm}$, of order one per cent of $d^*$, whereas the measured excursions of 30 nm to 70 nm cost $\sim 10^2\,k_{\text{B}}T$ and are thermally unreachable in either direction: the bath can neither create nor heal them. The 1–2 nm figure is therefore not a thermal width the system exhibits but the ceiling on what temperature could ever do to the spacing, the quantitative licence for treating the spacing coordinate as athermal. The same area factor that renders the registry barriers athermal (tens to hundreds of $k_{\text{B}}T$; "What replaces temperature" above) thus renders the spacing equilibrium thermally sharp: the spacing exhibits stability without stiffness, surviving not because its well is deep but because it is an occupied equilibrium in a system with nothing to evict it. In particular the dimers, the minimal coloured objects (Secs. S3 and S10), persist at $d^*$ because their single gallery occupies

its equilibrium. The equilibrium is furthermore orientation-blind: $\Pi_{\text{osm}}$ depends on the gallery's ion content and the reservoir salinity, $\Pi_{\text{ext}}$ on the local height through the buoyancy integral, and neither depends on the orientation of the stacking axis; rotating a domain by its mosaicity $\beta$ changes nothing in the balance of any gallery it contains, and a registry fault in one gallery does not alter the equilibrium condition of the next. Domains at every orientation therefore answer to the same reservoir, and the salinity-set spacing scale is common to the ensemble, the microscopic basis of the common reflection wavelength invoked in Sec. S3, while the registry, which has no reservoir-anchored equilibrium value, is left to the flat landscape described above. This is visible directly in the two-dimensional SAXS (Fig. S4), where the lamellar scattering forms sharp, tilted lobes at a common radial $|q^*|$: the tilt encodes the mosaicity $\beta$, the shared radius the single, orientation-independent spacing. The measured $\Delta d/d \approx 0.2$–$0.3$ is itself principally a spread of equilibria: structurally inequivalent stacks (see "Origin of the ruggedness" below) equilibrate to slightly different spacings under the same reservoir, each occupied and force-free, with a smaller loading-induced component that heals only over the slow timescale $\tau \sim \zeta/|k|$ set by the sub-pascal slope; this static, first-kind character of the disorder is consistent with the survival of the second-order peak (Sec. S3). The statement uses only F1 (athermal), F2 (weak restoring response) and F4 (spacing set by confinement and osmotics).

**Order of magnitude of $\Delta\Pi$.** Equation (S23) gives the local restoring slope at $d^*$. For the corresponding restoring-pressure amplitude $\Delta\Pi \approx |k|\,\Delta d$ (used in Fig. 5b and in Eq. (5) of the main text), the relevant $\Delta d$ is the measured equilibrium spread $\Delta d/d \approx 0.2$–$0.3$ from the Förster fits (Sec. S3). Because the balance $\Pi_{\text{osm}} = \Pi_{\text{ext}}$ that yields the compact slope holds only at $d^*$, and the distortion takes the spacing a finite distance $\Delta d/d$ off equilibrium (where $\Pi_{\text{osm}} \neq \Pi_{\text{ext}}$), $\Delta\Pi \approx k\Delta d$ should be read as an order-of-magnitude estimate of the restoring-pressure scale opposing the measured distortion rather than as a strict linear-response calculation. The conclusion, $\Delta\Pi$ of order pascal, is robust: with $|k| \sim 10^{-2}$ Pa/nm and $\Delta d \sim 30$–$70$ nm, $\Delta\Pi \sim$ several tenths of a pascal across the series.

**Inputs to the restoring slope.** The restoring slope $|k|$ depends on three independently anchored inputs: the empirical buoyancy term (visible directly as the slope of $\Pi(z)$ in Fig. 4c and Fig. S13, with no fit), the independently measured swelling exponent $\alpha$, and the Bragg-measured $d$. Because $k(d)$ is the *derivative* of the pressure balance with respect to $d$, not the balance itself, the effective charge $\sigma_{\text{eff}}$ cancels from it algebraically at balance and does not enter the result; the $\sigma_{\text{eff}}$ reconstruction of Fig. S15 is a separate internal-consistency check on the Donnan picture.

For weak confinement, the Donnan osmotic-pressure expression implies the approximate relation $\sigma_{\text{eff}}\phi_{\text{prep}} \propto \sqrt{n_0}$. The reconstructed data follow this expected trend (main-text Fig. 5c), consistent with the osmotic pressure being set primarily by the reservoir salinity rather than by modest variations in clay concentration.

**Magnitude consistency with the asymptotic-saturation regime.** The reconstructed $\sigma_{\text{eff}}$ at sample-specific mid-cell loading (Eq. (S24)), with $\Pi_{\text{ext}} = \frac{1}{2}\Delta\rho\, g\, \phi_{\text{prep}} H_{\text{T}}$, takes the explicit form

$$\sigma_{\text{eff}} = \frac{2eh}{\sqrt{k_{\text{B}}T}} \sqrt{\frac{\Delta\rho\, g\, H_{\text{T}}}{2}} \cdot \frac{\sqrt{n_0}}{\sqrt{\phi_{\text{prep}}}} \,. \tag{S26}$$

This formula does not contain the bare structural charge $\sigma_{\text{bare}}$ as a variable, so the reconstruction *by construction* returns a $\sigma_{\text{eff}}$ that depends only on $(h, \phi_{\text{prep}}, H_{\text{T}}, n_0)$. The reconstruction's main testable output is therefore the *magnitude* of $\sigma_{\text{eff}}$: whether the reconstructed value lies in the

range expected for a strongly renormalised effective charge in the asymptotic-saturation regime, where the screening length $\lambda_D$ rather than $\sigma_{bare}$ sets the effective-charge scale.

For an isolated planar surface in the strongly-charged limit, the Trizac–Bocquet–Aubouy saturation prediction is[19, 20, 21]

$$\sigma_{eff}^{sat} \approx \frac{4\,\varepsilon\,\varepsilon_0\,k_B T\,\kappa}{e}, \tag{S27}$$

the asymptotic value reached when the bare charge exceeds the saturation level (and the effective coupling parameter $\gamma_{eff} \to 1$). At our reservoir salinities $n_0 = 3 \times 10^{-4}$ M to $1.3 \times 10^{-3}$ M, the inverse Debye length $\kappa$ runs from $0.057\,\mathrm{nm}^{-1}$ to $0.119\,\mathrm{nm}^{-1}$, and Eq. (S27) evaluates to $\sigma_{eff}^{sat} \approx 0.025\,e\,\mathrm{nm}^{-2}$ to $0.053\,e\,\mathrm{nm}^{-2}$ across the salinity series.

The reconstructed $\sigma_{eff}$ values, $0.003\,e\,\mathrm{nm}^{-2}$ to $0.010\,e\,\mathrm{nm}^{-2}$, are about a factor of 3–5 below this isolated-planar prediction. This offset should not be over-interpreted as a single geometric correction. Our $\sigma_{eff}$ is an *osmotic* (Donnan) effective charge, defined through the pressure as the osmotically active counterion charge that reproduces $\Pi_{osm}$, whereas the Trizac–Bocquet–Aubouy value is a *far-field* effective charge, defined by matching the linearised potential far from an isolated plate. These are different operational definitions of effective charge and coincide only in limiting cases; the osmotic charge counts the free, osmotically active counterions and is generally the smaller, so a factor of a few in this direction is expected without invoking any geometric mechanism. The confined-gap geometry does differ from the isolated-plate reference, but at our $\kappa d$ of order 5–30 the two double layers have decayed to a few percent of their wall value by mid-gap and barely overlap, so gap-overlap renormalisation is small here and is not the origin of the factor. A further part of the offset is simply reconstruction latitude: $\sigma_{eff}$ scales as $h\sqrt{n_0/\phi_{prep}}$ with the assumed mid-cell loading and column height $H_T$ (Eq. (S26)), so an $O(1)$ change in $\Pi_{ext}$ shifts it by a comparable factor. Both reconstructed and saturation-predicted $\sigma_{eff}$ are roughly two orders of magnitude below the bare structural charge of fluorohectorite ($\sigma_{bare} \sim 1\,e\,\mathrm{nm}^{-2}$ from the Mg-for-Si substitution stoichiometry); both scale as $\sqrt{n_0}$ with reservoir salinity. What the osmotic balance constrains is the product $\sigma_{eff}\,\phi_{prep}$, through $\Pi_{osm} \propto (\sigma_{eff}\,\phi_{prep})^2$; factorising it into a charge and a volume fraction is a matter of convention, which we fix by taking $\phi_{prep}$ from the preparation. A statement about $\sigma_{eff}$ alone therefore carries this split as a latitude: a salt- or $d$-dependence reassigned to $\phi_{prep}$ would move $\sigma_{eff}$ correspondingly while leaving the product, and the data, unchanged. The restoring slope $|k|$ is unaffected, since $\sigma_{eff}$ cancels at mechanical balance and the $\phi_{prep}$ entering $|k|$ is the independently measured buoyancy value.

**SGL versus DBL: real charge difference, prefactor-dominated reconstruction.** The two architectures carry genuinely different layer charges. Zeta-potential measurements in Methods give SGL (Li form) $\zeta = -70.9\,\mathrm{mV}$ and DBL ($NH_4$/Na form) $\zeta = -63.3\,\mathrm{mV}$, so the diffuse-layer potential at the outer DBL surfaces is ~10 % lower than at the SGL surfaces, the consequence of Cs-intercalation in the non-swelling inner DBL interlayer reducing the charge presented to the external electrolyte (Methods). This difference is real but modest, and *both* bare charges remain about two orders of magnitude above the saturation level Eq. (S27). In the asymptotic-saturation regime[19, 20, 21], any bare charge well above saturation renormalises to the same saturated $\sigma_{eff}$, so the physical prediction is that SGL and DBL saturate to similar effective charge densities despite the ~10 % bare-charge difference, which is exactly the parameter-free $\sqrt{n_0}$ collapse seen in main-text Fig. 5c.

This makes it essential to read Fig. S15 correctly. There the SGL points sit systematically *below* the DBL points, but this vertical offset is not a readout of the physical effective-charge difference; if anything it runs opposite to it, since SGL presents the higher physical charge. The

offset is fixed instead by the geometric prefactors of the reconstruction. From Eq. (S24), with $\Pi_{\rm ext} = \frac{1}{2}\Delta\rho\, g\, \phi_{\rm prep} H_{\rm T}$, the reconstructed value depends on architecture only through

$$\sigma_{\rm eff} \;\propto\; h\sqrt{\frac{n_0}{\phi_{\rm prep}}}, \tag{S28}$$

i.e. through the assumed stacking-unit thickness $h$ and the prepared volume fraction $\phi_{\rm prep}$. The two architectures differ in both: $h = 1\,\rm nm$ (SGL, single sheet) versus $h = 2\,\rm nm$ (DBL, paired unit), and the DBL stocks sit at lower $\phi_{\rm prep}$ (∼1.5 vol%) than the SGL stocks (∼3.6 vol%). At matched salinity these give a reconstructed DBL/SGL ratio of about 3, of which a factor ∼2 comes from $h$ and a factor ∼1.5 from the $\phi_{\rm prep}$ difference, both unrelated to surface charge. The reconstructed values for the two architectures consequently lie within the same $0.003\, e\,\rm nm^{-2}$ to $0.010\, e\,\rm nm^{-2}$ band, and the absolute vertical ordering in Fig. S15 should not be interpreted as a physical $\sigma_{\rm eff}$ difference; the prefactor-free statement is the Fig. 5c collapse, in which the common geometric factor is divided out and SGL and DBL fall on the same Donnan line.

This is a magnitude-level consistency check, not a strong test: the Donnan formula Eq. (S26) that we invert is not bare-charge sensitive in its structure, so the reconstruction cannot distinguish, on its own, between a genuinely saturated system and one that is not. The defensible conclusion is the weaker one, that the reconstructed $\sigma_{\rm eff}$ sits in the asymptotic-saturation range across both architectures, within an order of magnitude of the Trizac–Bocquet–Aubouy isolated-planar prediction and roughly two orders of magnitude below $\sigma_{\rm bare}$.

**Robustness with respect to the in-cell height.** The reconstruction above uses the mid-cell value ($z = H_{\rm T}/2$), which corresponds to a local $\Pi_{\rm ext} = \Pi_{\rm max}/2$. Because the actual mechanical balance can in principle be realised anywhere in the cell, we assess the sensitivity of $\sigma_{\rm eff}$ to this choice by reconstructing the same data at $z = H_{\rm T}/4$ (where $\Pi_{\rm ext} = \frac{3}{4}\Pi_{\rm max}$) and $z = 3H_{\rm T}/4$ (where $\Pi_{\rm ext} = \frac{1}{4}\Pi_{\rm max}$). Since $\sigma_{\rm eff} \propto \sqrt{H_{\rm T} - z}$, changing $z$ by these amounts rescales the reconstructed values by common multiplicative factors of about 1.22 and 0.71, respectively, shared by all data points; the salinity clustering is preserved. Eq. (S23) itself scales linearly with the local $\Pi$, so across the full in-cell range $|k(d^*, z)|$ varies from zero (meniscus) to twice the mid-cell value (cell bottom), and $\Delta\Pi_i \approx |k(d_i^*, H_{\rm T}/2)|\,\Delta d_i$ stays of order 1 Pa or below across the whole range: the central conclusion does not depend on the precise representative height.

## S9. Order-of-magnitude estimate of the mechanical energy injected during syringe loading

This section supports the physical picture proposed in the main text: the swollen lamellar state already possesses intrinsically weak restoring response together with substantial spacing disorder, while mechanical loading can further perturb the weakly coupled structure. Because the restoring slope governing interlayer registry is exceptionally small, these distortions relax only inefficiently.

Here we estimate the corresponding mechanical energy on the scale of a correlated lamellar patch of area $A$.

An upper-bound estimate can be obtained by expressing the mechanical work performed during injection as pressure work,

$$E_{\rm inj} \sim \Delta P_{\rm inj}\, \Delta V, \tag{S29}$$

where $\Delta P_{\rm inj}$ is the injection-induced pressure drop and $\Delta V$ is the displaced volume associated with the local lamellar region under consideration. For a stack patch of area $A$ and interlayer

spacing $d$,

$$\Delta V \sim Ad, \tag{S30}$$

so that

$$E_{\text{inj}} \sim \Delta P_{\text{inj}}\, A\, d. \tag{S31}$$

The injected mechanical energy per unit area is therefore

$$\frac{E_{\text{inj}}}{A} \sim \Delta P_{\text{inj}}\, d. \tag{S32}$$

For large interlayer spacings of order

$$d \sim 10^{-7}\text{–}10^{-6}\,\text{m}, \tag{S33}$$

and modest injection pressure drops in the range

$$\Delta P_{\text{inj}} \sim 10\text{–}10^{3}\,\text{Pa}, \tag{S34}$$

one obtains

$$\frac{E_{\text{inj}}}{A} \sim 10^{-6}\text{–}10^{-3}\,\text{J}\,\text{m}^{-2}. \tag{S35}$$

This estimate can be compared with the elastic work required to heal a local spacing mismatch $\Delta d$,

$$\frac{E_{\text{elastic}}}{A} \sim \Delta\Pi\, \Delta d, \tag{S36}$$

where $\Delta\Pi$ is the restoring-pressure amplitude inferred from the reconstructed pressure landscape. Using sub-Pascal restoring pressures together with representative spacing distortions

$$\Delta d \sim 10^{-8}\text{–}10^{-7}\,\text{m}, \tag{S37}$$

gives

$$\frac{E_{\text{elastic}}}{A} \sim 10^{-8}\text{–}10^{-7}\,\text{J}\,\text{m}^{-2}. \tag{S38}$$

The mechanical energy injected during syringe loading can therefore exceed the local elastic work required to heal registry distortions by one to several orders of magnitude. This supports the physical picture proposed in the main text: sample loading introduces spacing and registry disorder, while the exceptionally weak restoring response in the swollen lamellar regime is insufficient to efficiently heal these mechanically introduced distortions.

The purpose of this estimate is not to model the injection hydrodynamics quantitatively, but rather to demonstrate that even modest loading stresses are energetically large compared with the weak elastic restoring response associated with the unusually flat mechanical restoring landscape reconstructed in the present work.

## S10. Domain-size statistics and scattering weighting

As a simple statistical illustration of the finite-domain lamellar state discussed in the main text, we consider a one-dimensional weakly coupled layered system in which each interlayer link has a finite probability of losing registry. If registry-breaking events are treated as approximately independent from layer to layer, the resulting domain-size distribution becomes geometric.

The same mathematical form appears in the cluster statistics of weakly coupled one-dimensional Ising or lattice-gas models, for which exact results are available[26]. The purpose of

the present section is therefore illustrative rather than predictive: it shows how a short-stack population and a scattering-weighted stacking-domain size can naturally arise in a weakly coupled layered system.

The exact cluster-size distribution for the one-dimensional Ising (or lattice-gas) model was derived by Yilmaz and Zimmermann[27]. In their formulation, the normalised number of clusters of length $l$ per lattice site is

$$a_l = \frac{a^2}{u_1}\left(1 - \frac{a}{u_1}\right)^{l-1}, \tag{S39}$$

where $a = q/n$ is the cluster density, $u_1 = n_1/n$ is the fraction of occupied sites, $q$ is the total number of clusters, and $n$ is the total number of lattice sites.

Since $a_l$ is a cluster density rather than a probability distribution, the probability that a randomly selected cluster has length $l$ is obtained by normalising with the total cluster density $a$,

$$P(l) = \frac{a_l}{a}. \tag{S40}$$

Substitution yields the geometric distribution

$$P(l) = p(1-p)^{l-1}, \tag{S41}$$

with

$$p = \frac{a}{u_1}. \tag{S42}$$

Here $p$ represents the probability that a domain terminates at the next layer.

The corresponding mean domain size is

$$\langle l \rangle = \frac{1}{p} = \frac{u_1}{a}. \tag{S43}$$

The same distribution can be written in exponential form,

$$P(l) = p\,\exp\left[-\frac{l-1}{\xi_{\mathrm{dom}}}\right], \tag{S44}$$

with characteristic domain length

$$\xi_{\mathrm{dom}} = -\frac{1}{\ln(1-p)}. \tag{S45}$$

For $p \ll 1$, one has $\ln(1-p) \approx -p$, giving

$$\xi_{\mathrm{dom}} \approx \frac{1}{p} = \langle l \rangle. \tag{S46}$$

Thus the geometric distribution corresponds to an exponential decay of domain sizes with characteristic length $\xi_{\mathrm{dom}}$, of the same order as the number-average stack size.

In the lamellar nanosheet system studied here, the integer $l$ corresponds to the number of nanosheets within a local stacking domain, i.e. singlets ($l = 1$), doublets ($l = 2$), triplets ($l = 3$), and so forth. The geometric distribution therefore provides a simple illustrative description of a weakly coupled finite-domain state.

The observed finite-domain statistics are interpreted as reflecting the combined effects of intrinsically weak restoring response in the swollen regime together with mechanically introduced registry perturbations that relax inefficiently after sample preparation and loading. Experi-

mentally extracted spacing disorder and stacking-domain sizes for the different samples are summarised in Table S2.

**Scattering weighting** In coherent scattering experiments, the amplitude scattered by a stack of $l$ coherently ordered layers scales approximately as

$$A_l \propto l, \tag{S47}$$

since the waves scattered from individual layers add coherently. The measured intensity is therefore

$$I_l \propto |A_l|^2 \propto l^2. \tag{S48}$$

The stack-size distribution sampled by scattering is thus weighted as

$$P_{\mathrm{scat}}(l) \propto l^2 P(l). \tag{S49}$$

The corresponding scattering-weighted average stack length is

$$\langle l \rangle_{\mathrm{scat}} = \frac{\sum_l l^3 P(l)}{\sum_l l^2 P(l)}. \tag{S50}$$

This weighting strongly biases scattering measurements towards longer stacks. Consequently, the stacking-domain size extracted from SAXS or SANS reflects a conditional, scattering-weighted average over the subset of stacks that contribute to the lamellar Bragg peak, rather than the number-average over all domains. This explains why experimentally observed stacking-domain sizes are systematically larger than the typical stack size in the full population.

Conceptually, this statistical picture is consistent with the mechanical weak-coupling interpretation discussed in the main text. When the restoring response against registry defects becomes very small, extended stacking order cannot be efficiently recovered once registry distortions are introduced, and the lamellar structure may fragment into finite correlated domains. We emphasise, however, that the geometric distribution derived here is intended only as a simple weak-coupling illustration and not as a unique or exclusive model of the experimental system.

The figure below preserves the fuller conceptual and statistical summary from the main text.

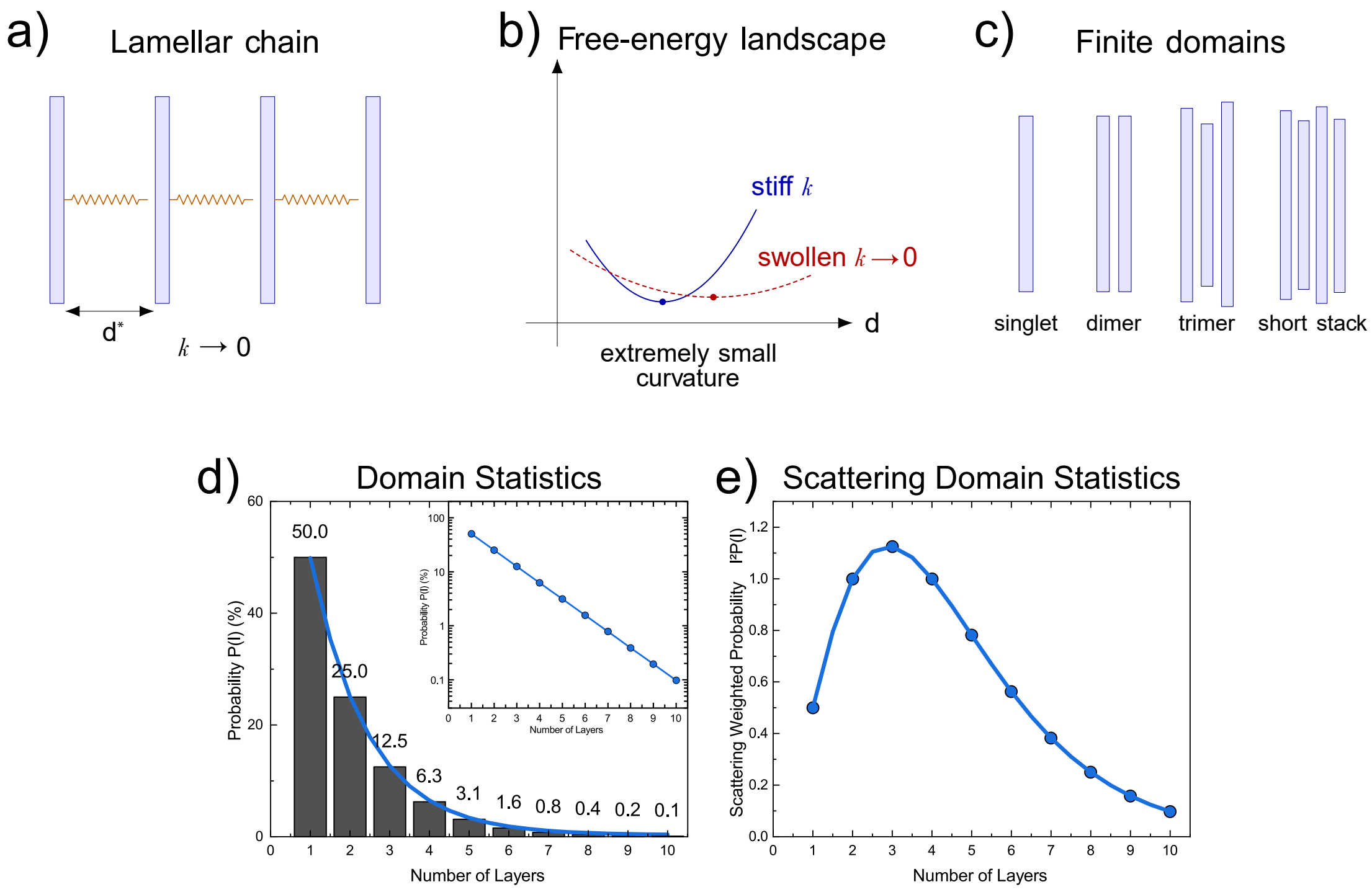


Figure S17: **Mechanical weak-coupling picture and finite-domain statistics in the swollen lamellar regime.** (a) The swollen lamellar structure can be represented schematically as a one-dimensional chain of layers coupled by a very weak effective restoring interaction. (b) In this limit the osmotic-pressure landscape becomes extremely flat, corresponding to a vanishing restoring slope. (c) Spacing mismatches are therefore only weakly penalised and cannot be efficiently healed, so the lamellar structure develops limited stacking order and finite correlated domains rather than robust long-range order. (d) Illustrative geometric distribution of stack lengths, $P(l) = p(1-p)^{l-1}$, as expected for a simple weakly coupled one-dimensional system. *Inset:* semi-logarithmic representation $\ln P(l)$ versus $l$, illustrating the equivalent exponential form of the distribution. The linear slope defines the characteristic domain length $\xi_{\mathrm{dom}}$. (e) Scattering weighting of stack sizes. Because the lamellar Bragg intensity scales as $I_{\mathrm{Bragg}} \propto l^2$, longer stacks contribute disproportionately strongly to the measured signal, so the extracted stacking-domain size reflects a scattering-weighted average. The schematic illustrates how weak interlayer coupling can give rise to finite domain sizes and a short-stack distribution in a simple one-dimensional picture.

**Finite-domain statistics and scattering weighting.** The lamellar structure may therefore be viewed as a weakly coupled finite-domain system in which extended stacking order is not efficiently maintained. The resulting stack-size distribution is dominated by short stacks, while longer stacks contribute disproportionately strongly to the scattering signal because the lamellar Bragg intensity scales approximately as $I_{\mathrm{Bragg}} \propto l^2$.

This provides a natural interpretation of the experimentally observed stacking-domain sizes. The SAXS/SANS stacking-domain sizes correspond to scattering-weighted averages over the subset of stacks contributing most strongly to the lamellar peak, rather than to number-averaged stack sizes over the full population.

As an order-of-magnitude illustration, the experimentally extracted stacking-domain sizes of 2–3 layers correspond in a simple number-average picture to

$$p \sim \frac{1}{\langle l \rangle} \approx 0.3\text{–}0.5. \tag{S51}$$

Thus the weak-coupling state may be viewed as one in which registry order has a substantial probability of terminating at each layer-to-layer step. Because the SAXS/SANS stacking-domain

sizes are scattering-weighted rather than number-averaged, this estimate should not be interpreted as a direct fitted domain-wall probability, but only as an illustrative scale for the density of registry interruptions.

A direct consequence of the geometric form $P(l) = p(1-p)^{l-1}$ is that the singlet ($l = 1$) is the single most probable species: for $p \approx 0.3$–$0.5$ the number fraction of singlets is itself $\sim 30$–$50\%$, with doublets $\sim 21$–$25\%$ and triplets $\sim 12$–$15\%$. Singlets possess no second reflecting interface and therefore contribute no interlayer interference; optically they are inactive and dilute the structural-colour signal generated by the doublets, triplets and longer stacks. The reflected colour of a freshly prepared homogeneous suspension is thus produced by the optically active minority of multilayer domains against a numerically dominant background of non-reflecting singlets. This dilution is partly reversible: because singlets are the lightest objects, they are preferentially retained in suspension while heavier multilayer stacks settle, so sedimentation (and, equivalently, centrifugation[1]) depletes the singlet population and enriches the optically active multilayer fraction. This is the microscopic origin of the brightening and colour purification observed in the sediment, quantified through the singlet-excluding conditional distribution $P_{\text{sed}}(l)$ in Sec. S11. The numerical dominance of optically inactive singlets and their partial removal by sedimentation are therefore fully consistent with both the limited mean domain size and the bright structural colour observed in the enriched sediment.

For schematic clarity, Fig. S17(c) illustrates spacing-registry defects primarily through locally expanded spacings ($d > d^*$). In practice, both locally expanded and locally compressed spacing distortions are expected within the broad shallow region around the selected spacing $d^*$. The absence of pronounced asymmetry in the experimentally observed first-order scattering peaks suggests that the distortions sampled within stacking domains remain approximately symmetric around $d^*$ over the experimentally relevant range.

This schematic is not intended as direct experimental evidence, but rather as a conceptual summary of the experimentally observed finite-domain weak-coupling state. A corresponding phenomenological weak-coupling Hamiltonian is discussed in Sec. S12.

## S11. Conditional stack distribution in the sediment

During sedimentation, isolated nanosheets remain suspended while stacks containing two or more layers sediment. The stack distribution in the sediment therefore corresponds to a conditional distribution obtained by excluding singlets,

$$P_{\text{sed}}(l) = \frac{P(l)}{1 - P(1)}, \qquad l \geq 2. \tag{S52}$$

This conditional selection enriches the sediment in multilayer stacks, i.e. in configurations that contribute more strongly to coherent optical reflection. As a consequence, the sediment exhibits enhanced structural colour compared with the original homogeneous suspension, in which singlets dominate and do not contribute efficiently to multilayer interference.

The sedimentation-induced redistribution of stack populations therefore has direct optical consequences, providing a simple physical explanation for the brighter and more saturated colours observed in the sediment compared with the initial homogeneous suspension.

In addition to depletion of singlets and enrichment of multilayer stacks, short finite-domain interruptions within the weak-coupling lamellar structure may also contribute to the observed optical response. Because optical interference depends primarily on locally preserved interlayer spacing rather than extended stacking order, fragmented multilayer domains separated by short registry disruptions (domain walls) can still contribute efficiently to structural coloration.

The enhanced blue contribution observed in the upper sediment region may therefore reflect not only population redistribution, but also the persistence of short weak-coupling lamellar domains with locally preserved spacing periodicity. Because such domains are bounded by registry disruptions (domain walls), a higher density of short domains, and hence of domain walls, near the top can contribute to the residual blue there, so it need not be ascribed to compositional fractionation alone. More generally, these observations illustrate that functional optical response can remain robust even when extended stacking order is lost and the lamellar structure fragments into finite domains.

These effects are analysed in more detail in Sec. S4 through the RGB characterisation of the sediment image.

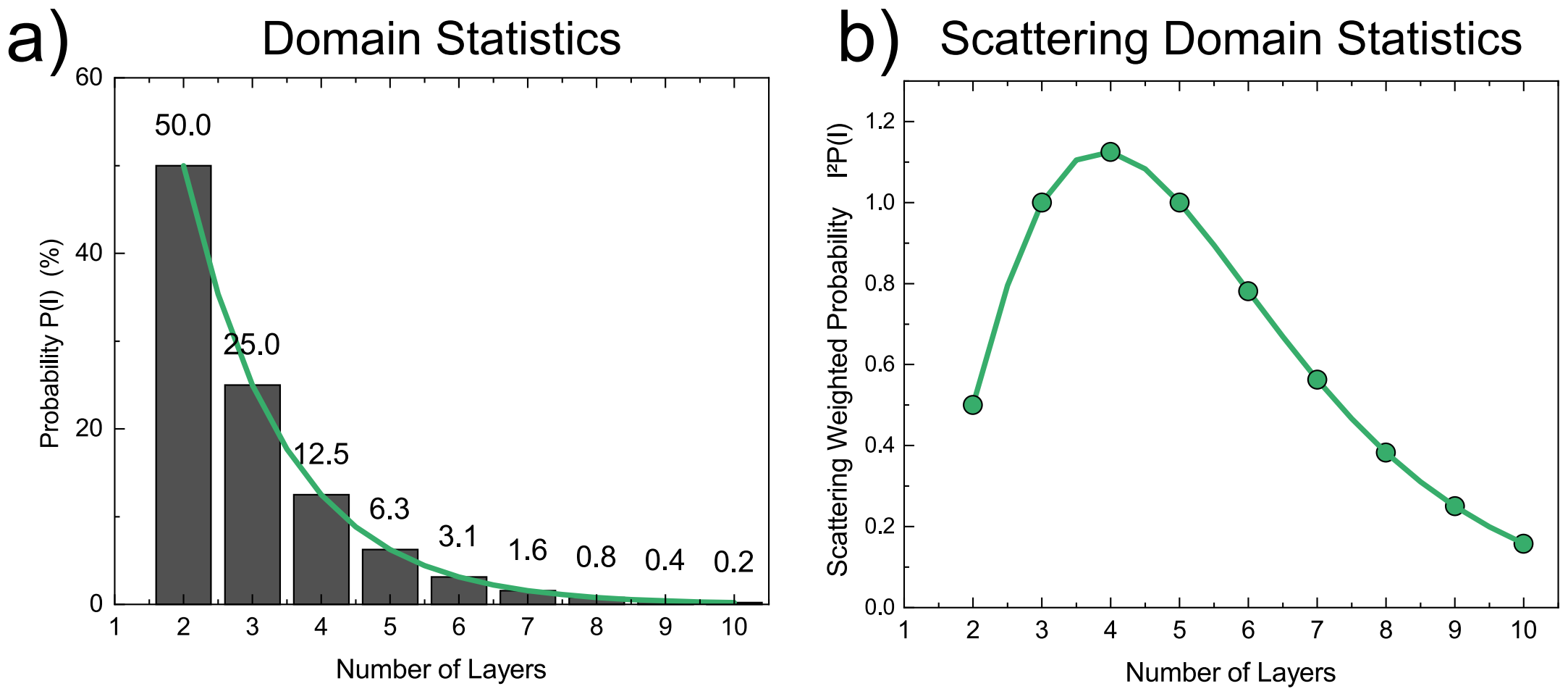


Figure S18: Conditional stack distribution in the sediment. During sedimentation, single nanosheets remain suspended while multilayer stacks settle, giving a conditional distribution that excludes singlets, $P_{\text{sed}}(l) = P(l)/[1 - P(1)]$ for $l \geq 2$. (a) The conditional distribution $P_{\text{sed}}(l)$; (b) the corresponding scattering-weighted distribution $l^2 P_{\text{sed}}(l)$, whose maximum shifts to $l \approx 4$ once singlets are removed. This redistribution enhances the relative population of coherently reflecting multilayer domains and provides a simple physical basis for the increased brightness and colour purity observed in the sediment compared with the initial homogeneous suspension.

## S12. Phenomenological weak-coupling elastic description

**What is established, and what is assumed.** Because this section connects our measurements to several established theoretical ideas, we first separate the experimental facts the description rests on (labelled **F**) from the modelling assumptions needed to make contact with those ideas (labelled **A**). The facts are properties established by the data of this work; the assumptions are invoked later in this section and are flagged again, by number, where they are used.

*Facts established for the present system.*

**F1** *Athermal.* Thermal energy $k_B T$ is negligible against the mechanical energies that displace the sheets, which are large enough to sediment under gravity; the area-weighted registry energies reach tens to hundreds of $k_B T$.

**F2** *Weak restoring response.* The reconstructed restoring slope $|k(d)|$ is sub-pascal per nanometre, so the spacing landscape is shallow.

**F3** *Quasi-one-dimensional order.* The periodic order lies along the single stacking axis, and it is this one-dimensional stacking (registry) order that is lost.

**F4** *Spacing set by confinement and osmotics.* The spacing scale $d^*$ is set primarily by the platelet volume fraction (confinement), with reservoir salinity tuning the osmotic balance at fixed loading.

**F5** *Finite, singlet-dominated domains.* The registered stacks are short, with a scattering-weighted size $N \approx 2$–3 layers and a number distribution dominated by singlets and pairs.

**F6** *Sharp spacing with lost long-range order; disorder mostly first-kind.* A sharp $d^*$ coexists with the loss of long-range stacking order. The way to read the disorder is the second-order lamellar peak, the harmonic at twice the first-order scattering angle: it is present but weak and broad, and that combination is diagnostic. Two kinds of disorder act differently on a stack. *First-kind* (static-displacement) disorder, in which each layer is scattered about a fixed ideal position, keeps the long-range lattice reference intact and merely dims the higher-order peaks; *cumulative* (second-kind, paracrystalline) disorder, in which spacing errors add up from one layer to the next, progressively broadens the higher orders and eventually washes them out. A purely cumulative $\Delta d/d$ of 0.2–0.3 would broaden the second order roughly fourfold and erase it, so the mere survival of a second-order peak shows the cumulative component is small and that the measured $\Delta d/d$ is dominated by first-kind static displacement, principally the spread of equilibrium spacings between different stacks. In short, order *within* a domain is clean; what is lost is the long-range registry *between* domains.

**F7** *DLVO/electrostatic forces negligible for our system.* In the loading and salinity regime studied here, the osmotic (Donnan) balance accounts for the data without an added electrostatic double-layer term. This is specific to our conditions and is not a general statement; it need not hold in regimes of stronger electrostatic coupling.

*Assumptions used below.*

**A1** The registry/stacking-phase degree of freedom can be modelled as a one-dimensional elastic chain carrying quenched (frozen-in) disorder, so that the Imry–Ma argument applies.

**A2** Registry-breaking defects occur essentially independently along the stack, with a per-junction probability $p$; this yields the geometric stack-size distribution and exponentially decaying stacking correlations given below.

**A3** The distribution of trap (barrier) depths has an exponential tail, $\rho(E) \sim \exp(-E/E_0)$, the standard input that produces a power-law distribution of waiting times.

**A4** The registered domains, being only two to three layers, are small enough to be treated as nearly independent objects, which justifies a mean-field trap-model description rather than an extended elastic-medium one.

**A5** The self-generated mechanical agitation $\Theta$ is approximately stationary, so that the scaled agitation $x = \Theta/E_0$ can be treated as constant and the ageing carried by the trap-depth statistics.

The finite-domain lamellar state discussed in the main text can be summarised by a minimal phenomenological description for the stacking direction. Let $d_i$ denote the local spacing between neighbouring lamellar units, $d^*$ the spacing selected by the osmotic–confinement balance (F4, F7), and $u_i = d_i - d^*$ the local spacing displacement. An effective one-dimensional weak-coupling elastic functional may then be written as

$$\mathcal{H}_{\text{eff}} = \sum_i \left[ \frac{k_i}{2} u_i^2 + \frac{J_i}{2} (u_{i+1} - u_i)^2 \right] + \sum_i V_i^{\text{prep}}. \tag{S53}$$

Here $k_i$ represents the local restoring curvature associated with the reconstructed mechanical restoring landscape, $J_i$ is an effective interlayer coupling that penalises spatial variations in spacing displacement, and $V_i^{\text{prep}}$ represents mechanically introduced distortions generated during sample preparation and injection.

The experimentally reconstructed restoring slope $k(d)$ provides an estimate of the local stiffness scale governing spacing stability in this effective description. The use of spatially varying $k_i$ and $J_i$ reflects the fact that local spacing variations generate corresponding local variations in restoring response. The swollen lamellar state is therefore more naturally viewed as a mechanically disordered weakly coupled elastic chain than as a perfectly uniform lamellar stack.

The present phenomenological description belongs broadly to the class of weakly coupled elastic-chain models related to generalised Frenkel–Kontorova and random elastic systems.[28, 29] More specifically, the osmotic-stress setup may be viewed as a fixed-stress ensemble, an idea borrowed from the statistical mechanics of athermal (granular and jammed) matter. In an ordinary thermal system it is the temperature that decides how the accessible configurations are weighted; in a granular pile, where thermal motion is negligible, Edwards proposed that the analogous role is played by the mechanical stress the system carries, with a stress-like quantity (*angoricity*) standing in for temperature[30]. Our suspension sits in the same athermal category: the external osmotic pressure $\Pi_{\text{ext}}$ holds the structure, and the configurations the system can occupy are those compatible with that imposed stress rather than those a temperature would explore.

Within this view it helps to separate two physically different things that the word "order" could refer to. The first is the *spacing*: how far apart neighbouring sheets sit, measured by the displacement $u_i = d_i - d^*$ of each gap from the mean. The second is the *registry*: how the sheets are lined up sideways relative to one another, the relative lateral offset or "stacking phase", together with the orientational misalignments (mosaicity) $\beta$ between neighbouring domains. These are independent degrees of freedom, a vertical one (separation) and a lateral one (alignment), and they need not order or disorder together. In our system the imposed pressure pins the spacing: $\Pi_{\text{ext}}$ fixes the mean gap, so the $u_i$ are held close to $d^*$ and remain sharp. The registry, by contrast, feels almost no restoring influence and is free to wander over a nearly flat landscape of equivalent configurations, in the spirit of the Edwards ensemble for athermal packings. Throughout this section, then, $u_i$ denotes the spacing, which stays ordered, while "loss of order" refers to this separate registry/stacking-phase field.

Spacing and registry may thus be regarded as conjugate to different effective ensembles, that is, governed by different bookkeeping: the spacing by the imposed stress, the registry by the flat count of equivalent lateral arrangements. This is what would rationalise the coexistence at the heart of this paper, a sharp, well-defined $d^*$ sitting alongside stacking that is broken into only a few registered layers. The two readings are complementary rather than competing. The Landau–Peierls/Imry–Ma instability discussed in the main text interprets the *loss of extended registry order*: in a marginal one-dimensional elastic chain, arbitrarily weak quenched disorder is enough to prevent long-range lateral alignment. The Edwards reading also interprets the *sharpness of $d^*$ itself*, as a value pinned by the mechanical stress balance rather than one set by averaging over thermal fluctuations.

It is worth saying plainly how this differs from the textbook model one might first reach for. The Frenkel–Kontorova model, the standard description of a chain of particles connected by

springs and sitting in a fixed periodic substrate potential, is the classic setting for objects that "want" to occupy a periodic lattice but are elastically coupled to their neighbours. The present system departs from it in three important ways. First, the restoring landscape we reconstruct experimentally is shallow and only weakly rugged, not the strong, deep periodic potential the Frenkel–Kontorova picture assumes. Second, the distortions that move the system between configurations are mechanically generated, by residual stresses, sedimentation and handling, rather than thermally activated by a heat bath. Third, the spacing displacements $u_i = d_i - d^*$ are continuous variables, the gap can take any value near $d^*$, rather than the discrete, spin-like states of the simplest lattice models. These differences are precisely why the dynamics are treated below in the athermal, trap-model language rather than with a thermally activated, strongly periodic model.

**Relation to the Mermin–Wagner, Landau–Peierls and Imry–Ma results.** It is worth being explicit about which of the classical "lower-critical-dimension" results is, and is not, the appropriate analogue for the present finite-domain state, as they differ in mechanism, symmetry and dimensionality. The *Mermin–Wagner* theorem states that thermal fluctuations forbid spontaneous breaking of a *continuous* symmetry in two dimensions with short-range interactions, so that no true long-range order survives at any finite temperature[31, 32]. The *Landau–Peierls* result is the corresponding statement for layered (smectic) order: thermal displacement fluctuations grow with system size and reduce one-dimensional stacking order to quasi-long-range, with algebraically decaying correlations, even in three dimensions[33, 34, 35]; a three-dimensional smectic is in this sense marginal, the layered cousin of a two-dimensional crystal. Both are *thermal, equilibrium* statements driven by fluctuations of a continuous elastic field. The *Imry–Ma* argument is different in origin: it concerns *quenched* random disorder coupled to such an elastic order parameter, which destroys long-range order below a critical dimension while leaving the local periodicity intact, and which sets a finite domain size through the balance between disorder energy and elastic cost[36].

The distinctions are summarised compactly below, by the disrupting agent (thermal fluctuations versus quenched disorder) and its effect:

| | **Mermin–Wagner** | **Landau–Peierls** | **Imry–Ma** |
|---|---|---|---|
| **Thermal fluctuations** | destroy long-range order of a continuous symmetry in 2D (short-range interactions) | reduce 1D (lamellar) stacking order to quasi-long-range (algebraic decay) | — |
| **Quenched disorder** | — | — | destroy long-range order while keeping the local periodicity; set a finite domain size |

Since our system is athermal (F1), the thermal-fluctuation row does not operate; the relevant entry is the lower-right one, Imry–Ma under quenched disorder, acting in the Landau–Peierls-marginal geometry (F3).

In the present system thermal energy is negligible relative to the mechanical energies governing the stiff nanosheets (see "What replaces temperature" below), so the system is *athermal* (F1), and this single fact decides which of the three results applies. Mermin–Wagner and Landau–Peierls are both *thermal* statements: they describe order being shaken apart by thermal fluctuations, and their conclusions follow from a thermal average over $\mathcal{H}_{\text{eff}}$ weighted by $k_{\text{B}}T$. When $k_{\text{B}}T$ is

effectively zero there are no such fluctuations to do the shaking, so these two mechanisms are not operative as written: in Eq. (S53) the configuration is fixed not by a thermal average but by the frozen-in (*quenched*) distortions $V_i^{\text{prep}}$ introduced during preparation, which do not average away. The result that *does* apply is the *Imry–Ma* one, because it is about exactly this situation, order disrupted by frozen-in randomness rather than by heat. It operates here in a stacking geometry that is already *Landau–Peierls-marginal* (F3), meaning the layered order is intrinsically on the edge of stability even before any disorder is added, so that only a weak push is needed to tip it; the push is supplied by the quenched disorder, and the weak elastic constants $k_i$ and $J_i$ are too soft to resist it (F2), so the registry/stacking phase loses long-range order while the sharp spacing $d^*$ survives.

A key qualification, which the scattering data force, concerns *how* the order is lost. It is not that each stack is internally scrambled; rather, the disorder *caps the domain size.* The registry-breaking is concentrated at the boundaries between domains, where the weak coupling cannot re-lock one block of registered sheets to the next, while *inside* each surviving domain the sheets stay cleanly registered, consistent with the sharp, finite-size first- and second-order lamellar peaks seen in both SAXS and SANS (Figs. S4–S5) (F6). The finite stacking-domain size $\xi$ of a few layers is then precisely the Imry–Ma length, the scale at which the disorder energy that wants to break registry first wins against the elastic cost of breaking it. The same weak restoring response shows up in two ways at once: it leaves isolated single sheets slow to re-stack, so that singlets and short stacks dominate by number (Secs. S10–S11), and it limits the registry of the stacks that do form. The scattering-weighted $N \approx 2$–$3$ is the coherent average over the multilayer fraction (F5) and sits above the singlet-dominated number average (Sec. S10). Mermin–Wagner is mentioned only as the familiar textbook statement that low-dimensional fluctuations suppress order; it is not directly applicable here, since it concerns the thermal breaking of a continuous symmetry in two dimensions, whereas our mechanism is the athermal, quenched-disorder destruction of registry in a marginally ordered three-dimensional stack.

Taken together, the experimentally observed finite-domain structure is interpreted primarily as the consequence of a weak restoring response combined with mechanically generated registry disorder, acting within a low-dimensional (lamellar) elastic chain. Put plainly: the sheets carry frozen-in misalignments from preparation, and the restoring forces are too weak to iron them out, so the stack settles into short registered domains rather than into one extended crystal.

This becomes most pronounced in the limit where both $k_i$ and $J_i$ are very small. There, a mechanically introduced spacing distortion costs almost nothing to leave in place, so the system has little incentive to heal it: finite-domain organisation is favoured and structural relaxation is slow. The resulting behaviour resembles that of weakly pinned low-dimensional elastic systems, such as the pinned charge-density waves and vortex lattices noted above, while remaining mechanical and athermal in origin rather than thermal.

We re-emphasise that Eq. (S53) is a *phenomenological* representation of the observed finite-domain regime, not a microscopic derivation from first principles. Its role is to make explicit, in a single expression, how the experimentally reconstructed restoring landscape, the effective interlayer coupling, and the observed finite-domain stacking statistics fit together.

In the extreme weak-coupling limit, where the effective restoring coefficients $k_i$ and interlayer couplings $J_i$ become very small, the model reduces to a weakly correlated one-dimensional elastic chain with finite-domain statistics. This is assumption A1. It is worth making explicit how the breaking into small domains, emphasised throughout this paper, is contained in the two equations that follow, because neither shows a "domain" variable directly. The physical content is a single statement: since the interlayer coupling is too weak to hold registry across a junction that carries enough quenched distortion, each junction between adjacent layers has some finite

probability $p$ of being a *registry-breaking defect*, a place where the lateral alignment between neighbours is lost. (assumption A2: the breaks are taken to be independent.) In one dimension this is decisive, a chain with any finite density of independent breaks is automatically cut into separate registered domains and no long-range order survives, so the breaking is an immediate consequence of $p > 0$ rather than an extra assumption.

The two equations simply count this in two complementary ways. Two layers separated by $n$ junctions stay in the same domain, and hence remain correlated, only if none of the $n$ intervening junctions is a break, which for independent junctions has probability $(1-p)^n$; the stacking correlation function is therefore exponential,

$$C(n) \sim e^{-n/\xi}, \qquad \xi = -1/\ln(1-p) \approx 1/p \;\; (p \text{ small}), \tag{S54}$$

with the correlation length $\xi$ equal to the stacking-domain size. Equivalently, asking how long an unbroken run of layers is, a domain of $l$ layers needs $l-1$ intact junctions followed by one break, which gives the geometric distribution

$$P(l) \sim p(1-p)^{l-1}, \tag{S55}$$

with mean domain size $\langle l \rangle = 1/p$. The two expressions thus carry the same physics through the same parameter: the domain size $\xi$ and the per-junction break probability $p$ are tied by $\xi \approx 1/p$, and an exponential $C(n)$, as opposed to the constant of true long-range order or the power law of quasi-long-range order, is precisely the mathematical signature of a finite-domain state. In short, the "breaking into small domains" is encoded as a non-zero break probability $p$ (equivalently a short $\xi$); these equations are that statement, not a description of an unbroken chain. The experimentally observed stacking-domain sizes of only 2–3 layers correspond to a large defect probability, $p \approx 0.3$–$0.5$, the strongly broken weak-coupling regime, and the same few-layer count appears here, through $\xi$ and $\langle l \rangle$, as it does in the scattering analysis.

**Origin of the ruggedness, and ageing.** The disorder this description encodes has an intrinsic structural origin alongside the mechanical history of cell filling. The nanosheets are laterally polydisperse, with lateral sizes expected to average roughly 10 µm to 20 µm for this material[37] and distributed broadly about that scale, so the delaminated accordions form an ensemble of structurally inequivalent stacks rather than near-identical objects. Because the equilibrium spacing is set by a local osmotic and electrostatic balance that depends on lateral extent, edge fraction and packing, accordions of different size equilibrate to slightly different spacings; the larger sheets, whose undulation amplitudes would grow with lateral size were they free, contribute the wider excursions, bounded by interlayer confinement rather than diverging. This distribution over inequivalent stacks is the rugged landscape sketched in Fig. S19: a manifold of shallow metastable states separated by small barriers within a weak global basin. It is measured by the quenched $\Delta d/d$, and is distinct from both the in-plane sheet size and the out-of-plane stacking correlation length; the polydispersity feeds the spacing distribution, not the layer count.

We further propose that the relaxation within this landscape is non-stationary; the picture set out here is motivated by the present structural observations and by analogy to disordered-elastic and granular systems, and is not established by the data presented here. The system heals the stacks under its own weak restoring tendency; because the restoring slope $k(d)$ is small this intrinsic healing is slow, but it proceeds even in the absence of gravity, for example for platelets freely suspended in water. This creep reshapes the landscape as it proceeds: in one region a barrier is lowered and healing advances, while in another a barrier emerges and healing

there arrests, and existing or newly formed barriers may in turn vanish as the process continues. External drives can accelerate the relaxation on top of this intrinsic channel, a *driven* healing: gravity through buoyant sedimentation is the route operative here, while sonication or imposed flow could serve similarly; conversely, sufficiently strong driving, as in the shear of cell filling, re-introduces distortions rather than removing them, so the same agent can heal or disorder according to its strength. A landscape that reshapes as the configuration explores it would be the microscopic basis of physical *ageing*: the response would then depend on the waiting time since preparation and slow as the system grows older, the athermal analogue of glassy and granular ageing[38, 39, 40, 41, 30]. The continued brightening of the sediment after its height has stabilised (Section S4) is consistent with such ageing, though it also carries a population-redistribution contribution (Sec. S11) that a clean dynamical test must separate from registry healing (below).

**Mapping to trap-model ageing.** This rugged landscape admits a more quantitative reading: we propose that the registry dynamics can be described by the *trap models* of glassy relaxation[38, 39, 40], from which testable predictions follow without a system-specific dynamical calculation. The mean-field character of those models, in which a region re-randomises on each escape, is appropriate here precisely because the registry domains span only two to three layers. They are too small to sustain the collective, spatially correlated displacements of an extended pinned elastic medium, so the suspension is better viewed as an ensemble of many nearly independent small objects, each hopping among its own metastable registry configurations.

**An alternative athermal-ageing scenario.** The trap-model mapping is not the only athermal route to slow dynamics of this kind, and it is worth naming the principal alternative so the two can be separated by experiment. In the *parking-lot* (adsorption–desorption) model of granular compaction[42, 43], hard objects repeatedly attempt to insert into, and occasionally leave, a one-dimensional line. Once the line is crowded a new insertion requires a rare, cooperative opening of a gap among the objects already present, so the packing fraction approaches its jammed value only logarithmically in time (as $1/\ln t$), with a terminal exponential stage; the same inverse-logarithmic compaction and ageing appear in geometric-frustration lattice models of granular packing, notably the Tetris model[44], in the Edwards-ensemble spirit invoked above. This is a genuinely athermal ageing model, driven by geometric exclusion rather than by temperature, and it admits the same Edwards/compactivity description invoked above for the spacing. In our context the parking-lot model maps naturally onto the *filling* of registry, the slow growth of the registered fraction and of the mean stack length as singlets and short stacks attempt to re-stack into a crowded column, and it carries a natural sedimentation–compaction flavour; it therefore speaks most directly to the population/re-stacking channel rather than to the within-domain registry hopping that the trap model describes.

The two scenarios make *different* predictions for the same measurements, which is what makes them separable. The parking-lot picture predicts a logarithmic approach of the registered fraction to a jammed value (compaction kinetics), whereas the trap model predicts a power-law distribution of waiting times and the age-rescaling collapse $\mathcal{F}(t/t_w^{\mu})$ of the two-time correlation. A long-baseline structural observable (the mean stack length, or the peak intensity) evolving as $1/\ln t$ would favour the parking-lot/compaction route, while a power-law, age-rescaled two-time XPCS signal would favour the trap-model route. The present, largely structural data do not distinguish the two. We develop the trap-model mapping in detail because the measured disorder reflects a rugged landscape of shallow metastable states, of the kind the trap model describes; this makes it a concrete, falsifiable example rather than a favoured outcome, and on present evidence neither route is preferred over the other. The measurements set out below can decide

between them. This near-independence of the few-layer domains is assumption A4. We identify each metastable registry configuration of a stack with a trap and the structural ruggedness with a distribution of rearrangement barriers $\rho(E)$; the lateral polydispersity established above supplies its breadth, since a structurally heterogeneous ensemble of stacks generates a correspondingly broad distribution of registry barriers. If that distribution has an exponential tail, $\rho(E) \sim \exp(-E/E_0)$ (assumption A3) (an assumption we make explicit and do not derive; it is the standard trap-model input that yields a power law), the trap model predicts a broad, power-law distribution of escape (waiting) times,

$$\psi(\tau) \sim \tau^{-(1+x)}, \qquad x = \Theta/E_0, \tag{S56}$$

where $\Theta$ is the effective agitation that drives escapes. Within this mapping, for $x < 1$ the mean waiting time diverges, ergodicity is weakly broken, and the system ages: the relaxation time grows with the age $t_w$ of the sample and two-time correlations are non-stationary, collapsing onto an age-rescaled form

$$C(t_w + t,\, t_w) = \mathcal{F}(t/t_w^{\mu})\,, \tag{S57}$$

with $\mu = 1$ (simple ageing) for the bare trap model and $\mu \neq 1$ (sub- or super-ageing) admissible for its variants.

**What replaces temperature, and why the dynamics are athermal.** Before identifying what does drive the dynamics, it is worth resolving an apparent paradox, since the osmotic interaction that sets the spacing is itself thermal. Temperature enters this system in two quite different roles, and calling the system *athermal* denies only the second. In its *thermodynamic* role, $T$ sets the scale of the interaction itself: the Donnan osmotic pressure is an entropic, ideal-gas-like pressure of the partitioned counterions, $\Pi \sim n_{\text{ion}} k_B T$, and $T$ likewise governs ion partitioning and screening. None of this is in question, and the counterions are fully thermalised. In its *dynamical* role, $T$ would be the agent that lets the *sheets* sample their configurations, kicking them between registry states; it is this role that fails here, because $k_B T$ is far too small to carry a sheet over the area-weighted barriers (below). The counterion bath thus sets a quasi-static force, not a thermal drive for the sheets: moving a ten-micron sheet across a barrier of tens to hundreds of $k_B T$ by thermal kicks alone would take a Kramers time that is effectively infinite on laboratory timescales. This is the same separation of scales that makes granular matter athermal, where grains sit in air at temperature $T$ yet do not thermally explore their packings because the relevant mechanical energies vastly exceed $k_B T$.

The mapping needs one physical substitution. In the thermal trap model a stack escapes a trap at a rate $\sim \exp(-E/T)$, with $x = T/T_g$ the scaled temperature; in our system $k_B T$ is tiny next to the mechanical energies, so heat does not drive the escapes. It is easy to see that the energies are mechanical, not thermal: the sheets are heavy enough to settle under gravity, and a single 10 µm to 20 µm sheet stores a gravitational energy across the cell of $10^2$–$10^3\, k_B T$. The energy needed to slide one sheet out of registry grows with its area in the same way (as $\Pi A \delta$), so although the restoring *slope* is sub-pascal, the absolute barrier to move a large sheet is still tens to hundreds of $k_B T$. A shallow landscape and large absolute barriers coexist simply because the sheets are large; only the smallest, most dilute sheets come close to $k_B T$, and that is the only place ordinary thermal motion could help.

What, then, supplies the agitation $\Theta$ in Eq. (S56) that lets a stack hop between registry states? Not heat, but the system's own unfinished relaxation. A stack that is out of registry is never quite at rest, because the restoring force is finite and grows the further the sheet is pushed from $d^*$ ($k(d)$ stiffens away from $d^*$); and because neighbouring sheets are coupled ($J$), every

slow rearrangement reshapes its neighbours' landscapes, opening one escape route as it closes another. Activity is thus passed from stack to stack, and it keeps regenerating for as long as any frustration remains, so $\Theta$ sustains itself and stays roughly constant in time rather than dying away (assumption A5), with no outside driving required (sedimentation, flow or sonication would add to it). This self-generated $\Theta$ is the athermal "noise temperature" of soft-glassy rheology[40], the dynamical counterpart of the stress-pinned (Edwards/angoricity) picture of the spacing introduced above[30].

Because $\Theta$ is roughly stationary, we treat the scaled agitation $x = \Theta/E_0$ as a fixed number and let the ageing come, as usual, from the spread of trap depths. Its value cannot be predicted in advance; it is the one quantity the mapping leaves to experiment, with $x < 1$ the condition for ageing. Whether $\Theta$ is truly constant or drifts slowly can itself be read off from the ageing exponent $\mu$: $\mu = 1$ signals a stationary $\Theta$ (simple ageing), $\mu \neq 1$ a slowly drifting one.

**Distinct from coarsening.** This is trap-like relaxation among pinned, finite domains, not domain growth. The registry length is capped at the finite Imry–Ma value $\xi$ of a few layers by the quenched disorder, so there is no growing length $L(t) \sim t^{1/z}$; what ages is the rearrangement dynamics at essentially fixed $\xi$. Framing the slow evolution as ordinary phase ordering would mis-state the mechanism.

**Naming the relaxation processes, and their thermal analogues.** Because there is no thermal bath here, we name each relaxation process by what it physically is and note the term used for its analogue in thermal colloidal dynamics, where the reader will have met these scalings. A sheet displaced from its registry is not at rest awaiting a kick: the finite restoring slope exerts a deterministic force $|k|\,\Delta d$ that drives it back, so an isolated out-of-registry stack undergoes overdamped, force-driven glide, $\zeta\,\dot{\Delta d} = -|k|\,\Delta d$, with a single-object time $\tau_0 \sim \zeta/|k|$ that is long only because $|k|$ is sub-pascal, not because motion is thermally rare. The dynamics are therefore deterministic relaxations of the registry/displacement field, not Brownian excursions, and the $q$-scaling of the measured relaxation time reports *how* that field relaxes. *Mobility-limited (local) relaxation* ($\tau \sim q^{-2}$): each region glides down its own restoring gradient quasi-independently, the rate set by the drag $\zeta$, with no propagation between regions; this is the scaling termed *diffusive* in thermal colloidal scattering, although no Brownian diffusion is involved. *Front-propagated relaxation* ($\tau \sim q^{-1}$): because the sheets are impenetrable and elastically coupled, one sheet's glide drives its neighbour through contact, so a rearrangement propagates along the stack at a finite speed $c$ set by the coupling, $\tau \sim 1/(cq)$; this is the scaling termed *ballistic*, and associated with compressed-exponential decays, in that literature, but it is force-driven propagation of a relaxation event rather than the free flight of a particle. Which of the two is seen depends on whether the elastic coupling or the local drag controls how a displacement spreads.

*Burst-and-stall dynamics* describe the time structure, a separate axis from the $q$-scaling above. The landscape is rugged and the sheets are coupled and impenetrable, so a gliding sheet does not relax freely: it advances until blocked by a neighbour, or stalls when a neighbour's own rearrangement shifts the minimum it was approaching, resuming only when the blocking configuration clears. The collective relaxation therefore proceeds as bursts of motion, in which a region unblocks and shunts, separated by stalls in which a stack waits on the gating neighbour or on stress to redistribute, with the broadly distributed waiting times of Eq. (S56). This is the analogue of the *intermittent* ageing dynamics seen in thermal colloidal glasses and gels, but the wait is not for a thermal activation: it is for the rest of the system to deliver the mechanical rearrangement that unblocks a given stack, and the agitation $\Theta$ above is precisely the rate at which that self-generated activity is supplied.

**Signatures.** Equations (S56)–(S57) are testable by waiting-time-resolved X-ray photon correlation spectroscopy, but the relevant observables are *ensemble-statistical*, not the rearrangements of any single stack. A coherent beam illuminates a very large number of accordions whose bursts are uncorrelated in time, so the time-averaged $g_2(t)$ is a smooth decay and the discrete steps of an individual stack are neither resolved nor expected to be. If the registry dynamics are of this kind, their intermittent, ageing character would survive the ensemble average in three quantities, none of which requires resolving a single stack. First, the two-time correlation $G(t_w, t)$ would be non-stationary, with the relaxation time growing with the age $t_w$ and the data collapsing onto the age-rescaled form $\mathcal{F}(t/t_w^{\mu})$ of Eq. (S57)[45]. Second, the temporal fluctuations of the instantaneous correlation, the dynamic susceptibility $\chi(t)$, are enhanced and peaked when the dynamics are intermittent and spatially heterogeneous[46], giving a direct ensemble measure of the burstiness even though no single burst is resolved. Third, the *shape* of the averaged decay is itself diagnostic (below). One practical condition follows: because $\chi(t)$ scales inversely with the number of independently fluctuating regions in the illuminated volume, the coherent scattering volume must be kept small, through a tightly focused beam and a thin or dilute cell, so that the heterogeneity is not itself averaged away; this is compatible with the density-matched or strongly confined geometry needed to suppress sedimentation (below). That XPCS can resolve such 1D-confined dynamics is established: it has measured one-dimensional nanoparticle motion in the channels of mesoporous silica, via the heterodyne mixing of mobile-scatterer and quasi-static-matrix scattering[47], a configuration close to ours. Our proposed regime differs in being the high-occupancy, athermal, periodically ordered limit, collective no-passing relaxation of a registry rather than the dilute, thermal, structureless single-particle diffusion measured there, which to our knowledge has not previously been accessed by XPCS. Such intermittent, ageing dynamics are well established by X-ray photon correlation spectroscopy and multispeckle light scattering in colloidal gels and glasses[48, 49, 46, 45, 50] and in synthetic-clay (Laponite) suspensions[51, 52, 53]. There, however, they are generally driven by internal-stress release or by attractive (Wigner-glass) arrest in low-aspect-ratio particles[48, 54], whereas here they are predicted to arise from the repulsive, rugged registry/displacement landscape of large, weakly coupled sheets. A separate, stress-release scenario makes a sharper but distinct prediction, not a generic trap-model result: if the self-generated agitation is dominated by elastic stress release, the decays would be compressed-exponential and ballistic ($C \sim \exp\left[-(t/\tau)^{\beta}\right]$, $\beta > 1$, with $\tau \sim q^{-1}$) rather than diffusive ($\tau \sim q^{-2}$), the established signature of stress-driven ageing in other arrested soft materials[48, 46]; this ballistic form is a property of the stress-release mechanism and should be tested separately from the trap-model power law. A further distinction is dimensional: those systems age as three-dimensional, amorphous arrested networks, whereas the present system would age within *one-dimensional*, Landau–Peierls-marginal periodic (lamellar) order. Athermal ageing of a low-dimensional *periodic* order sits between the well-studied thermal ageing of low-dimensional spin and coarsening models and the athermal ageing of amorphous granular and jammed matter. Periodic order pinned by quenched disorder has, of course, been studied extensively in other settings, notably vortex lattices, charge-density waves and related pinned elastic media[29]. What is distinctive here is the realisation of this physics in an *athermal, colloidal lamellar* system, in which the periodic order is fragmented to a few layers and the dynamics are driven by self-generated mechanical noise rather than by temperature. The present system would offer an experimental realisation of this regime. A clean test must separate two channels that both slow the sediment's evolution: the sedimentation-driven redistribution of the stack population (depletion of singlets, enrichment of multilayers; Secs. S4, S11), which is a transport process and not the registry ageing considered here, and the within-domain registry

healing to which Eqs. (S56)–(S57) apply. The photon-correlation measurement should therefore be made in a configuration that suppresses sedimentation transport, for example density-matched or strongly confined, or with the stack-size distribution tracked independently, so that any age-rescaling collapse reflects registry dynamics rather than population sorting. With that caveat, the same intermittency should also be observable in long-baseline sedimentation experiments that follow an internal structural observable after the sediment height has stabilised (main text).

**How the proposed measurement would turn assumptions into facts.** It is worth stating plainly, in the language of the lists at the start of this section, what such an experiment would settle. The facts F1–F7 fix the *state* of the system: they are structural and energetic, measured at rest, and not one of them is a dynamical measurement. Everything dynamical here, the ageing, the power-law waiting times, the exponent $\mu$, the intermittency, rests instead on the assumptions A1–A5 and is for that reason presented as proposed rather than established. A waiting-time-resolved XPCS measurement is precisely what would move these claims from the assumption list to the fact list, and it does so observable by observable.

Concretely: detecting a non-stationary two-time correlation that collapses onto the age-rescaled form $\mathcal{F}(t/t_w^\mu)$ would establish as a fact that the registry dynamics age, the central dynamical claim that is at present only proposed. The measured exponent $\mu$ would then test A5 directly, $\mu = 1$ confirming an effectively stationary self-generated agitation $\Theta$ and $\mu \neq 1$ showing that it drifts, which would falsify A5 as written. A broad, power-law distribution of waiting times, equivalently the specific shape of the averaged decay, would bear on A3, the assumed exponential tail of the barrier distribution, since that tail is what produces the power law; a consistent measurement would promote A3 from a stated input to an evidenced one, though, as with any such fit, it would support rather than uniquely prove the microscopic form. A peak in the dynamic susceptibility $\chi(t)$ would establish the intermittent, burst-and-stall character as a measured fact rather than an inference. And the wavevector scaling would fix the mechanism of an individual relaxation event: $\tau \sim q^{-1}$ would establish front-propagated (stress-release) relaxation and $\tau \sim q^{-2}$ mobility-limited relaxation, settling which of the two pictures distinguished above applies.

Two honest limits remain. First, a positive result would establish the *phenomenology*, the ageing, its exponent, the intermittency and the propagation mechanism, as fact, and would make the trap-model mapping (A1, A4) strongly evidenced; but trap models are only one of several frameworks that can generate ageing and power laws, so the mapping would be supported rather than proven unique. Second, the structural assumption A2 (independent registry-breaking defects) is tested not by XPCS but by the scattering line shapes already in hand, since XPCS speaks to the dynamics and not to the defect statistics. With these caveats, the measurement converts the dynamical part of this section from a set of proposals contingent on A1–A5 into measured facts, and is the single most informative next step.

a)

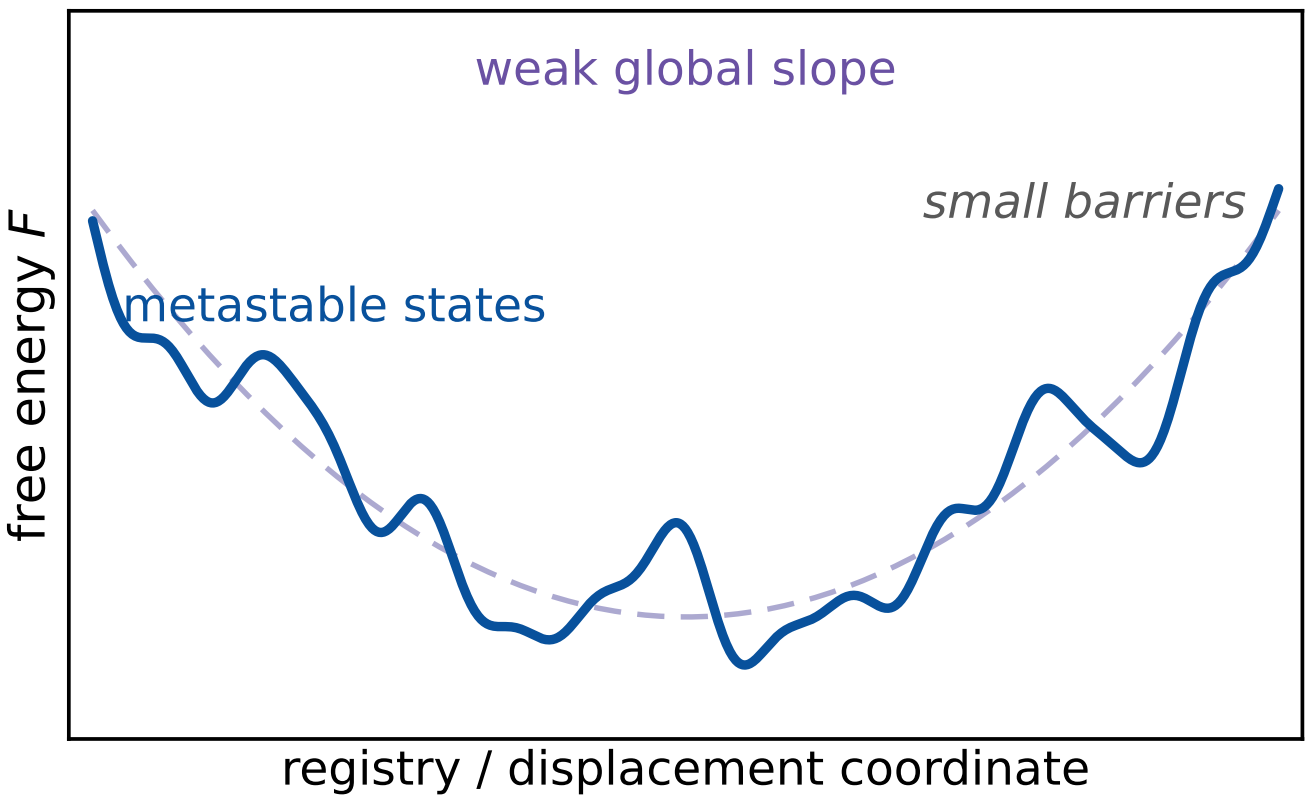


b)

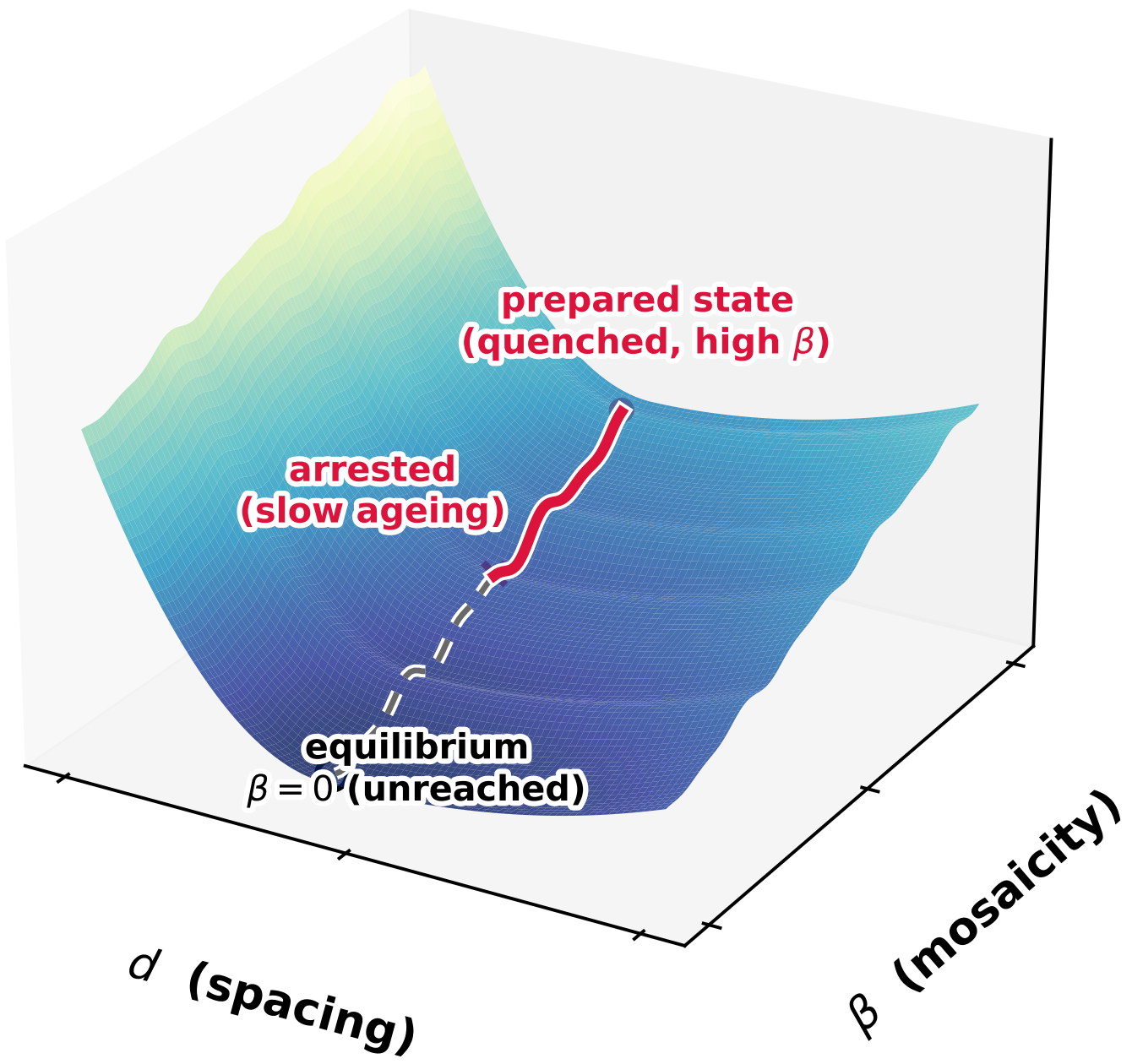


Figure S19: Schematic origin of the rugged stacking landscape. **a)** One-dimensional cut: shallow metastable states (local minima) separated by small barriers within a weak global basin (dashed), the sub-pascal restoring envelope. The horizontal axis is the collective registry/displacement coordinate of the whole stack, not the position of a particle. **b)** Proposed two-dimensional origin of the ruggedness, drawn over the interlayer spacing $d$ and the interdomain mosaicity $\beta$. Along $d$ the free energy is a single, sharply defined osmotic well, steep against compression (small $d$) and soft against expansion (large $d$), consistent with the well-defined, salinity-selected spacing. Equilibrium, corresponding to full registry, lies at $\beta = 0$; the free energy rises only weakly with mosaicity, and along $\beta$ the surface is finely corrugated into shallow, competing minima, because in the weak-coupling regime misaligning neighbouring domains costs little energy. Sample preparation quenches the system into a high-$\beta$, multidomain state (marked); because the restoring slope along $\beta$ is sub-pascal, it then descends the corrugated mosaicity direction toward $\beta = 0$ only very slowly, the athermal ageing, and remains arrested part-way (marked), far from full registry, on experimental timescales. The ruggedness that underlies this slow recovery therefore lives in the mosaicity direction, not in the spacing, which stays sharply defined throughout. Neither panel carries a calibrated scale: the ruggedness is inferred from the quenched $\Delta d/d$ and the slow, non-stationary (ageing) relaxation, and both surfaces are drawn schematically rather than computed. The relaxation dynamics within such a landscape are described in the text.

## S13. Outlook: from swollen suspensions to stiff composites

A practical motivation for characterising the finite-domain state so carefully is the prospect of converting these suspensions into solid, functional materials[55] while retaining the spacing-selected optical response. The central structural result, that the colour is governed by the well-defined mean interlayer spacing and survives as an incoherent sum over many domains of only two to three registered layers (Sec. S3), has a direct consequence for processing: the optical function does not require, and is not improved by, long-range stacking order. This decouples the property of interest from the structural quantity that is hardest to establish and maintain.

The envisaged route is to disperse the swollen nanosheets in a transparent, low-viscosity precursor and then solidify the precursor in place, for example by photopolymerisation (UV curing) or by chemical or thermal gelation, into a stiff matrix, ideally a sustainable one such as a hydrogel or a biopolymer. For the structural colour to carry over, two materials conditions must be met. First, the matrix must preserve the swollen interlayer spacing through the solidification step, rather than collapsing it as simple drying would. Second, the cured matrix must be optically transparent and present a sufficient refractive-index contrast to the embedded sheets for the spacing-set interference to remain visible; here the relevant scatterers are the double-layer (DBL) clay nanosheets, although nanosheets of other chemistries with intrinsically higher polarisability might reach the required contrast without recourse to the double-layer architecture.

Two features of the present system make this route physically reasonable rather than merely aspirational. (i) *Timescale separation.* Recovering long-range stacking order, re-establishing correlation across many layers once the stacks have been fragmented, is slow, because the restoring response is weak (main text; Secs. S9, S12). The metastable two-to-three-layer domain structure therefore persists far longer than the time required to disperse, shape and set a composite, so the spacing that generates the colour is still present when the matrix solidifies. One does not need to outrun a fast structural relaxation; on the processing timescale the structure is effectively quenched. (ii) *Short registry as an enabling feature.* It is advantageous, not merely tolerable, that the registry extends over only a few layers. A function set by the local spacing is compatible with arresting the structure in a rigid host, because solidification suppresses any further evolution of the stacking and none is needed. Had the optical (or other) function instead demanded long, registered stacks, the act of stiffening the matrix, which freezes the configuration and prevents the slow ordering from ever completing, would be self-defeating, and stiff functional composites would be exceedingly difficult to obtain. The weak-coupling, finite-domain character of the suspension is thus not only what we measure but also what would make a solid, spacing-controlled optical material practical to manufacture.

This argument is specific to colour, and the contrast with other target properties is instructive. Structural colour is set by the *local* spacing and is therefore insensitive to how much registry has re-formed at the instant of setting: it can be locked in at essentially any waiting time. Mechanical reinforcement and gas-barrier performance are different, because they depend not on the local spacing but on the *extent* of ordered, co-facial stacking, the tortuous-path length, the degree of sheet alignment, and the density of misaligned or defective domains[56, 57]. That structural order recovers only slowly after processing fragments it, so for these properties the achievable composite performance would depend on the waiting time between dispersion (or any shear-intensive processing step) and solidification: set the matrix early and a poorly aligned, fragmented network is frozen in; allow the microstructure to recover first and a better-templated composite results. For colour the slow relaxation is therefore not merely harmless but actively wanted; it guarantees that the spacing-bearing structure survives processing. For the stacking-extent-dependent functions it is an asset of a different kind: rather than fixing a single outcome,

it supplies a processing knob, since the attainable performance can be selected through the waiting time between processing and setting. In both regimes the weak-coupling, slow-recovering character of the suspension thus works for the material designer rather than against.

One qualifier prevents this from being over-read. In the *suspension*, stacking order is intrinsically capped at the few-layer domain scale by the weak registry-restoring coupling (main text); waiting allows the structure to recover *towards* that finite intrinsic state, not to grow extended stacks beyond it. The dried or cured state is a separate regime, in which polymer ultraconfinement can template much longer one-dimensional Bragg order than exists in the parent suspension[57]. The waiting-time dependence is therefore best understood not as growing unlimited stacking order in the fluid, but as controlling how faithfully the recovered suspension microstructure is captured at the moment the matrix is set.

We emphasise that this is a conceptual outlook. The present work characterises the suspension state; transferring the spacing-selected response into a specific cured matrix remains to be demonstrated, and the open questions are concrete, whether a given cure chemistry preserves the swollen spacing, what refractive-index contrast is achievable with sustainable transparent matrices, and whether spacing-controlled properties beyond colour (such as composite mechanical response[56] or gas-barrier performance[57]) transfer under the same decoupling. These define a clear experimental programme that the finite-domain picture established here is intended to support.

The transfer out of water is itself the central physical challenge, because the swollen spacing and its restoring slope are properties of the aqueous Donnan and electrostatic balance, not of the sheets alone. A curable precursor typically has a much lower dielectric constant than water, which shortens the Debye length and rescales the double-layer and osmotic pressures that set $d^*$ and $|k|$; the counterion solvation and the reservoir condition that fix the Donnan equilibrium may likewise not survive the exchange. Two outcomes are then possible in principle: the swollen spacing remains, at least kinetically, favoured in the new medium and is preserved, or the altered balance drives the structure toward collapse or restacking during infiltration. Because the restoring slope is weak, the structure re-equilibrates only slowly even in a low-viscosity precursor, so the practical question is one of competing timescales, whether infiltration and the onset of cure, which raises the viscosity and ultimately freezes the configuration, outrun the relaxation in the new medium. Identifying which curable, ideally sustainable, matrices preserve the aqueous spacing through this exchange, and with sufficient optical contrast, is the concrete next step that the present characterisation is intended to enable.

## S14. Data Tables

Table S1 compiles the calculated Debye lengths together with the estimated van der Waals and electrostatic interaction pressures (Secs. S5–S6); the fluctuation-induced Helfrich pressure is estimated separately in Sec. S7. Table S2 summarises the Förster-model fit parameters used in the discussion of spacing disorder and finite stacking-domain size in Secs. S3 and S10. Table S3 provides an overview of the sample conditions, measured spacings, geometries, and experimental techniques used throughout the manuscript and supplementary analysis.

| $\phi_{\text{prep}}$ | Container | $n_0$ (M) | $d^*$ (nm) | $\lambda_D$ (nm) | $\Pi_{\text{vdW}}$ (Pa) | $\Pi_{\text{Elec}}$ (Pa) | $\sigma_{\text{eff}}$ ($e$/nm$^2$) | Layer |
|---|---|---|---|---|---|---|---|---|
| 0.004 | Capillary | $3.0 \times 10^{-4}$ | 162 | 18 | – | – | $3.21 \times 10^{-3}$ | SGL |
| 0.004 | Capillary | $4.5 \times 10^{-4}$ | 150 | 14 | – | – | $3.94 \times 10^{-3}$ | SGL |
| 0.004 | Capillary | $6.0 \times 10^{-4}$ | 140 | 12 | – | – | $4.55 \times 10^{-3}$ | SGL |
| 0.004 | Capillary | $9.0 \times 10^{-4}$ | 135 | 10 | – | – | $5.57 \times 10^{-3}$ | SGL |
| 0.004 | Capillary | $1.3 \times 10^{-3}$ | 130 | 8 | – | – | $6.69 \times 10^{-3}$ | SGL |
| 0.0068 | Capillary | $3.0 \times 10^{-4}$ | 288 | 18 | $7.01 \times 10^{-7}$ | $3.58 \times 10^{-3}$ | $4.93 \times 10^{-3}$ | DBL |
| 0.0068 | Capillary | $4.5 \times 10^{-4}$ | 272 | 14 | $9.88 \times 10^{-7}$ | $4.10 \times 10^{-4}$ | $6.04 \times 10^{-3}$ | DBL |
| 0.0068 | Capillary | $6.0 \times 10^{-4}$ | 235 | 12 | $2.37 \times 10^{-6}$ | $5.70 \times 10^{-4}$ | $6.97 \times 10^{-3}$ | DBL |
| 0.0068 | Capillary | $9.0 \times 10^{-4}$ | 212 | 10 | $4.41 \times 10^{-6}$ | $1.14 \times 10^{-4}$ | $8.54 \times 10^{-3}$ | DBL |
| 0.0068 | Capillary | $1.3 \times 10^{-3}$ | 195 | 8 | $7.28 \times 10^{-6}$ | $1.15 \times 10^{-5}$ | $1.03 \times 10^{-2}$ | DBL |
| 0.0074 | Cuvette | $3.0 \times 10^{-4}$ | 254 | 18 | $1.49 \times 10^{-6}$ | $2.49 \times 10^{-2}$ | $4.73 \times 10^{-3}$ | DBL |
| 0.0074 | Cuvette | $4.5 \times 10^{-4}$ | 235 | 14 | $2.37 \times 10^{-6}$ | $5.43 \times 10^{-3}$ | $5.79 \times 10^{-3}$ | DBL |
| 0.0074 | Cuvette | $6.0 \times 10^{-4}$ | 206 | 12 | $5.23 \times 10^{-6}$ | $5.92 \times 10^{-3}$ | $6.68 \times 10^{-3}$ | DBL |
| 0.0074 | Cuvette | $9.0 \times 10^{-4}$ | 170 | 10 | $1.66 \times 10^{-5}$ | $7.42 \times 10^{-3}$ | $8.19 \times 10^{-3}$ | DBL |
| 0.0074 | Cuvette | $1.3 \times 10^{-3}$ | 150 | 8 | $3.51 \times 10^{-5}$ | $3.86 \times 10^{-3}$ | $9.84 \times 10^{-3}$ | DBL |
| 0.0082 | Capillary | $3.0 \times 10^{-4}$ | 227 | 18 | $2.92 \times 10^{-6}$ | $1.16 \times 10^{-1}$ | $4.49 \times 10^{-3}$ | DBL |
| 0.0082 | Capillary | $4.5 \times 10^{-4}$ | 214 | 14 | $4.16 \times 10^{-6}$ | $2.35 \times 10^{-2}$ | $5.50 \times 10^{-3}$ | DBL |
| 0.0082 | Capillary | $6.0 \times 10^{-4}$ | 196 | 12 | $7.06 \times 10^{-6}$ | $1.33 \times 10^{-2}$ | $6.35 \times 10^{-3}$ | DBL |
| 0.0082 | Capillary | $9.0 \times 10^{-4}$ | 175 | 10 | $1.39 \times 10^{-5}$ | $4.53 \times 10^{-3}$ | $7.78 \times 10^{-3}$ | DBL |
| 0.0082 | Capillary | $1.3 \times 10^{-3}$ | 163 | 8 | $2.13 \times 10^{-5}$ | $8.13 \times 10^{-3}$ | $9.35 \times 10^{-3}$ | DBL |

Table S1: Calculated Debye screening lengths, DLVO interaction pressure estimates (van der Waals and electrostatic) and reconstructed effective osmotic charge $\sigma_{\text{eff}}$ for the clay suspensions studied in this work. The DLVO entries are reported for the DBL series; for the SGL series the DLVO estimates depend on additional model details and are quoted only as $\Pi_{\text{vdW}} + \Pi_{\text{Elec}} \lesssim 0.1$ Pa in Section , well below the osmotic scale. $\sigma_{\text{eff}}$ is computed from the Donnan inversion with sample-specific mid-cell loading $\Pi_{\text{ext}} = \frac{1}{2}\Delta\rho\, g\, \phi_{\text{prep}} H_{\text{T}}$ (Eq. (S24)), taking $H_{\text{T}} = 30$ mm for both capillary and cuvette geometries as a representative sample column height (the actual values may differ between geometries and would rescale $\sigma_{\text{eff}}$ by a common factor $\sqrt{H_{\text{T}}/30 \text{ mm}}$ within each geometry, see Section S8). The $\phi_{\text{prep}}$ used in the $\sigma_{\text{eff}}$ column is the directly prepared volume fraction; the architecture- and container-specific swelling power laws of Sec.  are calibrated at zero added salt and are not used to derive $\phi_{\text{prep}}$ from the measured $d^*$, since salt screening modifies inter-platelet packing and the zero-salt swelling curves do not carry over uncorrected to the salt-added samples that constitute most of the present series. Sheet thickness $h = 1$ nm for SGL and $h = 2$ nm for DBL; $\Delta\rho = 1600\,\text{kg}\,\text{m}^{-3}$, $g = 9.81\,\text{m}\,\text{s}^{-2}$, $k_{\text{B}}T$ at room temperature.

| $\phi_{\text{prep}}$ | $n_0$ (M) | $d^*$ (nm) | $\Delta d$ (nm) | $\Delta d/d$ (%) | $d^2/\Delta d$ (nm) | $L_{\text{dom}}$ (nm) | Geometry | Layer |
|---|---|---|---|---|---|---|---|---|
| 0.004 | $3.0 \times 10^{-4}$ | 162 | 30 | 18.52 | 874.80 | 275 | Capillary | SGL |
| 0.004 | $4.5 \times 10^{-4}$ | 150 | 28.7 | 19.13 | 783.97 | 260 | Capillary | SGL |
| 0.004 | $6.0 \times 10^{-4}$ | 140 | 25 | 17.86 | 784.00 | 220 | Capillary | SGL |
| 0.004 | $9.0 \times 10^{-4}$ | 135 | 30 | 22.22 | 607.50 | 190 | Capillary | SGL |
| 0.004 | $1.3 \times 10^{-3}$ | 130 | 34.2 | 26.30 | 494.15 | 180 | Capillary | SGL |
| 0.0068 | $3.0 \times 10^{-4}$ | 288 | 62 | 21.53 | 1337.81 | 533 | Capillary | DBL |
| 0.0068 | $4.5 \times 10^{-4}$ | 272 | 53.6 | 19.71 | 1380.30 | 377 | Capillary | DBL |
| 0.0068 | $6.0 \times 10^{-4}$ | 235 | 60 | 25.53 | 920.42 | 315 | Capillary | DBL |
| 0.0068 | $9.0 \times 10^{-4}$ | 212 | 59 | 27.83 | 761.76 | 300 | Capillary | DBL |
| 0.0068 | $1.3 \times 10^{-3}$ | 195 | 70 | 35.90 | 543.21 | 302 | Capillary | DBL |
| 0.0074 | $3.0 \times 10^{-4}$ | 254 | 40 | 15.75 | 1612.90 | 379 | Cuvette | DBL |
| 0.0074 | $4.5 \times 10^{-4}$ | 235 | 45 | 19.15 | 1227.22 | 452 | Cuvette | DBL |
| 0.0074 | $6.0 \times 10^{-4}$ | 206 | 40.7 | 19.76 | 1042.65 | 322 | Cuvette | DBL |
| 0.0074 | $9.0 \times 10^{-4}$ | 170 | 35 | 20.59 | 825.71 | 290 | Cuvette | DBL |
| 0.0074 | $1.3 \times 10^{-3}$ | 150 | 25 | 16.67 | 900.00 | 220 | Cuvette | DBL |
| 0.0082 | $3.0 \times 10^{-4}$ | 227 | 49.3 | 21.72 | 1045.21 | 514 | Capillary | DBL |
| 0.0082 | $4.5 \times 10^{-4}$ | 214 | 38.5 | 18.00 | 1189.51 | 380 | Capillary | DBL |
| 0.0082 | $6.0 \times 10^{-4}$ | 196 | 39.4 | 20.10 | 975.03 | 296 | Capillary | DBL |
| 0.0082 | $9.0 \times 10^{-4}$ | 175 | 39.5 | 22.57 | 775.32 | 260 | Capillary | DBL |
| 0.0082 | $1.3 \times 10^{-3}$ | 163 | 40 | 24.54 | 664.23 | 227 | Capillary | DBL |

Table S2: Förster lamellar-model fit parameters extracted from SAXS and SANS. $\Delta d/d$ is the relative spacing disorder (the microstrain). $d^2/\Delta d$ is a length algebraically derived from the microstrain and is reported only as a fit-diagnostic; it is *not* the stacking-domain size (see Methods). $L_{\text{dom}}$ is the stacking-domain size obtained from the peak width. The stacking domain spans $L_{\text{dom}}/d^* \approx 1.3$–$2.3$ interlayer spacings, that is, two to three registered layers (a domain of $N$ layers spans $N-1$ spacings). Because coherent scattering is weighted towards the larger domains in the population, $L_{\text{dom}}$ is a scattering-weighted average rather than a number-average over all domains (Sec. S10).

| $\phi_{\text{prep}}$ | $n_0$ (M) | $d^*$ (nm) | Geometry | Layer | Technique |
|---|---|---|---|---|---|
| 0.004 | $3.0 \times 10^{-4}$ | 162 | Capillary | SGL | SAXS |
| 0.004 | $4.5 \times 10^{-4}$ | 150 | Capillary | SGL | SAXS |
| 0.004 | $6.0 \times 10^{-4}$ | 140 | Capillary | SGL | SAXS |
| 0.004 | $9.0 \times 10^{-4}$ | 135 | Capillary | SGL | SAXS |
| 0.004 | $1.3 \times 10^{-3}$ | 130 | Capillary | SGL | SAXS |
| 0.0068 | $3.0 \times 10^{-4}$ | 288 | Capillary | DBL | SAXS |
| 0.0068 | $4.5 \times 10^{-4}$ | 272 | Capillary | DBL | SAXS |
| 0.0068 | $6.0 \times 10^{-4}$ | 235 | Capillary | DBL | SAXS |
| 0.0068 | $9.0 \times 10^{-4}$ | 212 | Capillary | DBL | SAXS |
| 0.0068 | $1.3 \times 10^{-3}$ | 195 | Capillary | DBL | SAXS |
| 0.0072 | $3.0 \times 10^{-4}$ | 232 | Cuvette | DBL | Reflectivity |
| 0.0072 | $4.5 \times 10^{-4}$ | 213 | Cuvette | DBL | Reflectivity |
| 0.0072 | $6.0 \times 10^{-4}$ | 190 | Cuvette | DBL | Reflectivity |
| 0.0072 | $9.0 \times 10^{-4}$ | 161 | Cuvette | DBL | Reflectivity |
| 0.0072 | $1.3 \times 10^{-3}$ | 110 | Cuvette | DBL | Reflectivity |
| 0.0074 | $3.0 \times 10^{-4}$ | 254 | Cuvette | DBL | SANS |
| 0.0074 | $4.5 \times 10^{-4}$ | 235 | Cuvette | DBL | SANS |
| 0.0074 | $6.0 \times 10^{-4}$ | 206 | Cuvette | DBL | SANS |
| 0.0074 | $9.0 \times 10^{-4}$ | 170 | Cuvette | DBL | SANS |
| 0.0074 | $1.3 \times 10^{-3}$ | 150 | Cuvette | DBL | SANS |
| 0.0082 | $3.0 \times 10^{-4}$ | 227 | Capillary | DBL | SAXS |
| 0.0082 | $4.5 \times 10^{-4}$ | 214 | Capillary | DBL | SAXS |
| 0.0082 | $6.0 \times 10^{-4}$ | 196 | Capillary | DBL | SAXS |
| 0.0082 | $9.0 \times 10^{-4}$ | 175 | Capillary | DBL | SAXS |
| 0.0082 | $1.3 \times 10^{-3}$ | 163 | Capillary | DBL | SAXS |

Table S3: Summary of sample conditions and experimental techniques used for each dataset.



\end{document}